\documentclass[prd,twocolumn,floatfix,nofootinbib]{revtex4}
\usepackage{amssymb}
\usepackage{amsmath}
\usepackage{graphicx,subfigure,color,dcolumn,booktabs,bm}
\usepackage{longtable,lscape}
\usepackage{txfonts}
\usepackage{overpic}
\usepackage{indentfirst}
\usepackage{cases}
\usepackage{multirow}
\usepackage{ulem}
\usepackage[colorlinks,
            citecolor=blue,
            anchorcolor=red,
            menucolor=red,
            linkcolor=red,
            filecolor=red,
            runcolor=red,
            urlcolor=blue,
            frenchlinks=false]{hyperref}

\graphicspath{{Figures/}}
\allowdisplaybreaks
\begin{document}

\title{Compactness, mass spectra, and strong stability of singly heavy tetraquarks}

\author{Wen-Nian Liu$^{1,2}$}
\author{Kai-Kai Zhang$^{1}$}
\author{Duo-Jie jia$^{1,3}$}
\email{jiadj@nwnu.edu.cn(corresponding author  )}
\author{Fu-Quan Dou$^{1,4}$}
\email{doufq@nwnu.edu.cn(corresponding author  )}

\affiliation{$^1$College of Physics and Electronic
	Engineering, Northwest Normal University, Lanzhou 730070, China \\ 
$^2$Xinjiang Laboratory Phase Transitions and Microstructures in Condensed Matters,College of Physical Science and Technology, Yili Normal University, Yining, Xinjiang,835000,China\\
$^3$General Education Center,
Qinghai Institute of Technology,
Xining 810000, China \\
$^4$Gansu Provincial Research Center for Basic Disciplines of Quantum Physics, Lanzhou 730000, China }
\date{\today}

\begin{abstract}

We propose a Coulomb-like parameterization in terms of bag radius of the short-range chromo-electric interaction between heavy quarks and strange quarks within the framework of the MIT bag model for hadrons including multiquark systems and re-examine mass spectra of doubly and fully heavy baryons self-consistently and variationally. Building upon this, we apply this approach to systematically investigate the mass spectra and $S$-wave strong decay stability of singly heavy tetraquark systems, including $nQ\bar{n}\bar{n}$, $nQ\bar{s}\bar{n}$, $Qs\bar{n}\bar{n}$, $sQ\bar{s}\bar{n}$, and $sQ\bar{s}\bar{s}$ (with $Q=c,b$). Choosing the bag confinement energy  as a compactness criterion, the bag radius is shown to be close to the confinement uplimit of the radius, $R_c = 5.615\,\text{GeV}^{-1}$, for the compact singly charmed systems,and for these systems some high-spin states are unlikely to form compact structures, while several others do exhibit potential for compactnessNotably,  computation indicates that the state $T_{nc\bar{s}\bar{n}}(0^+,2.925)$  emerges as a compact tetraquark with relatively large strong decay width, a plausible candidate of the observed  tetrquark $T_{c\bar{s}0}^{a}(2900)$.

\end{abstract}

\maketitle
\date{\today}

\section{Introduction}

In addition to conventional baryons and mesons, QCD allows any color-neutral hadronic state to exist, including multiquark states (tetraquarks and pentaquarks) \cite{Gell-Mann:1964ewy,Zweig:1964ruk}. Since the first experimental report of the hidden-charm tetraquark candidate $X(3872)$ \cite{Belle:2003nnu} in 2003, subsequent experiments have revealed many more tetraquark candidates. The most representative ones include the first charged tetraquark $Z_c(3900)$ \cite{BESIII:2013ris}, the first open-heavy-flavor tetraquark candidate $T_{cc}(3875)$ \cite{LHCb:2021auc}, as well as the pentaquark $P_c$ states \cite{PhysRevLett.115.072001, LHCb:2019kea, LHCb:2020jpq, LHCb:2021chn, LHCb:2022ogu}. Meanwhile, theoretical discussions on the internal structures of these states have arisen, focusing mainly on two distinct pictures: the molecular configuration and the compact multiquark configuration \cite{Cheng:2026cgo,
	Shen:2025jmy,
	LIU20251,CHEN20161,BRAMBILLA20201,RevModPhys.90.015004,Lu:2020cns,Tiwari:2021tmz,Liu:2021rtn,Karliner:2020dta,Faustov:2020qfm,Chen:2022asf,Liu:2019zoy,Chen:2016qju,Karliner:2021xnq,Wang:2019nvm,Zou:2021sha,Deng:2022vkv,Liu:2024uxn}.

Recently, new breakthroughs have been achieved in the experimental study of singly-heavy tetraquark states. In 2023, the LHCb Collaboration observed two $J^P = 0^+$ resonant structures, denoted as $T_{c\bar{s}0}^{a}(2900)^{0}$ and $T_{c\bar{s}0}^{a}(2900)^{++}$ \cite{LHCb:2022sfr,LHCb:2022lzp}, in the $D_s\pi$ invariant mass spectra of the decays $B^0 \rightarrow D^0 D_s^+ \pi^-$ and $B^+ \rightarrow D^- D_s^+ \pi^+$. The mass of $T_{c\bar{s}0}^{a}(2900)^{0}$ is $M = 2892\pm 14\pm 15\,\mathrm{MeV}$ and its width is $\Gamma = 119\pm 26\pm 13\,\mathrm{MeV}$; while the mass of $T_{c\bar{s}0}^{a}(2900)^{++}$ is $M = 2921\pm 17\pm 20\,\mathrm{MeV}$ with a width $\Gamma = 137\pm 32\pm 17\,\mathrm{MeV}$. If these two states are considered as members of the same isospin triplet, their minimal quark compositions are $cu\bar{s}\bar{d}$ and $cd\bar{s}\bar{u}$, and the common mass and width are determined to be $M = 2908\pm 11\pm 20\,\mathrm{MeV}$ and $\Gamma = 136\pm 23\pm 13\,\mathrm{MeV}$, respectively. In 2024, the LHCb Collaboration further observed two resonance states, $T_{cs0}^{*}(2870)^{0}$ and $T_{cs1}^{*}(2900)^{0}$ \cite{LHCb:2024vfz}, in the $D^+ K^-$ invariant mass spectrum of the $B^- \rightarrow D^- D^+ K^-$ decay. The former has $J^{P} = 0^+$ and the latter $J^{P} = 1^-$, with the minimal quark content $cs\bar{u}\bar{d}$; they were identified as the previously reported $X_0(2900)$ and $X_1(2900)$ \cite{LHCb:2024vfz}. More recently, the LHCb Collaboration further confirmed the existence of the $T_{cs0}^{*}(2870)^{0}$ state in the $B^- \rightarrow D^- D^0 K_S^0$ decay process \cite{LHCb:2024xyx}, with its mass measured to be $M = 2883\pm 11\pm 8\,\mathrm{MeV}$ and width $\Gamma = 87\pm 17\,\mathrm{MeV}$.

The discovery of these charmed-strange spectra has triggered abundant theoretical discussions. For the $T_{c\bar{s}0}^{a}(2900)^{0}$ and $T_{c\bar{s}0}^{a}(2900)^{++}$ states, QCD sum rules support their interpretation as compact tetraquarks \cite{doi:10.1142/S0217751X23500562, Lian:2023cgs, PhysRevD.95.114005}; triangle singularities provide a possible non-resonant explanation \cite{Ge:2022dsp, Liu:2020orv}, while chiral effective theory suggests that they could be molecular states near the $D^{*}K^{*}$ threshold \cite{PhysRevD.109.034027}. Similarly, the interpretation of the $T_{cs0}^{*}(2870)^{0}$ state mainly focuses on either a compact tetraquark \cite{He:2020jna, PhysRevD.102.094016, PhysRevD.103.074011, PhysRevD.105.054018,Wang:2020prk} or a $D^*\bar{K}^*$ molecular state \cite{PhysRevD.102.091502, Wang:2020prk, PhysRevD.102.091502, PhysRevD.104.094012, doi:10.1142/S0217751X23500562, Wang:2021lwy}. The central question of the debate is whether these states are single confined objects or hadronic molecular states.

On the nonperturbative nature of color interactions, lattice QCD studies indicate that the string breaking distance between quarks is approximately $1.1\,\text{fm}$ to $1.3\,\text{fm}$ \cite{Bulava:2019iut,Bali:2005fu,Kou:2024dml}. This scale is close to the Compton wavelength corresponding to the typical nonperturbative QCD scale (about $200\,\text{MeV}$), namely $\hbar c/(200\,\text{MeV}) \approx 1\,\text{fm}$. This observation provides important clues for understanding the molecular and compact configurations of multiquark states. For example, the authors of Ref.~\cite{Ma:2024vsi, Wu:2024zbx, Yang:2025wqo, Zheng:2025uzy, Wu:2025nbu} employed the complex scaling method (CSM) and the Gaussian expansion method (GEM) to solve the complex-scaled four-body Schr\"odinger equation, and identified possible molecular states by examining whether the root-mean-square distance between any two quarks exceeds a certain value, as well as the proportion of the color-singlet $1_c \otimes 1_c$ component. Another insight comes from the bag confinement limit given by the MIT bag model \cite{Zhang:2025dsx,Liu:2025fbe}. This limit $R_{c}=1.1\,\text{fm}$ is almost identical to the string breaking distance obtained from lattice QCD. Therefore, for a multiquark system under a single confinement (i.e., a compact state), its actual equilibrium bag radius may depend on the number and flavor of the quarks involved \cite{Zhang:2025dsx,Barnes:1982tx,Chanowitz:1982qj}. Since the bag radius strongly depends on the number of light quarks contained \cite{Liu:2025fbe}, a question worth investigating is: For singly heavy tetraquark states and flavor configurations involving strange quarks, can their bag radii approach or exceed the bag confinement limit? Do these systems have the potential to form a single confined object (i.e., a compact state)?

We propose a unified variational formulation of the bag model that incorporates chromoelectric interactions between heavy quarks. First, using this unified variational framework, we refine the results for doubly heavy baryons previously obtained with perturbative methods, and further predict the masses of fully heavy baryons. Second, we estimate the confinement energies of tetraquark systems with different numbers of light quarks. Our results show that the bag confinement energy exhibits a strong dependence on the number of light quarks; for the $cqqq$ ($q = u, d, s$) system, it is already very close to the bag confinement limit. Subsequently, taking into account both chromoelectric and chromomagnetic interactions, we calculate the mass spectra of the $nQ\bar{n}\bar{n}$, $nQ\bar{s}\bar{n}$, $\bar{Q}\bar{s}nn$, $sQ\bar{s}\bar{n}$, and $sQ\bar{s}\bar{s}$ systems, and analyze their potential to form compact structures as well as their strong stability.

The remainder of this paper is organized as follows. Section~\ref{sec:bagmodel} presents the basic framework of the MIT bag model, in which Sec.~\ref{sec:VCEI} provides the unified variational form of the chromoelectric interaction, and Sec.~\ref{sec:CL} discusses the compactness of hadrons within the bag model. Section~\ref{multi-heavy hadrons} presents the calculated mass spectra of singly heavy tetraquark states, along with a discussion of their compactness potential and strong decay properties. Finally, Sec.~\ref{sec:summary} gives a brief summary of the computational results and analyses.

\section{Model and Method}
\label{sec:bagmodel}

\subsection{Chromomagnetic Bag Model Framework}
\label{sec:CMI}

The traditional MIT bag model is based on the mean-field approximation, confining quarks within a spherically symmetric cavity. Through variational methods, the optimal radius and mass spectrum of hadrons can be obtained \cite{Jaffe:1976ig,Jaffe:1976ih,DeGrand:1975cf}, and it was also one of the first models employed to study multiquark states \cite{Barnes:1982tx,Chanowitz:1982qj}. In recent years, within the conventional bag-model framework, Refs.~\cite{Zhang:2021yul,Yan:2023lvm,mwd4-l283} have further incorporated the chromomagnetic interactions among quarks, as well as the additional binding energies between heavy quarks. The resulting mass formula for hadrons is given by:

\begin{equation}
	M\left( R\right) =\sum_{i}\omega _{i}+\frac{4}{3}\pi R^{3}B-\frac{Z_{0}}{R}%
+H_{CMI}+H_{CEI},\label{M}
\end{equation}%
here 
\begin{equation}
	\omega _{i}=\left( m_{i}^{2}+\frac{x_{i}^{2}}{R^{2}}\right) ^{1/2}.
	\label{freq}
\end{equation}%
In the mass formula, the first term on the right-hand side represents the sum of the In the mass formula, the first term on the right-hand side represents the sum of the quark kinetic energies. The second term is the volume energy of the bag, where the parameter $B$ accounts for the difference in vacuum energy between the inside and outside of the bag. The third term contains the bag constant $Z_{0}$ \cite{Jaffe:1976ig,Jaffe:1976ih,DeGrand:1975cf}. $H_{CMI}$ denotes the chromomagnetic interaction (CMI) among quarks. $H_{CEI}$ denotes the short-range interaction between heavy quarks, which is usually parametrized in the literature \cite{Zhang:2021yul,Yan:2023lvm,mwd4-l283}, but is identified in our model as the short-range chromoelectric interaction, whose specific form will be given in the next subsection. In the kinetic term, $x_{i}$ is related to the quark momentum by $p = x_{i}/R$. Since quarks are confined within a spherical region of radius $R$, their spinor wave functions $\psi$ must satisfy a linear boundary condition (e.g., $i\gamma \cdot \hat{r}\psi = \psi$ at $r=R$) \cite{Jaffe:1976ig,Jaffe:1976ih,mwd4-l283}. This leads to a logarithmic derivative condition for the spherical Bessel functions and ultimately yields the following transcendental equation:
\begin{equation}
	\tan x_{i}=\frac{x_{i}}{1-m_{i}R-\left( m_{i}^{2}R^{2}+x_{i}^{2}\right)^{1/2}}. \label{transc}
\end{equation}
For each quark, given its mass $m_{i}$ and the bag radius $R$, the corresponding $x_{i}$ can be determined. Therefore, this equation and the total mass formula Eq.~(\ref{M}) can be solved self-consistently through iteration.

The Hamiltonian of the CMI can be expressed as \cite{Zhang:2021yul,Yan:2023lvm,PhysRevD.103.074006}

\begin{equation} 
			H_{CMI}=-\sum_{i<j}^{}(\lambda_{i}\cdot \lambda_{j})(\sigma_{i} \cdot \sigma_{j})C_{ij}. 
\label{eq:eqcmi} 
\end{equation}
here, $i, j$ denote the indices for quarks or antiquarks, and $\lambda$ and $\sigma$ represent the Gell-Mann matrices and Pauli matrices, respectively. Their expressions in terms of Casimir operators correspond to the color factor and the spin factor, respectively, and are calculated as follows:

\begin{equation}
	\begin{aligned}
		\left\langle\lambda_{i}\cdot\lambda_{j}\right\rangle_{nm}=
		\sum_{\alpha=1}^{8}\mathrm{Tr}(c_{in}^{\dagger}\lambda^{\alpha}c_{im})
		\mathrm{Tr}(c_{jn}^{\dagger}\lambda^{\alpha}c_{jm}),
	\end{aligned}
	\label{eq:eq7}    
\end{equation}

\begin{equation}
	\begin{aligned}
		\left\langle\sigma_{i}\cdot\sigma_{j}\right\rangle_{xy}=\sum_{\alpha=1}^{3}
		\mathrm{Tr}(\chi _{ix}^{\dagger}\sigma ^{\alpha}\chi_{iy})\mathrm{Tr}(\chi _{jx}^{\dagger}\sigma^{\alpha}\chi _{jy}),
	\end{aligned}
	\label{eq:eq8}    
\end{equation}
Here $n$, $m$ and $x$, $y$ label the color wavefunctions and spin wavefunctions of the hadron, respectively, while $c$ and $\chi$ are the basis vectors.

In Eq.~(\ref{eq:eqcmi}), $C_{ij}$ is the coupling constant of the chromomagnetic interaction, and its expression is given as follows \cite{Zhang:2021yul}

\begin{equation}
C_{ij} = 3 \frac{\alpha_s (R)}{R^3} \bar{\mu}_i \bar{\mu}_j I_{ij},
\label{Com}
\end{equation}
here
\begin{equation}
\alpha_s (R) = \frac{0.296}{\ln \left[ 1 + (0.281 R)^{-1} \right]},
\end{equation}
the $\bar{\mu}_{i}$ denotes the quark magnetic moment, and it is calculated as \cite{Jaffe:1976ig, Jaffe:1976ih, DeGrand:1975cf}
\begin{equation}
	\begin{aligned}
		\bar{\mu}_{i} &= \frac{Q_{i}}{2} \int_{bag}\mathrm{d}^{3}r\, 
		\bar{\psi_{i}}\left(\boldsymbol{r\times\gamma}\right)\psi_{i} \\
		&= \frac{Q_{i}}{2} \int_{0}^{R}\mathrm{d}r\,r^{2} \int\mathrm{d}\Omega\
		\bar{\psi_{i}}\left(\boldsymbol{r\times\gamma}\right)\psi_{i} \\
		&= Q_{i}\frac{R}{6}\frac{4\omega_{i}R+2m_{i}R-3}{2\omega_{i}R\left(\omega_{i}R-1\right)+m_{i}R},
	\end{aligned}\label{equ:mui}
\end{equation}
 
 here, $\psi_{i}(r)$ is the spinor wave function of the quark,
 
 \begin{equation}
 	\psi_{i}(r) = N_{i} \binom{j_{0}(x_{i}r/R)U}
 	{i\frac{x_{i}}{(\omega_{i}+m_{i})R}j_{1}(x_{i}r/R)\boldsymbol{\sigma}\cdot\boldsymbol{\hat{r}}U} e^{-i\omega_{i}t}. \label{spinor}
 \end{equation}

In Eq.~(\ref{Com}), $I_{ij}$ describes the interaction between the two magnetic moment sources and is expressed as
 
\begin{equation}
I_{ij} = 1 + 2 \int_0^R \frac{dr}{r^4} \bar{\mu}_i \bar{\mu}_j.
\end{equation}

The quark masses and bag parameters in the MIT bag model are set to the values from Refs. \cite{Zhang:2021yul,Yan:2023lvm,Zhu:2023lbx}:

\begin{equation}
	\begin{Bmatrix}
		Z_{0}=1.83,    & B^{1/4}=0.145\,\mathrm{GeV}, \\
		m_{n}=0\,\mathrm{GeV},    & m_{s}=0.279\,\mathrm{GeV}, \\
		m_{c}=1.641\,\mathrm{GeV},& m_{b}=5.093\,\mathrm{GeV}.
	\end{Bmatrix}
	\label{PR}
\end{equation} 

\subsection{Variational form of the CEI}
\label{sec:VCEI}

In the mean-field description of the MIT bag model, the long-range confinement of quarks is captured by the vacuum pressure and the Casimir energy, but the short-range dynamics among quarks is not fully accounted for. This is particularly relevant for systems containing heavy quarks, where the short-range chromoelectric interaction may need to be included as an additional term beyond the mean-field approximation. In other non-perturbative approaches, such effects have been parametrized through short-range binding energies between heavy quarks to reconcile the baryon and meson spectra~\cite{Karliner:2014gca,Karliner:2017elp,Karliner:2017qjm,Karliner:2020vsi,ljt6-cv33,PhysRevD.108.054019}. It was also observed that, for a given flavor configuration, the short-range binding energy in baryons is half of that in mesons, a relation that follows directly from the ratio of their color factors. The short-range interaction between heavy quarks is additionally enhanced relative to that between light quarks, and this enhancement arises from the chromoelectric interaction (CEI).

In Refs.~\cite{Zhang:2023teh,Zhang:2021yul}, binding-energy parameters were introduced within the MIT bag model to simultaneously describe the spectra of heavy mesons and heavy baryons. We adopt a similar procedure. Using the experimental masses of the vector mesons $\Upsilon$, $B_c^*$, $J/\psi$, $B_s^*$, and $D_s^*$ together with their corresponding scalar partners as inputs, we perform a fit using only the first three terms in Eq.~(\ref{M}). This yields two sets of binding-energy parameters, listed in Table~\ref{SCEI}: $B_M^*$ for the vector mesons and $B_M$ for the scalar mesons. The resulting binding energies show only minor differences between the vector and scalar channels, indicating that the binding energy is spin-independent here. Moreover, our $B_M^*$ values are very close to those adopted in Ref.~\cite{Zhang:2021yul}, indicating that our description of the heavy-meson spectrum is consistent with previous studies.

\renewcommand{\tabcolsep}{0.1cm} \renewcommand{\arraystretch}{1.4}
\begin{table}[htbp]
	\centering
	\caption{The bag radius \(R_{M}\) (in \(\mathrm{GeV}^{-1}\)) and the short-range chromoelectric interaction energy \(B_{M}\) (in \(\mathrm{MeV}\)) of heavy-flavor vector mesons.}
	\label{SCEI}
	\begin{tabular}{cccccc}
		\bottomrule[1.3pt]\bottomrule[0.5pt]
		Mesons & $\Upsilon(\eta_{b})$ & $B^*_{c}(B_{c})$ & $J/\Psi(\eta_{c})$ &  $B_{s}^*(B_{s})$ &$D_{s}^*(D_{s})$ \\
		\hline 
		$R_{M}^*(R_{M})$&$1.8(1.6)$&  $2.8(2.5)$& $3.5(3.2)$ & $3.6(3.4)$ & $4.2(3.8)$ \\
		$B_{M}^*(B_{M})$ &-259(-263) &-201(-206) &-154(-172) &-64(-48) &-53(-43)\\
		$\bar{R}_{M}$&$1.7$&  $2.7$& $3.4$ & $3.5$ & $4.0$ \\
		$\bar{B}_{M}$ &-261 &204 &-163 &-56 &-48\\
		\bottomrule[0.5pt]\bottomrule[1.3pt]
	\end{tabular}
\end{table}

\begin{figure}[h]
	\centering
	\includegraphics[width=0.45\textwidth]{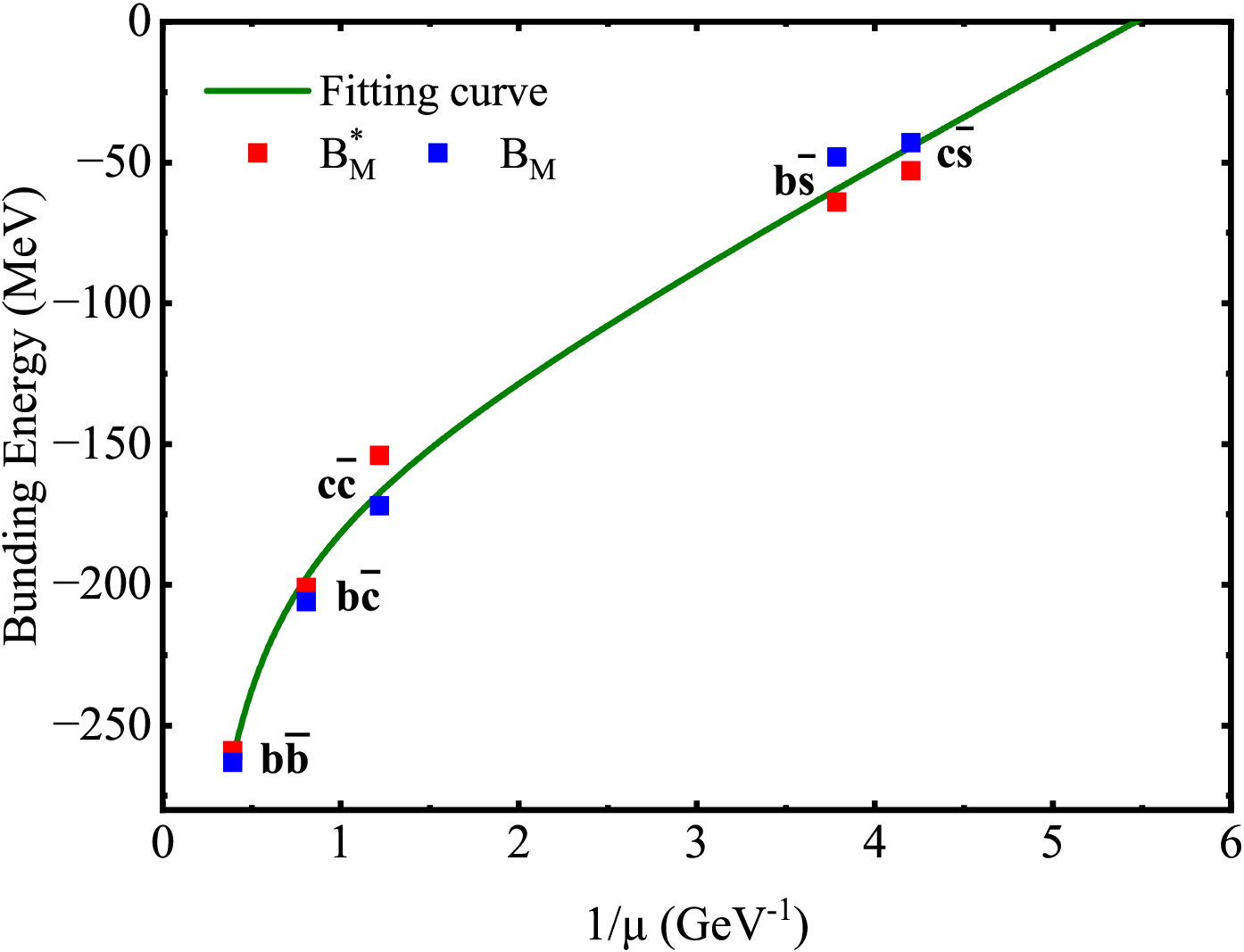}
	\caption{Red and blue points represent the binding energies $B_M^*$ and $B_M$ for vector and scalar mesons, respectively.}
	\label{fig:Bunding}
\end{figure}

We fit the correlation between the reduced quark mass $1/\mu$ and the corresponding binding energies $B_M^*$ and $B_M$ for each meson, with the quark masses taken from Eq.~(\ref{PR}), as shown in Fig.~\ref{fig:Bunding}. The fit yields the following relation with respect to the reduced mass:
\begin{equation}
	B(1/\mu) =\frac{a}{(1/\mu)}+ b(1/\mu)+c
	\label{CEI}
\end{equation}
where $a=-41$, $b=30$, and $c=-165$. In natural units ($\hbar = c = 1$), $1/\mu$ has the dimension of length. Kinematically, $1/\mu \propto r$, so the fitted form is consistent with a Cornell-type potential $V(r) = -\alpha/r + \sigma r + C$. This supports the interpretation that the binding energy between heavy quarks originates from a short-range CEI. Furthermore, the fit exhibits a clear scale dependence: in the small-$1/\mu$ region (corresponding to larger reduced masses), the Coulomb attraction dominates, as expected from the short-range CEI.

We therefore depart from the conventional practice of treating the binding energies as input parameters for baryons and multiquark systems \cite{Zhang:2021yul,Yan:2023lvm}. Inspired by the traditional quark model, we argue that the chromoelectric interaction in the bag model should contain a factor of $1/R$ associated with the bag radius, so that the mass formula possesses a unified variational parameter $R$ and allows for a unified variational determination of hadron masses. On this basis, the average short-range chromoelectric interaction strength $\bar{B}_M$ extracted from mesons, together with the corresponding average bag radius $\bar{R}_M$ (see Table~\ref{SCEI}), can be scaled to arbitrary multiquark systems through an appropriate combination of the color factor $\langle \lambda_i \cdot \lambda_j \rangle$ and the inverse bag-radius factor $1/R$. Specifically, we propose the following form for the chromoelectric interaction Hamiltonian
\begin{equation}
	H_{CEI} = -\sum_{i<j} \langle \lambda_i \cdot \lambda_j \rangle \frac{3\bar{B}_{M}\bar{R}_{M}}{16R},
	\label{BE}
\end{equation}
where $\langle \lambda_i \cdot \lambda_j \rangle$ is the color factor for an arbitrary where $\langle \lambda_i \cdot \lambda_j \rangle$ is the color factor for an arbitrary quark pair $(i,j)$, and $B_M$ and $R_M$ are the binding-energy parameter and bag radius obtained from fitting heavy-flavor vector mesons, respectively; their values are listed in Table~\ref{SCEI}~\cite{Zhang:2021yul}. The coefficient $-3/16$ originates from the color factor of the quark pair in the meson. Finally, Eq.~(\ref{BE}) can be substituted into the mass formula Eq.~(\ref{M}) for a unified variational solution.

Based on the unified variational form of Eq.~(\ref{M}), we present in Table~\ref{tab:nonchromomagnetic_doubly_heavy_baryons} the mass corrections for singly strange and doubly heavy baryons, which are within $\pm30$\,MeV~\cite{Zhang:2021yul}, and further predict the masses of fully heavy baryons. Our results are compared with those of the relativistic quark model~\cite{PhysRevD.66.014008} and lattice QCD~\cite{PhysRevD.90.094507}.

\renewcommand{\tabcolsep}{0.2cm} \renewcommand{\arraystretch}{1.35}
\begin{table}[htbp]
	\centering
	\caption{Calculated bag radii \(R_0\) (GeV\(^{-1}\)) and masses \(M_{\text{bag}}\) (GeV) for some singly heavy baryons, doubly heavy baryons, and fully heavy baryons.}
	\label{tab:nonchromomagnetic_doubly_heavy_baryons}
	\begin{tabular}{ccccccc}
		\bottomrule[1.5pt]\bottomrule[0.8pt]
		\toprule
		\multicolumn{3}{c}{ }&\multicolumn{3}{c}{Theoretical value}&\\
		\cline{4-6}
		Hadron & \( R_0 \) & \( M_{\text{bag}} \) &\cite{Zhang:2021yul}& \cite{PhysRevD.66.014008} &\cite{PhysRevD.90.094507}&$M_{exp}$ \\
		\hline
		$\Omega^*_c$&5.08&2.768&2.764&- &2.755&2.765\\
		$\Omega_c$&4.93&2.678&2.680&- &2.679&2.695\\
		$\Omega^*_b$&4.85&6.125&6.112&- &6.085&-\\
		$\Omega_b$&4.70&6.054&6.080&- &6.056&6.046\\
		\(\Xi^{++}_{cc}\),\(\Xi^{+}_{cc}\) & 4.48& 3.617 & 3.604 &3.620 &3.610&3.620\\
		\(\Xi^{*++}_{cc}\),\(\Xi^{*+}_{cc}\) & 4.59& 3.722 & 3.714 &3.727 & 3.692&-\\
		\(\Xi^{0}_{bb}\),\(\Xi^{-}_{bb}\) & 3.58& 10.298 & 10.311 &10.202 &10.143&-\\
		\(\Xi^{*0}_{bb}\),\(\Xi^{*-}_{bb}\) & 3.82& 10.401 & 10.360 &10.237 & 10.178&-\\
		\(\Omega_{cc}\) & 4.46 & 3.729 & 3.726 &3.778 & 3.738&-\\
		\(\Omega^{*}_{cc}\) & 4.65& 3.834 & 3.820 &3.872 &3.822&-\\
		\(\Omega_{bb}\) & 3.67& 10.375 & 10.408 &10.359 &10.273&-\\
		\(\Omega^{*}_{bb}\) & 3.83& 10.427 & 10.451 &10.389 & 10.308&-\\
		\(\Omega_{ccc}\) & 4.03 &4.816 & -  &- &4.796&- \\
		\(\Omega_{ccb}\) & 3.51 &8.039 & -  &- &8.007&-\\
		\(\Omega^{*}_{ccb}\) & 3.58 &8.090 &-&- &8.037&- \\
		\(\Omega_{cbb}\) & 2.86 & 11.273 &-&- &11.195&-\\
		\(\Omega^{*}_{cbb}\) & 3.07& 11.291 &-& - &11.229&- \\
		\(\Omega_{bbb}\) & 2.29 &14.414 & - &- &14.366&-\\
			\bottomrule
		\bottomrule[0.5pt]\bottomrule[1.5pt]
	\end{tabular}
\end{table}

\renewcommand{\tabcolsep}{0.2cm} \renewcommand{\arraystretch}{1.6}
\begin{table}[htbp]
		\centering
	\caption{The color factors of quark pairs in the tetraquarks.}
	\label{tab:color_factors_teraquarks}
	\begin{tabular}{ccccccc}
	\bottomrule[1.3pt]\bottomrule[0.5pt]
	    $6_c\otimes\bar{6}_c/\bar{3}_c\otimes 3_c$ & $q_{1} q_{2}$ & $q_{1} \bar{q_{3}}$ & $q_{1} \bar{q_{4}}$ &  $q_{2} \bar{q_{3}}$ & $q_{2} \bar{q_{4}}$ & $\bar{q_{3}}\bar{q_{4}}$ \\ \hline
		$\left\langle\phi_1^T,\phi_1^T\right\rangle$  &$\frac{4}{3}$&  $-\frac{10}{3}$& $-\frac{10}{3}$ & $-\frac{10}{3}$ & $-\frac{10}{3}$ & $\frac{4}{3}$ \\
		$\left\langle\phi_2^T,\phi_2^T\right\rangle$  & $-\frac{8}{3}$ &$-\frac{4}{3}$& $-\frac{4}{3}$ & $-\frac{4}{3}$ &$-\frac{4}{3}$ &$-\frac{8}{3}$ \\
		$\left\langle\phi_1^T,\phi_2^T\right\rangle$ & 0 & $-2\sqrt{2}$ & $2\sqrt{2}$ & $2\sqrt{2}$ & $-2\sqrt{2}$ & 0\\ \hline
        $8_c\otimes 8_c/1_c\otimes 1_c$ & $q_{1} \bar{q_{2}}$ & $q_{1} q_{3}$ & $q_{1} \bar{q_{4}}$ &  $ \bar{q_{2}}q_{3}$ & $\bar{q_{2}} \bar{q_{4}}$ & $q_{3}\bar{q_{4}}$ \\ \hline
        $\left\langle\phi_1^{T^{\prime}},\phi_1^{T^{\prime}}\right\rangle$  &$\frac{2}{3}$&  $-\frac{4}{3}$& $-\frac{14}{3}$ & $-\frac{14}{3}$ & $-\frac{4}{3}$ & $\frac{2}{3}$ \\
        $\left\langle\phi_2^{T^{\prime}},\phi_2^{T^{\prime}}\right\rangle$  &$-\frac{16}{3}$&  $0$& $0$ & $0$ & $0$ & $-\frac{16}{3}$ \\
        $\left\langle\phi_1^{T^{\prime}},\phi_2^{T^{\prime}}\right\rangle$  &$0$&  $\frac{4\sqrt{2}}{3}$& $-\frac{4\sqrt{2}}{3}$ & $-\frac{4\sqrt{2}}{3}$ & $\frac{4\sqrt{2}}{3}$ & $0$ \\
     \bottomrule[0.5pt]\bottomrule[1.5pt]
  \end{tabular}
  \end{table}

Next, we investigate tetraquark states using the modified CEI term in the bag model. Before that, we discuss the matrix elements of the CEI for tetraquark states. Based on $SU(3)_c$ color symmetry, the color wavefunctions of tetraquark states can be constructed in two ways. One is the $6_c \otimes \bar{6}_c$ and $\bar{3}_c \otimes 3_c$ representations, with the wavefunctions denoted as $\phi_1^T$ and $\phi_2^T$; the other is the $8_c \otimes8_c$ and $1_c \otimes 1_c$ representations, with the wavefunctions denoted as $\phi_1^{T^{\prime}}$ and $\phi_2^{T^{\prime}}$, as detailed in Appendix~\ref{apd:WF}. The color factors between quarks for these two constructions are listed in Table~\ref{tab:color_factors_teraquarks}.

Based on Eq.~(\ref{BE}) and the color factors given in Table~\ref{tab:color_factors_teraquarks}, we present the matrix elements of the chromoelectric interaction for the two color representations, taking $\phi_1^T$ as an example:
\begin{equation}
	\begin{split}
		\left\langle\phi_1^T|H_{CEI}|\phi_1^T\right\rangle
		=
		&
		-\Bigg[\frac{4}{3}E_{12} - \frac{10}{3}E_{13} - \frac{10}{3}E_{14}
		- \frac{10}{3}E_{23}
		\\
		&\hspace{0.8em}
		- \frac{10}{3}E_{24} + \frac{4}{3}E_{34}\Bigg]\cdot\frac{3B_{M}R_{M}}{16R},
	\end{split}	
\end{equation}
here the subscripts label the quarks, and the values for light-flavor combinations are set to zero. Finally, we substitute both the CMI and CEI interaction terms into Eq.~(\ref{M}) for a unified solution.

\subsection{The bag confinement limit and compact stability}
\label{sec:CL}

There are various models for discussing confinement, such as the lattice QCD string/flux-tube picture \cite{Bali:2000gf,Bali:2005fu,Bulava:2019iut,Kou:2024dml}, the soliton bag model \cite{Friedberg:1976eg}, the cloudy bag model \cite{Theberge:1980ye}, the NJL model \cite{Hatsuda:1994pi}, holographic QCD \cite{Erlich:2005qh}, and the dual superconductor model \cite{tHooft:1981bkw}, each of which describes the mechanism of quark confinement from its own perspective. In the MIT bag model, the long-range binding of quarks is encoded in the bag confinement energy $E_{\mathrm{CON}}$, which consists of two competing contributions \cite{Zhang:2025dsx,Liu:2025fbe}
\begin{equation}
	E_{\mathrm{CON}} = \frac{4}{3}\pi R^3 B - \frac{Z_0}{R},
\end{equation}
where the first term is the volume energy, originating from the difference in vacuum energy density between the interior and exterior of the bag, and the second term is the Casimir zero-point energy, which opposes the expansion of the bag and thus plays a stabilizing role \cite{Jaffe:1976ig,Jaffe:1976ih}. The parameters $B$ and $Z_0$ are determined by fitting the masses, magnetic moments, and charge radii of light-flavor hadrons \cite{DeGrand:1975cf}. In the variational procedure, the volume energy tends to enlarge the bag, while the Casimir term counteracts this expansion; their balance determines the optimal bag radius $R_0$ that minimizes the total mass.

From the perspective of confinement, the sign of $E_{\mathrm{CON}}$ has a clear physical meaning: when $E_{\mathrm{CON}}<0$, the bag experiences an effective binding potential and can maintain a stable compact configuration; conversely, if $E_{\mathrm{CON}}>0$, the bag is no longer energetically bound and tends to split into several color-singlet clusters. Setting $E_{\mathrm{CON}}=0$ yields the critical radius
\begin{equation}
	R_c = \left(\frac{3Z_0}{4\pi B}\right)^{1/4} \simeq 5.615\ \mathrm{GeV}^{-1} \simeq 1.12\ \mathrm{fm}.
\end{equation}
When the bag radius $R>R_c$, the confinement energy becomes positive, implying that the confinement mechanism of the bag model itself breaks down. It is worth noting that this critical scale is qualitatively consistent with the string-breaking distance obtained from lattice QCD (about $1.1$--$1.3\ \mathrm{fm}$) \cite{Bulava:2019iut,Bali:2005fu,Kou:2024dml}. Thus, the confinement limit of the bag model can be viewed as another description of the same nonperturbative energy scale \cite{Zhang:2025dsx}.

Here $R_c$ serves as the upper bound for compact multiquark states. Conventional hadrons---mesons and baryons---all have equilibrium bag radii well below $R_c$ \cite{Zhang:2021yul}. For tetraquark systems, however, the bag radius depends sensitively on the number of light quarks, since light quarks carry larger kinetic energies and tend to expand the bag. As shown in Table~\ref{Econ}, even before including chromoelectric and chromomagnetic interactions, the bag radius of the $cqqq$ system already approaches $R_c$. Once full interactions are switched on, some states may exceed this limit and thus lose the potential to form compact structures.

It should be clarified that a negative confinement energy $E_{\mathrm{CON}}<0$ may be a necessary but not sufficient condition for the existence of a compact tetraquark. A system satisfying $E_{\mathrm{CON}}<0$ could still manifest as a molecular state rather than a single confined bag, depending on the actual dynamics and the hadronization process \cite{Liu:2025fbe}. For instance, the $c\bar{c}q\bar{q}$ system, even if it has the potential to be compact, cannot be ruled out as a $D\bar{D}$ molecule. In the bag picture, a compact multiquark state is a single confined entity containing multiple valence quarks, whereas a molecular state consists of two or more color-singlet hadrons interacting weakly. These two scenarios are not mutually exclusive.

In this work, we use $E_{\mathrm{CON}}$ as a primary criterion to assess whether a singly heavy tetraquark state has the potential to form a compact structure. States with $E_{\mathrm{CON}}<0$ and $R_0<R_c$ are considered as compact candidates, while those with $E_{\mathrm{CON}}>0$ are disfavored.

\renewcommand{\tabcolsep}{0.05cm} \renewcommand{\arraystretch}{2}
\begin{table}[htbp]
		\centering
		\caption{The color-magnetic-interaction-independent bag radius and bag confinement energy for different numbers of light quarks (here $q = u, d,$ or $s$)}
		\label{Econ}
		\begin{tabular}{cccccc}
			\bottomrule[1.50pt]	\bottomrule[0.5pt]
			System & $qqqq$ & $cqqq/bqqq$ & $ccqq/bbqq$ &  $cccq/bbbq$ & $cccc/bbbb$ \\ \hline
			$R_{0}(GeV^{-1})$  &$5.98$&  $5.57/5.48$& $5.01/4.70$ & $4.86/4.145$ & $4.61/3.07$  \\
			$E_{CON}(MeV)$& $89$ &$-(10/31)$& $-(134/199)$ & $-(165/312)$ &$-(218/545)$  \\
			\bottomrule[0.5pt]\bottomrule[1.5pt]
		\end{tabular}
	\end{table}

\section{Mass spectrum and strong decay}
\label{multi-heavy hadrons}

For single-heavy tetraquark states, multiple color-flavor configurations exist. The systems we mainly consider are analogous to the mirror structures formed by inserting $n\bar{n}$ or $s\bar{s}$ into $DB/D_sB_s$ mesons, specifically $nQ\bar{n}\bar{n}$, $nQ\bar{s}\bar{n}$, $sQ\bar{s}\bar{n}$, $sQ\bar{s}\bar{s}$, and another color configuration $sQ\bar{n}\bar{n}$ (where $Q = b, c$ and $n = u, d$).

For tetraquark states, there are two types of color representations. The first consists of $6_c \otimes \bar{6}_c$ / $\bar{3}_c \otimes 3_c$, with the corresponding basis vectors denoted as $\phi_1^T$ / $\phi_2^T$; the second consists of $8_c \otimes 8_c$ / $1_c \otimes 1_c$, with the corresponding basis vectors denoted as $\phi_1^{T'}$ / $\phi_2^{T'}$ (see Appendix~\ref{apd:WF} for details). The spin basis vectors are given in Eq.~(\ref{Spin}). Accordingly, based on the group algebra of $SU(3)_c \otimes SU(2)_s$, the color-spin basis vectors for the two color representations of tetraquarks are given in Eq.~(\ref{sc36}) and Eq.~(\ref{sc18}), respectively. The color-spin basis vectors in both representations can be used to compute the tetraquark masses, and the results are strictly consistent. For the color representation $6_c \otimes \bar{6}_c$ / $\bar{3}_c \otimes 3_c$, certain basis vectors that violate the Pauli principle can be eliminated by flavor symmetry analysis. The basis vectors for different flavor combinations are listed in Table~\ref{tab:state_mixing}. For tetraquark states in the color representation $8_c \otimes 8_c$ / $1_c \otimes 1_c$, their strong decay properties can be further analyzed, as summarized in Table~\ref{tab:state_mixing}.

For tetraquark states in the color representations $6_c \otimes \bar{6}_c$ / $\bar{3}_c \otimes 3_c$, the eigenstates are CMI mixed states. The corresponding color-spin eigenstates can be written as
\begin{equation}
	\psi = c_{1} \phi_{1}^{T} \chi_{i}^{T} + c_{2} \phi_{2}^{T} \chi_{i}^{T} + \cdots + c_{3} \phi_{1}^{T} \chi_{j}^{T} + c_{4} \phi_{2}^{T} \chi_{j}^{T}+ \cdots,
\end{equation}
from which the weight $|c|^2$ of each color-spin configuration can be directly obtained.

Correspondingly, for the color representations $8_c \otimes 8_c$ / $1_c \otimes 1_c$, the eigenstates in the chromomagnetic mixed form can be expressed as
\begin{equation}
	\psi' = c_{1}' \phi_{1}^{T'} \chi_{i}^{T} + c_{2}' \phi_{2}^{T'} \chi_{i}^{T} + \cdots + c_{3}' \phi_{1}^{T'} \chi_{j}^{T} + c_{4}' \phi_{2}^{T'} \chi_{j}^{T}+ \cdots,
	\label{eq:eq18pr}
\end{equation}
where the weight of the color-singlet $1_c$ component is determined by the square of the coefficient of $\phi_{2}^{T'}$. This weight is particularly important for the partial width ratios to be discussed later.

To discuss the partial widths of tetraquark states, we follow the approach used in Refs.~\cite{Weng:2019ynv,Weng:2020jao,gaoc1992,Weng:2021ngd,An:2020vku}. Here, we consider only $S$-wave decays, and the decay width can be expressed as:
\begin{equation}
	\Gamma_{i}=\gamma_i\alpha k \cdot {|c_i|}^2,
	\label{eq:eqd}
\end{equation}
where $\Gamma_{i}$ is the strong decay width for the $i$-th channel. On the right-hand side, $\gamma_{i}$ is determined by the kinematics of the decay products, $\alpha$ is the coupling constant, $|c_{i}|^{2}$ corresponds precisely to the weight of the color-singlet $1_{c}$ component in the hadronic eigenstate, and $k$ is the momentum of the final-state hadrons. For a two-body strong decay process $A \to B + C$, energy-momentum conservation gives the relation
\begin{equation}\label{phasespace}
	m_A = \sqrt{m_B^2 + k^2} + \sqrt{m_C^2 + k^2},
\end{equation}
where $k$ is the momentum of the final-state hadrons. Therefore, once the masses of the initial and final states are fixed, the momentum $k$ can be readily obtained from the above equation.

For a tetraquark state with a specific flavor configuration, e.g., $nc\bar{s}\bar{n}$, in the color representations $8_c \otimes 8_c$ / $1_c \otimes 1_c$, there exist two possible flavor representations: $c\bar{s} \otimes n\bar{n}$ and $c\bar{n} \otimes n\bar{s}$. Under these two different flavor representations, the calculated mass spectra are exactly identical, but they correspond to different flavor final states. For example, when $J^P = 0$, for the $c\bar{s} \otimes n\bar{n}$ representation, the final-state particles corresponding to the color-singlet basis vector $\phi_2^{T^{\prime}} \chi_{3}^{T}$ in the eigenstate are uniquely identified as $D_s^*$ and $\omega$; whereas for the $c\bar{n} \otimes n\bar{s}$ representation, the final-state particles corresponding to the color-singlet basis vector $\phi_2^{T^{\prime}}$ in the eigenstate are $D^*$ and $K^*$. Therefore, when considering different decay channels, the appropriate flavor representation must be used accordingly. Second, states with $|c_i|^2$ close to 1 are treated as scattering states rather than genuine compact resonances. This criterion is consistent with the standard practice in bag-model studies of multiquark systems \cite{Weng:2019ynv,Weng:2020jao,Weng:2021ngd}.

\subsection{$nQ\bar{n}\bar{n}$}

\renewcommand{\tabcolsep}{0.2cm} \renewcommand{\arraystretch}{1.5}
\begin{table*}[!htb]
	\centering
	\caption{For the $nQ\bar{n}\bar{n}$ system, the color-magnetic eigenstates, bag radius $R_{0}$, confinement energy $E_{\mathrm{CON}}$, mass $M_{\mathrm{bag}}$, and the threshold difference with respect to two-meson thresholds are given respectively.}
		\label{mass:nQnn}
	\begin{tabular}{cccccc}
	\bottomrule[1.5pt]\bottomrule[0.5pt]	
	\toprule
$[I_{\bar{n}\bar{n}},I](J^{PC})$ & Eigenvector & \( R_0 \) (GeV\(^{-1}\))&$E_{CON}$(MeV) & $M_{bag}$ (GeV) & $\delta m=M-M_{Threshold}$ (MeV) \\
		\hline
		\multicolumn{6}{c}{ $nc\bar{n}\bar{n}$}\\
		\cline{1-6}
		\midrule
	\multirow{2}{*}{$[0,\frac{1}{2}](0^+)$} & -0.813,0.582 & 5.203 & -93 & 2.278 &$\delta_{D^*\omega} = -514;  \delta_{D\pi} = 270 $\\
	     & 0.582,0.813 &5.524 &-21 &2.793 &
	    $\delta_{D^*\omega} =1;  \delta_{D\pi} = 785 $\\
	
	    \midrule
	\multirow{3}{*}{$[0,\frac{1}{2}](1^+)$}& -0.680,0.451,0.578 &5.459 & -36 &2.443 &$\delta_{D^*\omega} =-349;\delta_{D\omega} =-208;  \delta_{D^*\pi} = 294 $\\
	      & 0.698,0.156,0.699 &5.412 & -46 &2.757 &$\delta_{D^*\omega} = -35;\delta_{D\omega} =106; \delta_{D^*\pi} = 608 $ \\
	      &-0.232,-0.880,0.415 & 5.586 & -6 & 2.913 &$\delta_{D^*\omega} =121; \delta_{D\omega} =262; \delta_{D^*\pi} = 764 $ \\
	    
	    \midrule
		$[0,\frac{1}{2}](2^+)$ & 1 & 5.612 & -1 &2.948 &$\delta_{D^*\omega} = 192$ \\
	
		\midrule
		\multirow{2}{*}{$[1,\frac{1}{2}/\frac{3}{2}](0^+)$}&-0.813,0.582& 5.395 & -50 & 2.578 &$\delta_{D^*\omega} = -214;  \delta_{D\pi} =570 $ \\
	   & 0.582,0.813 & 5.688 &17 & 3.087 &$\delta_{D^*\omega} = 295;  \delta_{D\pi} = 1079 $\\
		
		\midrule
		\multirow{3}{*}{$[1,\frac{1}{2}/\frac{3}{2}](1^+)$}&-0.540,0.604,0.586 & 5.473 & -33 &2.643 & $\delta_{D^*\omega} = -149;  \delta_{D\omega^*} = -8;\delta_{D^*\pi} = 494 $ \\
	  &0.782,0.618,0.085 &5.485 & -30 & 2.869 & $\delta_{D^*\omega} =77;\delta_{D\omega} =218;  \delta_{D^*\pi} =720 $\\
	  &0.313,-0.503,0.806 & 5.577 & -9 & 3.015 &  $\delta_{D^*\omega} =223;\delta_{D\omega} =364;  \delta_{D^*\pi} =866 $\\
	   
	    \midrule
		$[1,\frac{1}{2}/\frac{3}{2}](2^+)$  & 1 & 5.612 & -6 & 2.948 & $\delta_{D^*\omega} =192 $ \\
	    \hline
		\multicolumn{6}{c}{ $nb\bar{n}\bar{n}$}\\
	    \cline{1-6}
		\midrule
	  \multirow{2}{*}{$[0,\frac{1}{2}](0^+)$}  & -0.813,0.582 & 5.146 & -105 &5.750 &$\delta_{B^*\omega} = -358;  \delta_{B\pi} = 330 $\\
	  & 0.582,0.813 &5.367 &-56 &6.192 &$\delta_{B^*\omega} = 84;  \delta_{B\pi} = 772 $ \\
	
		\midrule
		\multirow{3}{*}{$[0,\frac{1}{2}](1^+)$} & -0.668,0.463,0.582 &5.261 &-80 &5.810 & $\delta_{B^*\omega} =-298;\delta_{B\omega} =-253;  \delta_{B^*\pi} = 345 $ \\
		 & 0.593,-0.141,0.793 &5.350 &-60 &6.188 & $\delta_{B^*\omega} =80;\delta_{B\omega} =125;  \delta_{B^*\pi} = 723 $  \\
		&-0.451,-0.874,0.179 &5.396 &-50 &6.303 & $\delta_{B^*\omega} =195;\delta_{B\omega} =240;  \delta_{B^*\pi} =838 $  \\
	
		\midrule
	   $[0,\frac{1}{2}](2^+)$ & 1 & 5.430 & -42 &6.325 & $\delta_{B^*\omega} =217$\\
	
		\midrule
		\multirow{2}{*}{$[1,\frac{1}{2}/\frac{3}{2}](0^+)$}&-0.813,0.582& 5.278 &-76 & 6.007 &$\delta_{B^*\omega} = -101;  \delta_{B\pi} = 587 $ \\
		 & 0.582,0.813 & 5.484 &-30& 6.445 &$\delta_{B^*\omega} = 337;  \delta_{B\pi} = 1025 $\\
	
		\midrule
		\multirow{3}{*}{$[1,\frac{1}{2}/\frac{3}{2}](1^+)$} &-0.629,0.514,0.583 & 5.322 &-67 &6.032 & $\delta_{B^*\omega} =-76;\delta_{B\omega} =-31;  \delta_{B^*\pi} =567 $\\
		&0.658,0.751,0.049 &5.376 & -55 & 6.297 & $\delta_{B^*\omega} =189;\delta_{B\omega} =234;  \delta_{B^*\pi} =832 $ \\
		&0.413,-0.414,0.811 & 5.435 & -41 & 6.417 & $\delta_{B^*\omega} =309;\delta_{B\omega} =354;  \delta_{B^*\pi} =952 $ \\
		
		\midrule
		$[1,\frac{1}{2}/\frac{3}{2}](2^+)$& 1 & 5.430 &-42 & 6.325 &  $\delta_{B^*\omega} =217$ \\
		\bottomrule
	\bottomrule[0.5pt]\bottomrule[1.5pt]	
	\end{tabular}
\end{table*}

\renewcommand{\tabcolsep}{0.1cm} \renewcommand{\arraystretch}{1.5}
\begin{table*}[!htb]
	\centering
	\caption{The table lists the \( |c_{ij}|^2 \cdot k \) (in \(\mathrm{GeV}\)) values for the \( nQ\bar{n}\bar{n} \) system; the numbers in parentheses can be regarded as branching ratios. A ``\(-\)'' indicates that the mass is below the threshold, and ``\(*\)'' denotes a scattering state.}
	\label{decay:nQnn}
	\begin{tabular}{cccccccccc}
		\bottomrule[1.5pt]\bottomrule[0.5pt]	
		\toprule
		$[I_{\bar{n}\bar{n}},I](J^{PC})$ & $M_{bag}$ (GeV) & $B^*/D^*\rho$ &$B/D\rho$ & $B^*/D^*\pi$ &$B/D\pi$& $B^*/D^*\omega$ &  $B/D\omega$& $B^*/D^*\eta$ &$B/D\eta$\\
		\hline
		\multicolumn{10}{c}{ $nc\bar{n}\bar{n}$}\\
		\cline{1-10}
		\midrule
		\multirow{2}{*}{$[0,\frac{1}{2}](0^+)$}  & 2.278 &- & & & 0.192(1)&- & & &- \\
		& 2.793 &0.063(1)& & &0.024(0.38) & & & & 0.019(0.30)\\
		
		\midrule
		\multirow{3}{*}{$[0,\frac{1}{2}](1^+)$}  & 2.443 &-&- &0.208(1) & &- &- &- & \\
		& 2.757 &-&0.168(1) &0.012(0.07) & &- &0.155(0.92) &0.008(0.05) & \\
		&  2.913 &0.185(1)&0.081(0.44) &0.008(0.04) & &0.172(0.93)&0.078(0.42) &0.007(0.04) &\\
		
		\midrule
		$[0,\frac{1}{2}](2^+)$ &2.948 &*& & & &*& & &\\
		
		\midrule
		\multirow{2}{*}{$[1,\frac{1}{2}/\frac{3}{2}](0^+)$} & 2.578 &- & & &0.249(1) &- & & & 0.160(0.64)\\
		&  3.087 &0.343(1)& & &0.002(0.01) &0.334(0.97) & & &$<10^{-4}$\\
		
		\midrule
		\multirow{3}{*}{$[1,\frac{1}{2}/\frac{3}{2}](1^+)$} &2.643 &-&0.004(0.02) &0.218(1) & &- &- &0.111(0.51) & \\
		& 2.869 &0.014(0.09)&0.149(1) &0.008(0.05) & &0.012(0.08)&0.144(0.97) &0.006(0.04) &\\
		& 3.015 &0.242(1)&0.066(0.27) &0.003(0.01) & &0.233(0.96)&0.065(0.27) & 0.002(0.01)& \\
		
		\midrule
		$[1,\frac{1}{2}/\frac{3}{2}](2^+)$ & 2.948 &*& & & &*& & & \\
		\hline
		\multicolumn{10}{c}{ $nb\bar{n}\bar{n}$}\\
		\cline{1-10}
		\midrule
		\multirow{2}{*}{$[0,\frac{1}{2}](0^+)$}& 5.750 & -& & &0.237(1) &- & & &- \\
		&6.192 &0.158(1) & & &0.026(0.16) &0.152(0.96) & & &0.021(0.13)\\
		
		\midrule
		\multirow{3}{*}{$[0,\frac{1}{2}](1^+)$} &5.810 &- &- &0.246(1) & &- &- &- & \\
		&6.188 &0.061(0.51) &0.120(1) &0.023(0.19) & &0.055(0.46) &0.112(0.93) &0.018(0.15) & \\
		&6.303 &0.188(0.94) &0.200(1) &0.002(0.01) & &0.180(0.90) &0.193(0.97) &0.002(0.01) & \\
		
		\midrule
		$[0,\frac{1}{2}](2^+)$ &6.352 &* & & & &* & & & \\
		
		\midrule
		\multirow{2}{*}{$[1,\frac{1}{2}/\frac{3}{2}](0^+)$}  &6.007 &- & & &0.278(1) &- & & &0.185(0.67) \\
		&6.445 & 0.412(1)& & &0.002(0.00) &0.396(0.96) & & &$<10^{-4}(0)$ \\
		
		\midrule  
		\multirow{3}{*}{$[1,\frac{1}{2}/\frac{3}{2}](1^+)$}  &6.032 & - &- &0.270(1) & &- &-  & 0.174(0.64) & \\
		& 6.297& 0.043(0.27) &0.158(1) &0.001(0.01) & & 0.041(0.26)& 0.152(0.96) & 0.001(0.01) & \\
		&6.417 & 0.289(1) &0.111(0.38)  &0.002(0.01)  &  &0.282(0.98)&0.109(0.38)  & 0.002(0.01) &\\
		
		\midrule
		$[1,\frac{1}{2}/\frac{3}{2}](2^+)$ &6.352& * &  &  &  &*  &  &  &\\
		\bottomrule
		\bottomrule[0.5pt]\bottomrule[1.5pt]	
	\end{tabular}
\end{table*}

For the $nQ\bar{n}\bar{n}$ systems with $Q=c,b$, we investigate two cases distinguished by the isospin of the light antiquark pair, $I_{\bar{n}\bar{n}}=0$ and $1$. The $[I_{\bar{n}\bar{n}},I]=[0,\frac{1}{2}]$ configuration can be isolated as a distinct sector, while the $I=\frac{1}{2}$ and $I=\frac{3}{2}$ sectors are mass-degenerate. Table~\ref{mass:nQnn} summarizes the resulting mass spectra. In the charm sector, the $I_{\bar{n}\bar{n}}=0$ states lie between $2.278$ and $2.948$~GeV, whereas the $I_{\bar{n}\bar{n}}=1$ states span $2.578$ to $3.015$~GeV. For the bottom sector, the corresponding ranges are $5.750$--$6.325$~GeV and $6.007$--$6.325$~GeV, respectively. All computed masses exceed the lowest meson-meson thresholds ($J^p=0^+(B/D\pi)$, $1^+(B^{*}/D^{*}\pi)$, and $2^+(B^{*}/D^{*}\omega)$), indicating that these tetraquarks are above the strong-decay thresholds. We also note that some $I=\frac{1}{2}$ states share quantum numbers with established $D$-meson resonances such as $D_{0}^{*}(2300)$ and $D_{1}(2420)$; while mixing may occur in principle, it is not addressed in the present work.

We now turn to the confinement energy $E_{\text{CON}}$ as a diagnostic for compactness. Several states—$T_{nc\bar{n}\bar{n}}(1^{+},2.913)$, $T_{nc\bar{n}\bar{n}}(2^{+},2.948)$, $T_{nc\bar{n}\bar{n}}(1^{+},3.015)$, and $T_{nc\bar{n}\bar{n}}(0^{+},3.087)$—exhibit $E_{\text{CON}}$ values that are only marginally negative or even positive, suggesting that they are unlikely to form compact single-bag configurations. In contrast, states such as $T_{nc\bar{n}\bar{n}}(0^{+},2.278)$, $T_{nc\bar{n}\bar{n}}(1^{+},2.757)$, and $T_{nc\bar{n}\bar{n}}(1^{+},2.578)$ show significantly more negative confinement energies, close to or below $-50$~MeV, indicating a greater potential for compactness. The bottom-sector analogs tend to have even more negative $E_{\text{CON}}$ values, reflecting the stronger suppression of the bag radius by the heavier $b$ quark.

The strong decay stability of the $nc\bar{n}\bar{n}$ system is analyzed in Table~\ref{decay:nQnn}. There we list the weight $|c_i|^2$ of the color-singlet component in the $8_c \otimes 8_c$ and $1_c \otimes 1_c$ representations—i.e., the coefficient of $\phi_{2}^{T'}$ in Eq.~(\ref{eq:eq18pr})—with the final-state momentum $k$ determined via Eq.~(\ref{eq:eqd}). While the strong coupling $\alpha$ and the kinematic factor $\gamma_i$ are not known a priori, two observations justify the subsequent analysis. First, in the quark model, spin does not affect the spatial dynamics \cite{Weng:2019ynv}; second, $\alpha$ is scale-dependent and can be taken as common for identical flavor configurations \cite{Zhang:2023hmg,Zhang:2023teh}. Under these assumptions, the similarity relation \cite{Weng:2019ynv,Weng:2020jao,Weng:2021ngd}
\begin{equation}
	\gamma_{B/D\pi} \simeq\gamma_{B/D\eta}\simeq\gamma_{B^*/D^*\omega}\simeq \gamma_{B^*/D^*\rho}\simeq\gamma_{B/D\omega} \simeq\gamma_{B/D\rho},
	\label{gamma}
\end{equation}
holds for final states in the $c\bar{n}\otimes n\bar{n}$ flavor representation. This allows us to compare the factors $k\cdot |c_i|^2$ across different channels and thereby obtain the ratios of partial widths.

In Table~\ref{tab:bottom_charmed_decay}, we list the $k\cdot |c_i|^2$ values for each state and provide the corresponding partial-width ratios in parentheses. It is worth noting that for states such as $T_{nc\bar{n}\bar{n}}(\frac{1}{2}/\frac{3}{2},2^{+},2.948)$, the $1_c$ component exceeds $80\%$ of the total ($|c_i|^2>0.8$), which qualifies them as scattering states following the standard criterion in bag-model studies of multiquark systems \cite{Weng:2019ynv,Weng:2020jao}.

Although some $nc\bar{n}\bar{n}$ states satisfy the compactness criteria—$R_0<R_c$ and $E_{\text{CON}}<0$—they often have large $k\cdot |c_i|^2$ values, implying strong decay widths, particularly for OZI-superallowed modes. Conversely, phase-space suppression can significantly reduce the decay momentum $k$ for certain states, leading to smaller $k\cdot |c_i|^2$ values and hence narrower widths. This is the case for $T_{nc\bar{n}\bar{n}}(0^{+},2.278)$, $T_{nc\bar{n}\bar{n}}(1^{+},2.443)$, $T_{nc\bar{n}\bar{n}}(0^{+},2.578)$, $T_{nb\bar{n}\bar{n}}(0^{+},5.750)$, and $T_{nb\bar{n}\bar{n}}(1^{+},5.810)$, all of which exhibit suppressed decay factors. These relatively narrow compact tetraquark candidates are expected to be of particular interest for future experimental searches.

\subsection{$nQ\bar{s}\bar{n}$ and $sQ\bar{n}\bar{n}$}

\renewcommand{\tabcolsep}{0.2cm} \renewcommand{\arraystretch}{1.6}
\begin{table*}[!htb]
	\centering
	\caption{For the $nQ\bar{s}\bar{n}$ system, the color-magnetic eigenstates, bag radius $R_{0}$, confinement energy $E_{\mathrm{CON}}$, mass $M_{\mathrm{bag}}$, and the threshold difference with respect to two-meson thresholds are given respectively.}
	\label{tab:bottom_charmed_tetraquarks}
	\begin{tabular}{cccccc}
		\bottomrule[1.5pt]\bottomrule[0.5pt]	
		\toprule
	   \( J^{PC} \) & Eigenvector & \( R_0 \) (GeV\(^{-1}\))&$E_{CON}$(MeV) & $M_{bag}$ (GeV) & $\delta m=M-M_{Threshold}$ (MeV) \\
		\hline
		\multicolumn{6}{c}{ $nc\bar{s}\bar{n}$}\\
		\cline{1-6}
	    \multirow{4}{*}{\( 0^{+} \)} & 0.599,-.0019,-0.014,0.800 & 5.667&12 & 3.194 & $\delta_{D^*K^*} = 291;  \delta_{DK} = 830 $ \\
		 & -0.047,-0.827,-0.599,0.008 & 5.456&-36 & 2.925 &  $\delta_{D^*K^*} = 22;  \delta_{DK} = 561 $  \\
		 & -0.797,0.079,-0.045,0.597 & 5.324&-66 & 2.744 & $\delta_{D^*K^*} = -159;  \delta_{DK} = 380 $ \\
		 & -0.061,-0.557,0.827,0.047 & 5.101&-115 & 2.472 & $\delta_{D^*K^*} = -431;  \delta_{DK} = 108 $ \\
		\midrule
		 \multirow{6}{*}{\( 1^{+} \)} & 0.107,0.520,-0.106,-0.701,-0.119,0.449 & 5.131&-108 & 2.624 & $\delta_{D^*K^*}=-279;\delta_{D^*K} = 812;  \delta_{DK^*} = 555 $\\
		 &0.498,-0.187,-0.589,0.065,-0.597,-0.097& 5.477&-31 & 2.805 &$\delta_{D^*K^*}=-98;\delta_{D^*K} = 517;  \delta_{DK^*} = 260 $ \\
		 &0.327,-0.037,-0.517,-0.020,0.789,0.020 &5.735&28 &3.126 &$\delta_{D^*K^*}=223;\delta_{D^*K} = 480;  \delta_{DK^*} = 223 $ \\
		 & -0.246,-0.439,-0.150,0.196,-0.033,0.828 &5.683&16  &3.051 & $\delta_{D^*K^*}=148;\delta_{D^*K} = 403;  \delta_{DK^*} = 146 $\\
		 &-0.171,-0.668,0.001,-0.682,0.029,-0.242 &5.565&-11 & 2.916 &  $\delta_{D^*K^*}=13;\delta_{D^*K} = 544;  \delta_{DK^*} = 287 $\\
		 &0.737,-0.235,0.593,-0.027,0.066,0.211 & 5.597&-4 & 3.007 &  $\delta_{D^*K^*}=104;\delta_{D^*K} = 400;  \delta_{DK^*} = 143 $\\
		\midrule
		 \multirow{2}{*}{\( 2^{+} \)} &0.457,0.889 & 5.649&8 & 3.066 & $\delta_{D^*K^*} = 163$ \\
		 & -0.889,0.456 &5.638&5 & 3.087 &$\delta_{D^*K^*} = 184$   \\
		\hline
		\multicolumn{6}{c}{ $nb\bar{s}\bar{n}$}\\
		\cline{1-6}
		\midrule
		 \multirow{4}{*}{\( 0^{+} \)} & 0.607,-0.027,-0.019,0.794 & 5.473&-32 & 6.547 & $\delta_{B^*K^*} = 328; \delta_{BK} = 771 $ \\
		 & -0.063,-0.833,-0.550,0.009 &5.307&-70 & 6.345 & $\delta_{B^*K^*} = 126; \delta_{BK} = 569 $  \\
		 & -0.787,0.110,-0.067,0.603 & 5.211&-91 & 6.163 & $\delta_{B^*K^*} = -56; \delta_{BK} = 387 $  \\
		 & -0.088,-0.543,0.832,0.069 &5.055&-125 & 5.929 & $\delta_{B^*K^*} = -290; \delta_{BK} = 153 $  \\
		\midrule
		 \multirow{6}{*}{\( 1^{+} \)} & 0.098,0.529,-0.077,-0.687,-0.099,0.472 &5.155&-103 & 5.986 & $\delta_{B^* K^*} =-233;\delta_{B^*K} =929; \delta_{BK^*} =576 $\\
		 &-0.598,0.147,0.496,-0.072,0.604,0.062 & 5.236&-85 &6.188 &$\delta_{B^*K^*} =-31;\delta_{B^*K} =615; \delta_{BK^*} = 262 $ \\
		 &0.431,-0.035,-0.435,-0.019,0.789,0.015& 5.514&-23 &6.521 &$\delta_{B^*K^*} =302;\delta_{B^*K} = 544; \delta_{BK^*} =191 $ \\
		 &-0.541,-0.108,-0.596,0.268,-0.042,0.517 &5.467& -34   &6.438 & $\delta_{B^*K^*} =219;\delta_{B^*K} =438; \delta_{BK^*} =85 $\\
		 &0.379,-0.231,0.451,0.309,0.025,0.709 &5.442&-39 & 6.413 &$\delta_{B^*K^*} =194;\delta_{B^*K} =644; \delta_{BK^*} =311 $ \\
		 &-0.100,-0.795,0.010,-0.596,0.010,0.048 & 5.408&-47  & 6.342 & $\delta_{B^*K^*} =123;\delta_{B^*K} = 508; \delta_{BK^*} =155 $\\
		\midrule
		 \multirow{2}{*}{\( 2^{+} \)} &0.432,0.902 & 5.472&-33 & 6.435 & $ \delta_{B^*K^*} = 216 $ \\
		 & -0.902,0.432 &5.462&-35 & 6.462 &$ \delta_{B^*K^*} = 243 $  \\
		
		\bottomrule
		\bottomrule[0.5pt]\bottomrule[1.5pt]	
	\end{tabular}
\end{table*}

Table~\ref{tab:bottom_charmed_tetraquarks} summarizes the bag radius $R_0$, confinement energy $E_{\text{CON}}$, mass spectrum, and the mass difference $\delta m = M - M_{\text{Threshold}}$ for each eigenstate of the $nQ\bar{s}\bar{n}$ system. The basis vectors are ordered consistently with those in Table~\ref{tab:state_mixing} of Appendix~\ref{apd:WF}. For the $nc\bar{s}\bar{n}$ system, the $I=0$ and $I=1$ states are degenerate, with masses spanning $2.450$--$3.200$~GeV, while the $nb\bar{s}\bar{n}$ system exhibits masses in the range $5.925$--$6.547$~GeV. All computed states lie above their respective lowest thresholds. The bag radii of these systems are found to be close to the confinement limit. Notably, the states $T_{nc\bar{s}\bar{n}}(0^{+},3.194)$, $T_{nc\bar{s}\bar{n}}(1^{+},3.126)$, $T_{nc\bar{s}\bar{n}}(1^{+},3.051)$, and the two highest-spin states $T_{nc\bar{s}\bar{n}}(2^{+},3.066)$ and $T_{nc\bar{s}\bar{n}}(2^{+},3.087)$ yield positive $E_{\text{CON}}$, indicating that compact structures are not supported for these states. In contrast, states such as $T_{nc\bar{s}\bar{n}}(0^{+},2.472)$ and $T_{nc\bar{s}\bar{n}}(1^{+},2.624)$ exhibit $E_{\text{CON}} \simeq -100$~MeV, suggesting a viable potential for compactness.

Secondly, among our results, a state that deserves particular attention is $T_{nc\bar{s}\bar{n}}(0^{+},2.925)$ (the second state in Table~\ref{tab:bottom_charmed_tetraquarks}), which may correspond to the experimentally observed $T_{c\bar{s}}(2900)$. The latter has $J^P = 0^{+}$, a mass of $2908$\,MeV, and isospin $I=1$. In our calculation, the confinement energy of $T_{nc\bar{s}\bar{n}}(0^{+},2.925)$ is $E_{\text{CON}} = -36$\,MeV, indicating its potential to form a compact structure. The experimentally observed $T_{c\bar{s}}(2900)$ lies close to but slightly above the $D^*(2007)K^*(892)$ threshold, with a decay width of $\Gamma = 136\pm 23\pm 13\,\mathrm{MeV}$. Such a large width makes its interpretation as a hadronic molecule problematic, particularly because the $D_s\pi$ decay channel is expected to be dynamically suppressed for a molecular state composed of $D^*K^*$. Notably, the experimentally observed decay channel is $D_s \pi$, which may suggest short-range dynamical behavior, especially if the CEI between the $c$ and $\bar{s}$ quarks is attractive, potentially leading to $D_s \pi$ as the dominant decay channel in the quark rearrangement process.

\renewcommand{\tabcolsep}{0.2cm} \renewcommand{\arraystretch}{1.6}
\begin{table*}[!htb]
	\centering
	\caption{The table lists the \( |c_{ij}|^2 \cdot k \) (in \(\mathrm{GeV}\)) values for the  $nQ\bar{s}\bar{n}$ system; the numbers in parentheses can be regarded as branching ratios. A ``\(-\)'' indicates that the mass is below the threshold, and ``\(*\)'' denotes a scattering state.}
	\label{tab:bottom_charmed_decay}
	\begin{tabular}{ccccccccccc}
		\bottomrule[1.5pt]\bottomrule[0.5pt]	
		\toprule
		\toprule
		\multicolumn{2}{c}{ }&\multicolumn{4}{c}{$c\bar{n} \otimes n\bar{s}$ } &&\multicolumn{4}{c}{$c\bar{s} \otimes n\bar{n}$}\\
		\cline{3-6}
		\cline{8-11}
		\( J^{PC} \) & $M_{bag}$ (GeV) & $B/DK$ & $B^*/D^*K$ & $B/DK^*$ & $B^*/D^*K^*$ && $B_{s}/D_{s}\pi$ & $B^*_{s}/D^*_{s}\pi$ & $B_{s}/D_{s}\rho/\omega$ & $B^*_{s}/D^*_{s}\rho/\omega$ \\
		\hline
		\multicolumn{11}{c}{ $nc\bar{s}\bar{n}$}\\
		\cline{1-11}
		\multirow{4}{*}{\( 0^{+} \)} &3.194&$<10^{-4}(0)$&&&0.390(1)&&$<10^{-4}(0)$&&&0.382(1)\\
		&2.925&0.185(1)&&&0.026(0.14)&&0.211(1)&&&0.047(0.22)\\
		&2.744&0.349(1)&&&-&&0.376(1)&&&-\\
		&2.472 &0.002(1)&&&-&&0.172(1)&&&-\\
		\multirow{6}{*}{\( 1^{+} \)} &2.624&&0.086(1)&-&&&&0.367(1)&-&-\\
		&2.805&&0.355(1)&0.011(0.03)&-&&&0.089(1)&0.010(0.11)&-\\
		&3.126&&$<10^{-4}(0)$&0.046(0.13)&0.349(1)&&&0.003(0.02)&0.076(0.56)&0.136(1)\\
		&3.051&&0.009(0.07)&0.130(1)&0.115(0.89)&&&0.006(0.02)&0.021(0.08)&0.266(1)\\
		&2.916&&0.011(0.06)&0.170(1)&0.002(0.01)&&&0.003(0.01)&0.251(1)&0.004(0.02)\\
		&3.007&&0.020(0.13)&0.153(1)&0.006(0.04)&&&0.006(0.04)&0.139(1)&0.036(0.26)\\
		\multirow{2}{*}{\( 2^{+} \)} & 3.058&&&&*&&&&&\\
		& 3.095&&&&&&&&&*\\
		\hline
		\multicolumn{11}{c}{ $nb\bar{s}\bar{n}$}\\
		\cline{1-11}
		\multirow{4}{*}{\( 0^{+} \)} & 6.547&$<10^{-4}(0)$&&&0.499(1)&&0.002(0.01)&&&0.310(1)\\
		& 6.345&0.127(1)&&&0.028(0.22)&&0.255(1)&&&0.012(0.05)\\
		& 6.163&0.433(1)&&&-&&0.148(1)&&&-\\
		& 5.929&0.130(1)&&&-&&0.394(1)&&&-\\
		\multirow{6}{*}{\( 1^{+} \)} & 5.986&&0.108(1)&-&-&&&0.431(1)&-&-\\
		&6.188&&0.448(1)&0.003(0.01)&-&&&0.117(1)&0.004(0.03)&-\\
		&6.521&&$<10^{-4}(0)$&0.107(0.27)&0.395(1)&&&0.002(0.01)&0.095(0.51)&0.185(1)\\
		&6.438&&$<10^{-4}(0)$&0.336(1)&0.159(0.47)&&&0.001(0.06)&0.003(0.17)&0.018(1)\\
		&6.413&&0.008(0.18)&0.004(0.09)&0.045(1)&&&0.001(0.00)&0.277(0.94)&0.295(1)\\
		&6.342&&0.021(0.20)&0.106(1)&0.033(0.31)&&&0.009(0.04)&0.225(1)&0.079(0.35)\\
		\multirow{2}{*}{\( 2^{+} \)} & 6.423 &&&&*&&&&&\\
		& 6.473 &&&&&&&&&*\\
		\bottomrule
		\bottomrule[0.5pt]\bottomrule[1.45pt]	
	\end{tabular}
\end{table*}

In Table~\ref{tab:bottom_charmed_decay}, we further analyze the strong decay stability of the $nc\bar{s}\bar{n}$ system. Within the color representations $8_c \otimes 8_c$ and $1_c \otimes 1_c$, the weight of the color-singlet component in each state is given by the coefficient of $\phi_{2}^{T'}$ in Eq.~(\ref{eq:eq18pr}), and the final-state momentum is determined by Eq.~(\ref{eq:eqd}). According to the similarity relations
\begin{equation}
	\gamma_{B/DK} = \gamma_{B^*/D^*K} = \gamma_{B/DK^*} = \gamma_{B^*/D^*K^*},
\end{equation}
\begin{equation}
	\gamma_{B_{s}/D_{s}\pi} = \gamma_{B_{s}^*/D_{s}^*\pi} = \gamma_{B_{s}/D_{s}\omega} = \gamma_{B_{s}^*/D_{s}^*\omega},
\end{equation}
we can compare the $k \cdot |c_i|^2$ values for different final states to obtain the ratios of their partial widths. Moreover, the $k \cdot |c_i|^2$ value itself serves as a useful reference for the partial width.

For the $T_{nc\bar{s}\bar{n}}(0^{+},2.925)$ state, its dominant decay channel is $D_s \pi$, with $k \cdot |c_i|^2 = 211$~MeV. This indicates that it has an expected large width in the $D_s \pi$ channel. Experimentally, the observed $T_{c\bar{s}}(2900)$ also exhibits a large width in the $D_s \pi$ channel, $\Gamma = 136\pm 23\pm 13\,\mathrm{MeV}$. Although $k \cdot |c_i|^2$ is not directly equal to the decay width, the latter depends strongly on it. Therefore, if we interpret $T_{c\bar{s}}(2900)$ as a candidate for a compact tetraquark state, its $D_s \pi$ decay channel can naturally be understood as the dissociation of a compact tetraquark state, and this process is OZI-superallowed~\cite{PhysRevD.15.267}. Consequently, its large decay width receives a self-consistent explanation.

In addition, these states should also attract experimental attention, such as $T_{nc\bar{s}\bar{n}}(0^{+},2.472)$, $T_{nc\bar{s}\bar{n}}(1^{+},2.624)$, $T_{nb\bar{s}\bar{n}}(0^{+},5.929)$, $T_{nb\bar{s}\bar{n}}(0^{+},6.163)$, and $T_{nb\bar{s}\bar{n}}(1^{+},5.986)$. These states not only have greater potential to form compact structures, but also, due to threshold suppression, their $k \cdot |c_i|^2$ values are small (i.e., they have narrow widths), making them easier to identify in experiments.

\renewcommand{\tabcolsep}{0.4cm} \renewcommand{\arraystretch}{1.5}
\begin{table*}[!htb]
	\centering
	\caption{For the $Qs\bar{n}\bar{n}$ system, the color-magnetic eigenstates, bag radius $R_{0}$, confinement energy $E_{\mathrm{CON}}$, mass $M_{\mathrm{bag}}$, and the threshold difference with respect to two-meson thresholds are given respectively.}
	\label{tab:Qsnn_bottom_charmed}
	\begin{tabular}{cccccc}
		\bottomrule[1.5pt]\bottomrule[0.5pt]	
		\toprule
	 $I(J^{P})$ & Eigenvector & \( R_0 \) (GeV\(^{-1}\))&$E_{CON}$(MeV) & $M_{bag}$ (GeV) & $\delta m=M-M_{Threshold}$ (MeV) \\
		\hline
		\multicolumn{6}{c}{ $cs\bar{n}\bar{n}$}\\
		\cline{1-6}
		\midrule
		\multirow{2}{*} {$0(0^+)$}  & 0.762,0.647 & 5.194 &-95 &2.511 &$\delta_{D^*K^*} =-392 ;  \delta_{DK} =147 $  \\
		  & 0.651,0.758 &5.587 &-6  & 2.919& $\delta_{D^*K^*} =16 ;  \delta_{DK} =555 $ \\
		\multirow{3}{*}{$0(1^+)$}  & -0.576,0.419,0.702 &5.350 & -60 &2.672 & $\delta_{D^*K^*}=-231;\delta_{D^*K} =167 ;  \delta_{DK^*} =-90 $  \\
		  & 0.782,0.032,0.622 &5.397 &-50  &2.894 & $\delta_{D^*K^*}=-9;\delta_{D^*K} =389 ;  \delta_{DK^*} =132 $  \\
		  &  -0.245,-0.908,0.340 &5.565 & -11 &3.060 & $\delta_{D^*K^*}=157;\delta_{D^*K} =555 ;  \delta_{DK^*} =298 $ \\
		$0(2^+)$ &1  &5.656 &10 &3.050 & $\delta_{D^*K^*} =147$\\
		
		\multirow{2}{*}{$1(0^+)$} & -0.846,0.532 & 5.389&-52&2.769&$\delta_{D^*K^*} =-134 ;  \delta_{DK} =405 $\\
	    & 0.530,0.848 &5.729&27&3.219&$\delta_{D^*K^*} =316 ;  \delta_{DK} =855 $\\
		\multirow{3}{*}{$1(1^+)$} &-0.546,0.660,0.515 &5.379&-54&2.838& $\delta_{D^*K^*}=-65;\delta_{D^*K} =333 ;  \delta_{DK^*} =76 $ \\
		&0.801,0.589,0.100 &5.457&-36&2.989& $\delta_{D^*K^*}=86;\delta_{D^*K} =484 ;  \delta_{DK^*} =227 $ \\
		 &0.239,-0.465,0.852&5.570&-10&3.155&  $\delta_{D^*K^*}=252;\delta_{D^*K} =650 ;  \delta_{DK^*} = 393$ \\
		$1(2^+)$ & 1 &5.643&7 & 3.046& $\delta_{D^*K^*} =143$ \\
		
		\hline
		\multicolumn{6}{c}{ $bs\bar{n}\bar{n}$}\\
		\cline{1-6}
		
		\multirow{2}{*}{$0(0^+)$} & -0.735,0.678 &5.140 &-106&5.978&$\delta_{B^*K^*} =-241 ;  \delta_{BK} =202 $ \\
		&0.682,0.731 &5.406&-48&6.323&$\delta_{B^*K^*} =104 ;  \delta_{BK} =547 $\\
		\multirow{3}{*}{$0(1^+)$} &-0.581,0.414,0.700 &5.146&-105&6.035& $\delta_{B^*K^*}=-184;\delta_{B^*K} =214 ;  \delta_{BK^*} =-139 $ \\
		&0.669,-0.252,0.699 &5.325&-66&6.319&$\delta_{B^*K^*}=100;\delta_{B^*K} =498 ;  \delta_{BK^*} =145 $\\
		&-0.467,-0.874,0.129 &5.382&-53&6.460&$\delta_{B^*K^*}=241;\delta_{B^*K} =639 ;  \delta_{BK^*} =286 $\\
		$0(2^+)$ &1 &5.479&-31&6.433&$\delta_{B^*K^*} =214$\\
		
		\multirow{2}{*} {$1(0^+)$}  &-0.861,0.509 &5.276 &-76 &6.194 &$\delta_{B^*K^*} =-25 ;  \delta_{BK} =418 $   \\
	    &0.507,0.862 &5.531 &-19 & 6.581& $\delta_{B^*K^*}=362;\delta_{B^*K} =760 ;  \delta_{BK^*} =407 $  \\
		\multirow{3}{*} {$1(1^+)$}  &-0.668,0.553,0.497 & 5.226 &-88 & 6.221&  $\delta_{B^*K^*}=2;\delta_{B^*K} =400 ;  \delta_{BK^*} =47 $ \\
		&0.658,0.751,0.053& 5.348& -61& 6.407&  $\delta_{B^*K^*}=188;\delta_{B^*K} =586 ;  \delta_{BK^*} =233 $ \\
	    &0.344,-0.361,0.867 &5.426&-43  &6.555&  $\delta_{B^*K^*}=336;\delta_{B^*K} =734 ;  \delta_{BK^*} =381 $ \\
		$1(2^+)$ &1& 5.466& -34&6.485 &$\delta_{B^*K^*} =266$\\
		\bottomrule
		\bottomrule[0.5pt]\bottomrule[1.5pt]	
	\end{tabular}
	\label{tab:12x7}
\end{table*}

Table~\ref{tab:Qsnn_bottom_charmed} presents the mass eigenstates, bag radius $R_0$, confinement energy $E_{\text{CON}}$, mass spectrum, and the mass difference $\delta m = M - M_{\text{Threshold}}$ for the $Qs\bar{n}\bar{n}$ system, for both isospin cases $I=0$ and $I=1$.

For the $I=0$ sector, the masses range from $2.511$ to $3.060$~GeV in the charm sector, and from $5.978$ to $6.460$~GeV in the bottom sector. In the charm sector, the $T_{cs\bar{n}\bar{n}}(0,2^{+},3.050)$ state has a positive confinement energy ($E_{\text{CON}} = 10$~MeV), indicating that this high-spin state is unlikely to form a compact structure. For the $I=1$ charm states, the masses are about $200$~MeV higher than their $I=0$ counterparts, and their compactness potential is correspondingly reduced. In particular, the $T_{cs\bar{n}\bar{n}}(1,0^{+},3.219)$ and $T_{cs\bar{n}\bar{n}}(1,2^{+},3.046)$ states both have positive $E_{\text{CON}}$ ($27$ and $7$~MeV, respectively). For the $bs\bar{n}\bar{n}$ system, the $I=0$ masses range from $5.978$ to $6.460$~GeV. Except for $T_{bs\bar{n}\bar{n}}(0,2^{+},6.433)$, which has $E_{\text{CON}} = -31$~MeV, the other $I=0$ states have confinement energies between $-50$ and $-110$~MeV, suggesting good compactness potential. For the $I=1$ bottom states, the masses are again about $200$~MeV higher than the $I=0$ ones, and their compactness potential is correspondingly diminished.

\renewcommand{\tabcolsep}{0.7cm} \renewcommand{\arraystretch}{1.6}
\begin{table*}[!htb]
	\centering
	\caption{The table lists the \( |c_{ij}|^2 \cdot k \) (in \(\mathrm{GeV}\)) values for the \( Qs\bar{n}\bar{n} \) system; the numbers in parentheses can be regarded as branching ratios. A ``\(-\)'' indicates that the mass is below the threshold, and ``\(*\)'' denotes a scattering state.}
	\label{tab:Qsnn_bottom_charmed_decay}
	\begin{tabular}{cccccc}
		\bottomrule[1.5pt]\bottomrule[0.5pt]	
		\toprule
		$I(J^{PC})$ &$M_{bag}$ (GeV) &$B^*/D^*K^*$&$B^*/D^*K$&$B/DK^*$&$B/DK$  \\
		\hline
		\multicolumn{6}{c}{ $cs\bar{n}\bar{n}$}\\
		\cline{1-6}
		\midrule
		
		\multirow{2}{*}{$0(0^+)$}  & 2.511 &- &&&0.098(1) \\
		&  2.919&0.068(0.72)&&&0.094(1)  \\
		\multirow{3}{*}{$0(1^+)$}  &2.672 &-&0.190(1)&-&  \\
		& 2.894 &-&0.042(0.25)&0.167(1)&  \\
		&3.060 &0.197(1)&0.013(0.07)&0.107(0.54)&\\
		$0(2^+)$ &3.050&*&& &\\
		
		\multirow{2}{*}{$1(0^+)$} & 2.769&-&&&0.258(1)\\
		&3.219&0.369(1)&&&0.007(0.02)\\
		\multirow{3}{*}{$1(1^+)$} &2.838&-&0.220(0.72)&0.306(1)&\\
		&2.989&0.011(0.07)&0.009(0.06)&0.156(1)&\\
		&3.155&0.269(1)&0.012(0.04)&0.071(0.26)& \\
		$1(2^+)$ &3.046&*&&& \\
		
		\hline
		\multicolumn{6}{c}{ $bs\bar{n}\bar{n}$}\\
		\cline{1-6}
		\multirow{2}{*}{$0(0^+)$} & 5.978&-&&&0.169(1) \\
		&6.323&0.168(1)&&&0.061(0.36)\\
		\multirow{3}{*}{$0(1^+)$} &6.035&-&0.240(1)&-&\\
		&6.319&0.074(0.65)&0.066(0.58)&0.113(1)&\\
		&6.460&0.197(0.81)&0.002(0.01)&0.243(1)&\\
		$0(2^+)$ &6.433&*&&&\\
		
		\multirow{2}{*}{$1(0^+)$}  &6.194 &-&&&0.289(1)  \\
		& 6.581&0.455(1)&&&0.010(0.02)  \\
		\multirow{3}{*}{$1(1^+)$}   &6.221&$<10^{-4}$&0.279(1)&0.002(0.01)&  \\
		&6.407&0.042(0.26)&0.001(0.01)&0.160(1)&  \\
		&6.555&0.321(1)&0.012(0.04)&0.124(0.39)&  \\
		$1(2^+)$ &6.485&*&&&\\
		\bottomrule
		\bottomrule[0.5pt]\bottomrule[1.5pt]	
	\end{tabular}
	\label{tab:12x7}
\end{table*}

For the $T_{cs\bar{n}\bar{n}}(0,0^{+},2.919)$ state, its mass is close to the experimentally observed $T_{cs0}^{*}(2870)^{0}$ ($M = 2883$~MeV) \cite{LHCb:2024xyx}, and their isospin and $J^P$ assignments are consistent. The confinement energy of this state is only $-6$~MeV, which still allows for the possibility of forming a compact structure, but its small magnitude makes the compact interpretation less convincing in our model.

The $S$-wave strong decay stability within the $c\bar{n}\otimes n\bar{s}$ flavor representation is analyzed in Table~\ref{tab:Qsnn_bottom_charmed_decay}. Several states exhibit small $k\cdot |c_i|^2$ values due to phase-space suppression, indicating narrow decay widths. In particular, the states $T_{cs\bar{n}\bar{n}}(0^{+},2.511)$, $T_{cs\bar{n}\bar{n}}(0^{+},2.672)$, and $T_{bs\bar{n}\bar{n}}(0^{+},5.978)$ deserve experimental attention, as they combine narrow widths with significant compactness potential.

\subsection{$sQ\bar{s}\bar{n}$}

\renewcommand{\tabcolsep}{0.2cm} \renewcommand{\arraystretch}{1.6}
\begin{table*}[!htb]
	\centering
	\caption{For the $sQ\bar{s}\bar{n}$  system, the color-magnetic eigenstates, bag radius $R_{0}$, confinement energy $E_{\mathrm{CON}}$, mass $M_{\mathrm{bag}}$, and the threshold difference with respect to two-meson thresholds are given respectively.}
	\label{tab:bottom_charmedscsn}
	\begin{tabular}{cccccc}
		\bottomrule[1.5pt]\bottomrule[0.5pt]	
		\toprule
		 $ J^{PC}$ & Eigenvector & $ R_0$ (GeV\(^{-1}\))&$E_{CON}$(MeV) & $M_{bag}$ (GeV) & $\delta m=M-M_{Threshold}$ (MeV) \\
		\midrule
		\hline
		\multicolumn{6}{c}{ $sc\bar{s}\bar{n}$}\\
		\cline{1-6}
		\multirow{4}{*}{\( 0^{+} \)} &-0.562,0.013,0.012,0.827 &5.652&9 &3.328 & $\delta_{D_{s}^*K^*} =322;  \delta_{D_{s}K} =864 $ \\
		& -0.004,0.793,0.608,0.010& 5.434 &-41  &3.092 & $\delta_{D_{s}^*K^*} =86;  \delta_{D_{s}K} =628 $  \\
		& -0.824,-0.070,0.025,0.562&5.246&-83 &2.936 & $\delta_{D_{s}^*K^*} =-70;  \delta_{D_{s}K} = 472 $\\
		& -0.046,0.604,-0.795,0.034 &4.996&-137 &2.699 & $\delta_{D_{s}^*K^*} =-307;  \delta_{D_{s}K} = 235 $ \\
		
		\midrule
	     \multirow{6}{*}{\( 1^{+} \)} & -0.100,-0.610,-0.100,-0.641,-0.100,-0.432 &5.208&-92 &2.856 & $\delta_{D_{s}^*K^*} =-150;\delta_{D_{s}^*K} =248;  \delta_{D_{s}K^*} =-6 $\\
		&0.480,-0.263,0.630,0.053,0.541,-0.090 & 5.452&-37  &3.001 &$\delta_{D_{s}^*K^*} = -5;\delta_{D_{s}^*K} = 393;  \delta_{D_{s}K^*} =139 $ \\
		&0.256,-0.024,0.504,0.024,-0.824,0.013& 5.673& 14 & 3.268 &$\delta_{D_{s}^*K^*} = 262;\delta_{D_{s}^*K} =660;  \delta_{D_{s}K^*} = 406 $ \\
		&0.217,0.614,0.096,-0.736,0.089,0.132&5.518&-22  & 3.058 & $\delta_{D_{s}^*K^*} = 52;\delta_{D_{s}^*K} =450;  \delta_{D_{s}K^*} = 196 $\\
		&-0.184,-0.390,0.100,-0.207,0.023,0.872 &5.624&3  & 3.195 & $\delta_{D_{s}^*K^*} = 189;\delta_{D_{s}^*K} =587;  \delta_{D_{s}K^*} =333 $\\
		&0.783,-0.171,-0.566,0.033,-0.094,0.164 &5.571& -10 &3.142& $\delta_{D_{s}^*K^*} = 136;\delta_{D_{s}^*K} =534;  \delta_{D_{s}K^*} = 280 $\\
		
		\midrule
		 \multirow{2}{*}{\( 2^{+} \)} &-0.924,0.334 & 5.678&14 & 3.209 & $\delta_{D_{s}^*K^*} = 203$ \\
		& 0.334,0.943&5.675&14 & 3.227 &$\delta_{D_{s}^*K^*} = 221$   \\
		\hline
		\multicolumn{6}{c}{ $sb\bar{s}\bar{n}$}\\
		\hline
		\midrule
		\multirow{4}{*}{\( 0^{+} \)} 
		 & 0.550,0.019,0.019,0.834 &5.458&-36 &6.684 & $\delta_{B_{s}^*K^*} = 375; \delta_{B_{s}K} = 821 $ \\
		&-0.074,0.778,0.623,0.018&5.283&-75 &6.478 & $\delta_{B_{s}^*K^*} = 169; \delta_{B_{s}K} = 615 $ \\
		& -0.827,-0.115,0.034,0.549 &5.131&-108 & 6.352 & $\delta_{B_{s}^*K^*} = 43; \delta_{B_{s}K} = 489 $  \\
		& -0.077,0.615,-0.783,0.054 &4.943&-149 &6.153 & $\delta_{B_{s}^*K^*} =-156; \delta_{B_{s}K} = 290 $  \\
		
		\midrule
		\multirow{6}{*}{\( 1^{+} \)} &-0.092,-0.621,-0.073,-0.632,-0.082,-0.441 &5.046& -127 &6.217 & $\delta_{B_{s}^*K^*} = -92;\delta_{B_{s}^*K} =306; \delta_{B_{s}K^*} = -44 $\\
		&0.628,-0.179,0.529,-0.021,0.539,-0.036&5.163& -102 & 6.381&$\delta_{B_{s}^*K^*} = 72;\delta_{B_{s}^*K} =470; \delta_{B_{s}K^*} = 120 $ \\
		&0.373,-0.025,0.401,0.023,-0.836,0.013 & 5.482& -31 &6.659 &$\delta_{B_{s}^*K^*} = 350;\delta_{B_{s}^*K} =748; \delta_{B_{s}K^*} = 395 $ \\
		&0.119,0.741,0.042,-0.637,0.031,-0.167 & 5.281& -76 &6.475 &$\delta_{B_{s}^*K^*} = 166;\delta_{B_{s}^*K} =564; \delta_{B_{s}K^*} = 214 $\\
		&-0.246,-0.152,0.259,-0.419,0.020,0.819 &5.447& -39 &6.580&$\delta_{B_{s}^*K^*} = 271;\delta_{B_{s}^*K} =669; \delta_{B_{s}K^*} = 319 $ \\
		& 0.618,-0.097,-0.696,-0.132,-0.053,0.322 &5.435&-41  & 6.550 &$\delta_{B_{s}^*K^*} = 241;\delta_{B_{s}^*K} =639; \delta_{B_{s}K^*} = 289 $\\
		
		\midrule
		\multirow{2}{*}{\( 2^{+} \)} &-0.941,0.339 &5.507&-25 &6.577& $ \delta_{B_{s}^*K^*} = 268$ \\
		& 0.339,0.941 &5.503&-26 & 6.601 &$ \delta_{B_{s}^*K^*} = 292 $  \\
		
		\bottomrule
		\bottomrule[0.5pt]\bottomrule[1.5pt]	
	\end{tabular}
\end{table*}

Table~\ref{tab:bottom_charmedscsn} lists the mass eigenstates, bag radius $R_0$, confinement energy $E_{\text{CON}}$, and masses for the $sQ\bar{s}\bar{n}$ system. In the charm sector, the masses of the $sc\bar{s}\bar{n}$ states range from $2.699$ to $3.328$~GeV, all lying above the $D_s K$ threshold. The $0^+$ states span $2.699$--$3.328$~GeV, with the lightest state $T_{sc\bar{s}\bar{n}}(0^+,2.699)$ having $E_{\text{CON}} = -137$~MeV, while the heaviest $T_{sc\bar{s}\bar{n}}(0^+,3.328)$ has $E_{\text{CON}} = 9$~MeV. For the $1^+$ states, the mass range is $2.856$--$3.268$~GeV; among these, $T_{sc\bar{s}\bar{n}}(1^+,2.856)$ shows $E_{\text{CON}} = -92$~MeV, whereas $T_{sc\bar{s}\bar{n}}(1^+,3.268)$ has a positive value of $14$~MeV. The two $2^+$ states at $3.209$ and $3.227$~GeV both have $E_{\text{CON}} = 14$~MeV. States with positive confinement energies—namely $T_{sc\bar{s}\bar{n}}(0^+,3.328)$, $T_{sc\bar{s}\bar{n}}(1^+,3.268)$, $T_{sc\bar{s}\bar{n}}(1^+,3.195)$, and the two $2^+$ states—are unlikely to form compact structures. In contrast, $T_{sc\bar{s}\bar{n}}(0^+,2.699)$ and $T_{sc\bar{s}\bar{n}}(1^+,2.856)$ exhibit $E_{\text{CON}}$ around $-100$~MeV, suggesting viable compactness potential.

For the $sb\bar{s}\bar{n}$ system, the masses lie between $6.153$ and $6.684$~GeV. The $0^+$ states range from $6.153$ to $6.684$~GeV, with $E_{\text{CON}}$ values between $-36$ and $-149$~MeV; the $1^+$ states span $6.217$--$6.659$~GeV with $E_{\text{CON}}$ from $-31$ to $-127$~MeV; and the two $2^+$ states at $6.577$ and $6.601$~GeV have $E_{\text{CON}} = -25$ and $-26$~MeV, respectively. All states in the bottom sector have negative confinement energies, owing to the suppression of the bag radius by the heavier $b$ quark. This implies that the $sb\bar{s}\bar{n}$ system generally possesses the potential to form compact structures.

\renewcommand{\tabcolsep}{0.3cm} \renewcommand{\arraystretch}{1.5}
\begin{table*}[!htb]
	\centering
	\caption{The table lists the \( |c_{ij}|^2 \cdot k \) (in \(\mathrm{GeV}\)) values for the  $sQ\bar{s}\bar{n}$ system; the numbers in parentheses can be regarded as branching ratios. A ``\(-\)'' indicates that the mass is below the threshold, and ``\(*\)'' denotes a scattering state.}
	\label{tab:bottom_charmedscsn_decay}
	\begin{tabular}{ccccccccc}
		\bottomrule[1.5pt]\bottomrule[0.5pt]	
		\toprule
		\toprule
		\toprule
		\multicolumn{2}{c}{ }&\multicolumn{4}{c}{$c\bar{s} \otimes s\bar{n}$ } &&\multicolumn{2}{c}{$c\bar{n} \otimes s\bar{s}$}\\
		\cline{3-6}
		\cline{8-9}
		
		\( J^{PC} \) & $M_{bag}$ (GeV) &$B_{s}/D_{s}K$&$B^*_{s}/D^*_{s}K$&$B_{s}/D_{s}K^*$& $B^*_{s}/D^*_{s}K^*$&&$B/D\phi$&$B^*/D^*\phi$\\
		\hline
		\multicolumn{9}{c}{ $sc\bar{s}\bar{n}$}\\
		\cline{1-9}
		\multirow{4}{*}{\( 0^{+} \)} &3.328 &0.005(0.02)&&&0.311(1)&&&0.425(1)\\
		&3.092 &0.022(0.13)&&&0.168(1)&&&0.095(1)\\
		& 2.936 &0.186(1)&&&-&&&-\\
		&2.699&0.319(1)&&&-&&&-\\
		\multirow{6}{*}{\( 1^{+} \)} &2.856&&0.383(1)&-&-&&-&-\\
		&3.001 &&0.110(1)&0.006(0.05)&-&&0.039(1)&-\\
		&3.268&&0.004(0.02)&0.066(0.38)&0.176(1)&&0.040(0.10)&0.398(1)\\
		&3.058 &&0.001(0.00)&0.292(1)&0.007(0.02)&&0.143(1)&0.004(0.03)\\
		&3.195 &&0.007(0.02)&0.031(0.10)&0.314(1)&&0.166(1)&0.132(0.80)\\
		&3.142 &&0.009(0.06)&0.163(1)&0.021(0.13)&&0.179(1)&0.004(0.02)\\
		\multirow{2}{*}{\( 2^{+} \)} &3.197&&&&&&&*\\
		&3.240&&&&*&&&\\
		\hline
		\multicolumn{9}{c}{ $sb\bar{s}\bar{n}$}\\
		\cline{1-9}
		\multirow{4}{*}{\( 0^{+} \)} & 6.684 &0.008(0.02)&&&0.349(1)&&&0.552(1)\\
		&6.478&0.020(0.07)&&&0.293(1)&&&0.132(1)\\
		&6.352&0.164(1)&&&0.001(0.01)&&&0.005(1)\\
		&6.153&0.431(1)&&&-&&&-\\
		\multirow{6}{*}{\( 1^{+} \)} &6.217 &&0.466(1)&-&-&&-&-\\
		&6.381&&0.150(1)&0.005(0.03)&0(0.00)&&0.011(1)&0.006(0.55)\\
		&6.659&&0.002(0.01)&0.099(0.44)&0.227(1)&&0.108(0.26)&0.419(1)\\
		&6.475&&0.010(0.04)&0.241(1)&0.090(0.37)&&0.104(1)&0.053(0.51)\\
		&6.580&&0.002(0.07)&0.030(1)&0.003(0.10)&&0.454(1)&0.209(0.46)\\
		&6.550&&0.001(0.00)&0.297(0.89)&0.334(1)&&0.019(1)&0.002(0.11)\\
		\multirow{2}{*}{\( 2^{+} \)} &6.577&&&&&&&*\\
		&6.601&&&&*&&&\\
		
		\bottomrule
		\bottomrule[0.5pt]\bottomrule[1.5pt]	
	\end{tabular}
\end{table*}

Table~\ref{tab:bottom_charmedscsn_decay} presents the $k\cdot |c_i|^2$ values for the $sc\bar{s}\bar{n}$ system, obtained from the $8_c\otimes8_c$ and $1_c\otimes 1_c$ representations. We analyze the color-singlet fractions in the $c\bar{s}\otimes n\bar{s}$ and $c\bar{n}\otimes s\bar{s}$ flavor configurations to evaluate the $S$-wave strong decay stability. Since the scalar $s\bar{s}$ state is typically mixed with other mesons, we only consider the vector $\phi$ meson in the final state. The relevant similarity relations, analogous to those in Eq.~(\ref{gamma}) for the $nc\bar{s}\bar{n}$ system, are
\begin{equation}
	\gamma_{B_{s}/D_{s}K} = \gamma_{B_{s}^*/D_{s}^*K} = \gamma_{B_{s}/D_{s}K^*} = \gamma_{B_{s}^*/D_{s}^*K^*}, \quad 
	\gamma_{B/D\phi} = \gamma_{B^*/D^*\phi}.
	\label{gammas}
\end{equation}
These relations allow us to compare $k\cdot |c_i|^2$ across different final states and thereby determine the partial-width ratios.

Our results indicate that several states retain significant compactness potential while also having threshold-suppressed decay channels with small $k\cdot |c_i|^2$ values. Notable examples include $T_{sc\bar{s}\bar{n}}(0^+,2.699)$, $T_{sc\bar{s}\bar{n}}(0^+,2.936)$, $T_{sc\bar{s}\bar{n}}(1^+,2.856)$, $T_{sb\bar{s}\bar{n}}(0^+,6.153)$, and $T_{sb\bar{s}\bar{n}}(1^+,6.217)$. These states are expected to be of experimental interest.

\subsection{$sQ\bar{s}\bar{s}$}

\renewcommand{\tabcolsep}{0.6cm} \renewcommand{\arraystretch}{1.5}
\begin{table*}[!htb]
	\centering
	\caption{For the $sQ\bar{s}\bar{s}$ system, the bag radius $R_{0}$, the color-magnetic eigenstates, the mass $M_{\mathrm{bag}}$, and the threshold difference $\delta m = M - M_{\mathrm{Threshold}}$ with respect to two-meson thresholds are given respectively.}
	\label{tab:hidden_heavy_flavor_tetraquarks}
	\begin{tabular}{ccccccc}
		\bottomrule[1.5pt]\bottomrule[0.5pt]	
		\toprule
	    \( J^{PC} \) & Eigenvector & \( R_0 \) (GeV\(^{-1}\))&$E_{CON}$(MeV) & $M_{bag}$ (MeV)&$B_{s}/D_{s}\phi$ & $B_{s}^*/D_{s}^*\phi$ \\
		
		\hline
		\multicolumn{7}{c}{ $sc\bar{s}\bar{s}$}\\
		\cline{1-7}
		\midrule
     	\multirow{2}{*}{\( 0^{+} \)} & -0.813,0.582 &5.202&-22 &3.097 & &- \\
		& -0.582,0.813 &5.629&3 &3.441 & &0.373 \\
		\multirow{3}{*}{ \( 1^{+} \)} &0.271,-0.528,0.805 &5.532 &-19 & 3.384 &0.063 &0.057 \\
		&0.832,0.548,0.083 &5.406 &-48 &3.285&0.185 &0.015 \\
		&-0.481,0.650,0.588&5.291 &-73 &3.149 &0.025 &0.001\\
		\( 2^{+} \) & 1.00 &5.702 &21 &3.351& & *\\
		\hline
		\multicolumn{7}{c}{ $sb\bar{s}\bar{s}$}\\
		\cline{1-7}
		\multirow{2}{*}{\( 0^{+} \)} & -0.813,0.582 & 5.086&-118 &6.501 & &0.011 \\
		& -0.813,0.582 &5.435&-41 &6.792 & & 0.458\\
		\multirow{3}{*}{ \( 1^{+} \)}& 0.396,-0.432,0.810&5.389&-52 &6.769& 0.116 &0.332 \\
		&0.680,0.730,0.059&5.292&-73 &6.690 &0.198 &0.048\\
		&-0.616,0.529,0.583&5.133 &-108 &6.523 &0.010 &0.006 \\
		\( 2^{+} \) & 1.00 &5.536&-18  &6.716 & & *\\
		
		\bottomrule
		\bottomrule[0.5pt]\bottomrule[1.5pt]
	\end{tabular}
\end{table*}

Table~\ref{tab:hidden_heavy_flavor_tetraquarks} lists the eigenstates, bag radius, confinement energy, and mass spectrum for the $sQ\bar{s}\bar{s}$ system, along with the $k\cdot |c_i|^2$ factors for the two decay channels involving the $\phi$ meson.

In the charm sector, the $cs\bar{s}\bar{s}$ masses range from $3.097$ to $3.441$~GeV. The $0^+$ states are found at $3.097$ and $3.441$~GeV, with $E_{\text{CON}} = -22$ and $3$~MeV, respectively; the latter has a positive confinement energy and is therefore disfavored as a compact candidate. The $1^+$ multiplet consists of three states at $3.149$, $3.285$, and $3.384$~GeV, with $E_{\text{CON}}$ values of $-73$, $-48$, and $-19$~MeV, respectively, indicating progressively weaker binding. The $2^+$ state at $3.351$~GeV has $E_{\text{CON}} = 21$~MeV, which is positive and thus unfavorable for compactness. Among these, $T_{sc\bar{s}\bar{s}}(1^+,3.149)$ and $T_{sc\bar{s}\bar{s}}(1^+,3.285)$ exhibit the strongest binding and are the most promising compact candidates in this sector.

For the $sb\bar{s}\bar{s}$ system, the masses lie between $6.501$ and $6.792$~GeV. All states have negative confinement energies except the $2^+$ state at $6.716$~GeV, which has $E_{\text{CON}} = -18$~MeV. The $0^+$ states at $6.501$ and $6.792$~GeV have $E_{\text{CON}} = -118$ and $-41$~MeV, respectively; the $1^+$ states at $6.523$, $6.690$, and $6.769$~GeV have $E_{\text{CON}} = -108$, $-73$, and $-52$~MeV. In particular, $T_{sb\bar{s}\bar{s}}(1^+,6.523)$ has a confinement energy of about $-108$~MeV, indicating strong compactness potential. Overall, the bottom sector exhibits significantly deeper binding than its charm counterpart, owing to the suppression of the bag radius by the heavier $b$ quark.

The last two columns of Table~\ref{tab:hidden_heavy_flavor_tetraquarks} provide the $k\cdot |c_i|^2$ values for decays into $D_s^{(*)}\phi$ and $B_s^{(*)}\phi$. Since the scalar $s\bar{s}$ meson is not well established as a pure state, we restrict our discussion to the vector $\phi$ meson in the final state. These factors can be used to estimate the relative partial widths for the corresponding decay channels.

Within the spherical cavity approximation of the MIT bag model, the bag radius of singly charmed systems already approaches $R_c$; any further increase would drive $E_{\mathrm{CON}}$ positive and trigger bag instability. Consequently, compact configurations in the singly charmed sector are inherently precarious: a modest shift in the variational radius can push the system past the confinement threshold. This is consistent with the finding that all high-spin $J^P=2^+$ states yield positive $E_{\mathrm{CON}}$, indicating that compact configurations, even when they occur, lie close to the confinement limit. In contrast, the heavier bottom quark suppresses relativistic effects and provides more negative confinement energies, making compactness in the bottom sector a more robust scenario.

A compact tetraquark has a larger bag radius than a typical meson \cite{Liu:2025fbe}, which already raises its volume energy. As $R_0$ approaches $R_c$, $E_{\mathrm{CON}}$ becomes barely negative or even positive, and the bag is only weakly bound. In this situation, the total mass of the compact state naturally exceeds the sum of the masses of two separated color-singlet mesons, each having a smaller radius and a more negative $E_{\mathrm{CON}}$. Hence, whenever a compact tetraquark exists, its mass in our model is generally higher than its corresponding lowest threshold.

\section{Summary}
\label{sec:summary}

In this work, based on the MIT bag model, we construct a unified variational formulation that incorporates a short-range chromoelectric interaction (CEI) explicitly dependent on the bag radius $R$. Using this model, we first verify the mass spectra of singly, doubly, and fully heavy baryons (Table~\ref{tab:nonchromomagnetic_doubly_heavy_baryons}), finding good agreement with lattice QCD results and experimental data.

Furthermore, we calculate the mass spectra of the singly heavy tetraquark states $nQ\bar{n}\bar{n}$, $nQ\bar{s}\bar{n}$, $Q\bar{s}\bar{n}\bar{n}$, $sQ\bar{s}\bar{n}$, and $sQ\bar{s}\bar{s}$, and analyze their stability against strong decays. In the calculations, we consider two color representations, namely $6_c\otimes\bar{6}_c/\bar{3}_c\otimes 3_c$ and $8_c\otimes8_c/1_c\otimes 1_c$, and the mass spectra obtained from these two representations are strictly consistent. Subsequently, we analyze the $S$-wave strong decay stability by evaluating the weight of the color-singlet component in the eigenstates of the $8_c\otimes8_c/1_c\otimes 1_c$ representation, and provide the partial width ratios.

We use the bag confinement energy $E_{\text{CON}}$ as a criterion for judging the formation of compact structures and analyze the compactness potential of each system. For tetraquark systems, the confinement energy exhibits a strong dependence on the number of light quarks. In particular, for the singly charmed tetraquark system, the bag radius has already approached the critical region near the confinement limit $R_c = 5.615\,\text{GeV}^{-1}$. Our results indicate that some singly heavy tetraquark states, especially the high-spin $S=2$ states in the singly charmed system, have positive confinement energies, which makes it difficult for them to form compact structures. Nonetheless, some states still possess the potential to form compact structures. For the singly bottom tetraquark system, due to the suppression effect of the bottom quark on the bag radius, the compactness is enhanced.

Our results indicate that for singly heavy systems, the masses tend to lie above their corresponding lowest thresholds, and such states are generally unstable against strong decays. In the $nc\bar{s}\bar{n}$ system, the state $T_{c\bar{s}n\bar{n}}(0^+,2.925)$ (the second state in Table~\ref{tab:bottom_charmed_tetraquarks}) has a mass of $2.925$~GeV, which is close to the experimentally observed $T_{c\bar{s}0}^{a}(2900)$ ($M = 2908$~MeV, $J^P = 0^+$). Its dominant decay channel is $D_s\pi$, with $k\cdot|c_i|^2 = 211$~MeV, a value comparable in magnitude to the large experimental width of $T_{c\bar{s}0}^{a}(2900)$ ($\Gamma = 136$~MeV). We therefore suggest that $T_{c\bar{s}0}^{a}(2900)$ could be a candidate for a compact singly heavy tetraquark state, with its large decay width naturally understood as arising from an OZI-superallowed decay mechanis.

Using the MIT bag model, we systematically predict the mass spectra and decay properties of various singly heavy tetraquark states, and identify a set of candidates with significant potential to form compact structures (e.g., $T_{nc\bar{s}\bar{n}}(0^+,2.472)$, $T_{nc\bar{s}\bar{n}}(1^+,2.624)$, $T_{sb\bar{s}\bar{n}}(1^+,6.523)$, etc.). In our model, the confinement energy is obtained under the spherical‑cavity approximation, which provides a macroscopic description. A more precise description would require a deeper understanding of the model and further investigation of the quark–quark interactions.

\medskip
\textbf{ACKNOWLEDGMENTS}

We thanks Wen-Xuan Zhang and Ming Zhu Liu for usefull discussings.
The work is supported by the National Natural Science Foundation of China (Grants No. 12475026 and No.12075193).

\appendix
\section{Color and Spin Wavefunctions}
\label{apd:WF}

Consider the two color wavefunctions in the $\bar{6}_c \otimes 6_c$ and $\bar{3}_c \otimes 3_c$ representations of a tetraquark system:
\begin{align*}
	\phi_1^T &= \frac{1}{\sqrt{6}} (rr\bar{r}\bar{r} + gg\bar{g}\bar{g} + bb\bar{b}\bar{b}) + \frac{1}{2\sqrt{6}} (rb\bar{b}\bar{r} + br\bar{b}\bar{r} + gr\bar{g}\bar{r} + rg\bar{g}\bar{r} \\
	&\quad + gb\bar{b}\bar{g} + bg\bar{b}\bar{g}+ gr\bar{r}\bar{g} + rg\bar{r}\bar{g} + gb\bar{g}\bar{b} + bg\bar{g}\bar{b} + rb\bar{r}\bar{b} + br\bar{r}\bar{b}),\\[6pt]
	\phi_2^T &= \frac{1}{2\sqrt{3}} (rb\bar{b}\bar{r} - br\bar{b}\bar{r} - gr\bar{g}\bar{r} + rg\bar{g}\bar{r} + gb\bar{b}\bar{g} - bg\bar{b}\bar{g} \\
	&\quad + gr\bar{r}\bar{g} - rg\bar{r}\bar{g} - gb\bar{g}\bar{b} + bg\bar{g}\bar{b} - rb\bar{r}\bar{b} + br\bar{r}\bar{b}).
\end{align*}
The two color wavefunctions in the $8_c \otimes 8_c$ and $1_c \otimes 1_c$ color representations are 
\begin{align*}
	\phi_1^{T^{\prime}} &= \frac{1}{3\sqrt{2}} (r\bar{r}r\bar{r} + g\bar{g}g\bar{g} + b\bar{b}b\bar{b}) + \frac{1}{2\sqrt{2}} (b\bar{r}r\bar{b} + g\bar{r}r\bar{g} + r\bar{g}g\bar{r} + g\bar{b}b\bar{g} + b\bar{g}g\bar{b})\\
	&-\frac{1}{6\sqrt{2}} (r\bar{r}g\bar{g} + g\bar{g}r\bar{r} + b\bar{b}g\bar{g} + g\bar{g}b\bar{b} + b\bar{b}r\bar{r} + r\bar{r}b\bar{b}),\\[6pt]
	\phi_2^{T^{\prime}} &= \frac{1}{3} (r\bar{r}r\bar{r} + b\bar{b}b\bar{b} + g\bar{g}g\bar{g} + r\bar{r}b\bar{b} + r\bar{r}g\bar{g} + b\bar{b}r\bar{r} \\
	&\quad + b\bar{b}g\bar{g} + g\bar{g}r\bar{r} + g\bar{g}b\bar{b}),
\end{align*}

For a tetraquark system, the spin basis states can be written as

\begin{equation}\label{Spin}
	\begin{aligned}
	\chi_1^{2,2} &= |(q_1 q_2)_1 (q_3 q_4)_1\rangle_2, 
	\chi_2^{1,1} = |(q_1 q_2)_1 (q_3 q_4)_1\rangle_1, \\
	\chi_3^{0,0} &= |(q_1 q_2)_1 (q_3 q_4)_1\rangle_0, 
	\chi_4^{1,1} = |(q_1 q_2)_1 (q_3 q_4)_0\rangle_1, \\
	\chi_5^{1,1} &= |(q_1 q_2)_0 (q_3 q_4)_1\rangle_1, 
	\chi_6^{0,0} = |(q_1 q_2)_0 (q_3 q_4)_0\rangle_0.\\
	\end{aligned}
\end{equation}
Here $q_1$, $q_2$, $q_3$, and $q_4$ on the right‑hand side denote the quark labels, the subscripts inside parentheses correspond to the spin of the diquark subsystems, and the outermost subscript on the ket represents the total spin of the tetraquark.

The color-spin basis vectors of tetraquarks are described by the algebra of the SU(3) $\otimes$ SU(2) group, ultimately yielding the basis vectors in the color representations $\bar{6}_c \otimes 6_c$ and $\bar{3}_c \otimes 3_c$, as well as in the color representations $8_c \otimes 8_c$ and $1_c \otimes 1_c$, as follows:

\begin{equation}\label{sc36}
	\begin{aligned}
				\phi_1^{T} \chi_{1}^{2,2} &= \left| [q_1q_2]^{6_c}_1 [\bar{q}_3\bar{q}_4]^{\bar{6}_c}_1 \right\rangle_2, \\
				\phi_2 \chi_{1}^{2,2} &= \left| \{q_1q_2\}^{\bar{3}_c}_1 \{\bar{q}_3\bar{q}_4\}^{3_c}_1 \right\rangle_2, \\
				\phi_1^{T} \chi_{2}^{1,1} &= \left| [q_1q_2]^{6_c}_1 [\bar{q}_3\bar{q}_4]^{\bar{6}_c}_1 \right\rangle_1, \\
				\phi_2^{T} \chi_{2}^{1,1} &= \left| \{q_1q_2\}^{\bar{3}_c}_1 \{\bar{q}_3\bar{q}_4\}^{3_c}_1 \right\rangle_1, \\
				\phi_1^{T} \chi_{3}^{0,0} &= \left| [q_1q_2]^{6_c}_1 [\bar{q}_3\bar{q}_4]^{\bar{6}_c}_1 \right\rangle_0, \\
				\phi_2^{T} \chi_{3}^{0,0} &= \left| \{q_1q_2\}^{\bar{3}_c}_1 \{\bar{q}_3\bar{q}_4\}^{3_c}_1 \right\rangle_0, \\
				\phi_1^{T} \chi_{4}^{1,1} &= \left| [q_1q_2]^{6_c}_1 \{\bar{q}_3\bar{q}_4\}^{\bar{6}_c}_0 \right\rangle_1, \\
				\phi_2^{T} \chi_{4}^{1,1} &= \left| \{q_1q_2\}^{\bar{3}_c}_1 [\bar{q}_3\bar{q}_4]^{3_c}_0 \right\rangle_1, \\
				\phi_1^{T} \chi_{5}^{1,1} &= \left| \{q_1q_2\}^{6_c}_0 [\bar{q}_3\bar{q}_4]^{\bar{6}_c}_1 \right\rangle_1, \\
				\phi_2^{T} \chi_{5}^{1,1} &= \left| [q_1q_2]^{\bar{3}_c}_0 \{\bar{q}_3\bar{q}_4\}^{3_c}_1 \right\rangle_1, \\
				\phi_1^{T} \chi_{6}^{0,0} &= \left| \{q_1q_2\}^{6_c}_0 \{\bar{q}_3\bar{q}_4\}^{\bar{6}_c}_0 \right\rangle_0, \\
				\phi_2 \chi_{6}^{0,0} &= \left| [q_1q_2]^{\bar{3}_c}_0 [\bar{q}_3\bar{q}_4]^{3_c}_0 \right\rangle_0,
	\end{aligned}
\end{equation}
here, $\{ \cdot \}$ and $[ \cdot ]$ denote symmetric and antisymmetric flavor configurations, respectively.

\begin{equation}\label{sc18}
	\begin{aligned}
\begin{aligned}
	\phi_1^{T^{\prime}} \chi_{1}^{2,2} &=\left|\left(q_{1} \bar{q}_{2}\right)_{1}^{8_c}\left(q_{3} \bar{q}_{4}\right)_{1}^{8_c}\right\rangle_{2}, \\
	\phi_2^{T^{\prime}} \chi_{1}^{2,2} &=\left|\left(q_{1} \bar{q}_{2}\right)_{1}^{1_c}\left(q_{3} \bar{q}_{4}\right)_{1}^{1_c}\right\rangle_{2}, \\
	\phi_1^{T^{\prime}}\chi_{2}^{1,1} &=\left|\left(q_{1} \bar{q}_{2}\right)_{1}^{8_c}\left(q_{3} \bar{q}_{4}\right)_{1}^{8_c}\right\rangle_{1}, \\
	\phi_2^{T^{\prime}} \chi_{2}^{1,1} &=\left|\left(q_{1} \bar{q}_{2}\right)_{1}^{1_c}\left(q_{3} \bar{q}_{4}\right)_{1}^{1_c}\right\rangle_{1}, \\
	\phi_1^{T^{\prime}} \chi_{3}^{0,0} &=\left|\left(q_{1} \bar{q}_{2}\right)_{1}^{8_c}\left(q_{3} \bar{q}_{4}\right)_{1}^{8_c}\right\rangle_{0}, \\
	\phi_2^{T^{\prime}} \chi_{3}^{0,0} &=\left|\left(q_{1} \bar{q}_{2}\right)_{1}^{1_c}\left(q_{3} \bar{q}_{4}\right)_{1}^{1_c}\right\rangle_{0}, \\
	\phi_1^{T^{\prime}} \chi_{4}^{1,1} &=\left|\left(q_{1} \bar{q}_{2}\right)_{1}^{8_c}\left(q_{3} \bar{q}_{4}\right)_{0}^{8_c}\right\rangle_{1}, \\
	\phi_2^{T^{\prime}} \chi_{4}^{1,1} &=\left|\left(q_{1} \bar{q}_{2}\right)_{1}^{1_c}\left(q_{3} \bar{q}_{4}\right)_{0}^{1_c}\right\rangle_{1}, \\
	\phi_1^{T^{\prime}} \chi_{5}^{1,1} &=\left|\left(q_{1} \bar{q}_{2}\right)_{0}^{8_c}\left(q_{3} \bar{q}_{4}\right)_{1}^{8_c}\right\rangle_{1}, \\
	\phi_2^{T^{\prime}} \chi_{5}^{1,1} &=\left|\left(q_{1} \bar{q}_{2}\right)_{0}^{1_c}\left(q_{3} \bar{q}_{4}\right)_{1}^{1_c}\right\rangle_{1}, \\
	\phi_1^{T^{\prime}} \chi_{6}^{0,0} &=\left|\left(q_{1} \bar{q}_{2}\right)_{0}^{8_c}\left(q_{3} \bar{q}_{4}\right)_{0}^{8_c}\right\rangle_{0}, \\
	\phi_2^{T^{\prime}} \chi_{6}^{0,0} &=\left|\left(q_{1} \bar{q}_{2}\right)_{0}^{1_c}\left(q_{3} \bar{q}_{4}\right)_{0}^{1_c}\right\rangle_{0}.
\end{aligned}
	\end{aligned}
\end{equation}

\renewcommand{\tabcolsep}{1.4cm} \renewcommand{\arraystretch}{1.6}
\begin{table*}[htbp]
	\centering
	\caption{Symmetry-adapted color-spin basis vectors for tetraquarks.}
	\label{tab:state_mixing}
	\begin{tabular}{ccc}
		\toprule[1.5pt]
		State & $J^{PC}$ & Allowed states for mixing \\
			\hline
		\multicolumn{3}{c}{The basis vectors of the color representations $6_c \otimes \bar{6}_c$ and $3_c \otimes \bar{3}_c$}\\
		\hline
		\addlinespace[3pt]
		\multirow{3}{*}{$Qs\bar{s}\bar{s}$} 
		& $0^{+}$ & $(\phi_{2}^{T}\chi_{3}^{0,0},\phi_{1}^{T}\chi_{6}^{0,0})$ \\
		& $1^{+}$ & $(\phi_{2}^{T}\chi_{2}^{1,1},\phi_{2}^{T}\chi_{5}^{1,1},\phi_{1}^{T}\chi_{4}^{1,1})$ \\
		
		& $2^{+}$ & $(\phi_{2}^{T}\chi_{1}^{2,2})$ \\
		
		\addlinespace[4pt]
		\multirow{3}{*}{$Qs\bar{s}\bar{n}$,$Qn\bar{s}\bar{n}$} 
		& $0^{+}$ & $(\phi_{2}^{T}\chi_{3}^{0,0},\phi_{2}^{T}\chi_{6}^{0,0},\phi_{1}^{T}\chi_{3}^{0,0},\phi_{1}^{T}\chi_{6}^{0,0})$ \\
		& $1^{+}$ & $(\phi_{2}^{T}\chi_{2}^{1,1},\phi_{2}^{T}\chi_{4}^{1,1},\phi_{2}^{T}\chi_{5}^{1,1},\phi_{1}^{T}\chi_{2}^{1,1},\phi_{1}^{T}\chi_{4}^{1,1},\phi_{1}^{T}\chi_{5}^{1,1})$ \\
    	& $2^{+}$ & $(\phi_{2}^{T}\chi_{1}^{2,2},\phi_{1}^{T}\chi_{1}^{2,2})$ \\
	
	\addlinespace[4pt]
	\multirow{3}{*}{$Qn\bar{n}\bar{n}/Qs\bar{n}\bar{n}(I_{n\bar{n}}=0)$} 
	& $0^{+}$ & $(\phi_{1}^{T}\chi_{3}^{0,0},\phi_{2}^{T}\chi_{6}^{0,0})$ \\
	& $1^{+}$ & $(\phi_{1}^{T}\chi_{2}^{1,1},\phi_{1}^{T}\chi_{4}^{1,1},\phi_{2}^{T}\chi_{5}^{1,1})$ \\
	& $2^{+}$ & $(\phi_{1}^{1}\chi_{1}^{2,2})$ \\
	
	\addlinespace[4pt]
	\multirow{3}{*}{$Qn\bar{n}\bar{n}/Qs\bar{n}\bar{n}(I_{n\bar{n}}=1)$} 
	& $0^{+}$ & $(\phi_{2}^{T}\chi_{3}^{0,0},\phi_{1}^{T}\chi_{6}^{0,0})$ \\
	& $1^{+}$ & $(\phi_{2}^{T}\chi_{2}^{1,1},\phi_{2}^{T}\chi_{4}^{1,1},\phi_{1}^{T}\chi_{5}^{1,1})$ \\
	& $2^{+}$ & $(\phi_{2}^{T}\chi_{1}^{2,2})$ \\
	
		\hline
	\multicolumn{3}{c}{The basis vectors of the color representations $8_c \otimes 8_c$ and $1_c \otimes 1_c$}\\
	\cline{1-3}
	\addlinespace[3pt]
	\multirow{3}{*}{$Qq\bar{q}\bar{q}(q=u,d,s)$} 
	& $0^{+}$ & $(\phi_1^{T^{\prime}} \chi_{3}^{0,0},\phi_2^{T^{\prime}}\chi_{3}^{0,0},\phi_1^{T^{\prime}}\chi_{6}^{0,0},\phi_2^{T^{\prime}}\chi_{6}^{0,0})$\\
	& $1^{+}$ & $(\phi_1^{T^{\prime}} \chi_{2}^{1,1},\phi_1^{T^{\prime}} \chi_{4}^{1,1},\phi_1^{T^{\prime}} \chi_{5}^{1,1},\phi_2^{T^{\prime}} \chi_{2}^{1,1},\phi_2^{T^{\prime}} \chi_{4}^{1,1},\phi_2^{T^{\prime}} \chi_{5}^{1,1})$ \\
	
	& $2^{+}$ & $(\phi_1^{T^{\prime}}\chi_{1}^{2,2},\phi_2^{T^{\prime}}\chi_{1}^{2,2})$ \\

		\bottomrule[1.5pt]
	\end{tabular}
\end{table*}

Finally, by taking into account flavor symmetry and the Pauli exclusion principle, we present the eigenbasis corresponding to different tetraquark systems, summarized in Table \ref{tab:state_mixing}.


\begin{thebibliography}{89}%
	\makeatletter
	\providecommand \@ifxundefined [1]{%
		\@ifx{#1\undefined}
	}%
	\providecommand \@ifnum [1]{%
		\ifnum #1\expandafter \@firstoftwo
		\else \expandafter \@secondoftwo
		\fi
	}%
	\providecommand \@ifx [1]{%
		\ifx #1\expandafter \@firstoftwo
		\else \expandafter \@secondoftwo
		\fi
	}%
	\providecommand \natexlab [1]{#1}%
	\providecommand \enquote  [1]{``#1''}%
	\providecommand \bibnamefont  [1]{#1}%
	\providecommand \bibfnamefont [1]{#1}%
	\providecommand \citenamefont [1]{#1}%
	\providecommand \href@noop [0]{\@secondoftwo}%
	\providecommand \href [0]{\begingroup \@sanitize@url \@href}%
	\providecommand \@href[1]{\@@startlink{#1}\@@href}%
	\providecommand \@@href[1]{\endgroup#1\@@endlink}%
	\providecommand \@sanitize@url [0]{\catcode `\\12\catcode `\$12\catcode
		`\&12\catcode `\#12\catcode `\^12\catcode `\_12\catcode `\%12\relax}%
	\providecommand \@@startlink[1]{}%
	\providecommand \@@endlink[0]{}%
	\providecommand \url  [0]{\begingroup\@sanitize@url \@url }%
	\providecommand \@url [1]{\endgroup\@href {#1}{\urlprefix }}%
	\providecommand \urlprefix  [0]{URL }%
	\providecommand \Eprint [0]{\href }%
	\providecommand \doibase [0]{http://dx.doi.org/}%
	\providecommand \selectlanguage [0]{\@gobble}%
	\providecommand \bibinfo  [0]{\@secondoftwo}%
	\providecommand \bibfield  [0]{\@secondoftwo}%
	\providecommand \translation [1]{[#1]}%
	\providecommand \BibitemOpen [0]{}%
	\providecommand \bibitemStop [0]{}%
	\providecommand \bibitemNoStop [0]{.\EOS\space}%
	\providecommand \EOS [0]{\spacefactor3000\relax}%
	\providecommand \BibitemShut  [1]{\csname bibitem#1\endcsname}%
	\let\auto@bib@innerbib\@empty
	\bibitem [{\citenamefont {Gell-Mann}(1964)}]{Gell-Mann:1964ewy}%
	\BibitemOpen
	\bibfield  {author} {\bibinfo {author} {\bibfnamefont {M.}~\bibnamefont
			{Gell-Mann}},\ }\href {\doibase 10.1016/S0031-9163(64)92001-3} {\bibfield
		{journal} {\bibinfo  {journal} {Phys. Lett.}\ }\textbf {\bibinfo {volume}
			{8}},\ \bibinfo {pages} {214} (\bibinfo {year} {1964})}\BibitemShut {NoStop}%
	\bibitem [{\citenamefont {Zweig}(1964)}]{Zweig:1964ruk}%
	\BibitemOpen
	\bibfield  {author} {\bibinfo {author} {\bibfnamefont {G.}~\bibnamefont
			{Zweig}},\ }\href@noop {} {\bibfield  {journal} {\bibinfo  {journal}
			{CERN-TH-401}\ } (\bibinfo {year} {1964})}\BibitemShut {NoStop}%
	\bibitem [{\citenamefont {Choi}\ \emph {et~al.}(2003)\citenamefont {Choi} \emph
		{et~al.}}]{Belle:2003nnu}%
	\BibitemOpen
	\bibfield  {author} {\bibinfo {author} {\bibfnamefont {S.~K.}\ \bibnamefont
			{Choi}} \emph {et~al.} (\bibinfo {collaboration} {Belle}),\ }\href {\doibase
		10.1103/PhysRevLett.91.262001} {\bibfield  {journal} {\bibinfo  {journal}
			{Phys. Rev. Lett.}\ }\textbf {\bibinfo {volume} {91}},\ \bibinfo {pages}
		{262001} (\bibinfo {year} {2003})},\ \Eprint
	{http://arxiv.org/abs/hep-ex/0309032} {arXiv:hep-ex/0309032} \BibitemShut
	{NoStop}%
	\bibitem [{\citenamefont {Ablikim}\ \emph {et~al.}(2013)\citenamefont {Ablikim}
		\emph {et~al.}}]{BESIII:2013ris}%
	\BibitemOpen
	\bibfield  {author} {\bibinfo {author} {\bibfnamefont {M.}~\bibnamefont
			{Ablikim}} \emph {et~al.} (\bibinfo {collaboration} {BESIII}),\ }\href
	{\doibase 10.1103/PhysRevLett.110.252001} {\bibfield  {journal} {\bibinfo
			{journal} {Phys. Rev. Lett.}\ }\textbf {\bibinfo {volume} {110}},\ \bibinfo
		{pages} {252001} (\bibinfo {year} {2013})},\ \Eprint
	{http://arxiv.org/abs/1303.5949} {arXiv:1303.5949 [hep-ex]} \BibitemShut
	{NoStop}%
	\bibitem [{\citenamefont {Aaij}\ \emph
		{et~al.}(2022{\natexlab{a}})\citenamefont {Aaij} \emph
		{et~al.}}]{LHCb:2021auc}%
	\BibitemOpen
	\bibfield  {author} {\bibinfo {author} {\bibfnamefont {R.}~\bibnamefont
			{Aaij}} \emph {et~al.} (\bibinfo {collaboration} {LHCb}),\ }\href {\doibase
		10.1038/s41467-022-30206-w} {\bibfield  {journal} {\bibinfo  {journal}
			{Nature Commun.}\ }\textbf {\bibinfo {volume} {13}},\ \bibinfo {pages} {3351}
		(\bibinfo {year} {2022}{\natexlab{a}})},\ \Eprint
	{http://arxiv.org/abs/2109.01056} {arXiv:2109.01056 [hep-ex]} \BibitemShut
	{NoStop}%
	\bibitem [{\citenamefont {Aaij}(2015)}]{PhysRevLett.115.072001}%
	\BibitemOpen
	\bibfield  {author} {\bibinfo {author} {\bibfnamefont {R.~e.~a.}\
			\bibnamefont {Aaij}} (\bibinfo {collaboration} {LHCb Collaboration}),\ }\href
	{\doibase 10.1103/PhysRevLett.115.072001} {\bibfield  {journal} {\bibinfo
			{journal} {Phys. Rev. Lett.}\ }\textbf {\bibinfo {volume} {115}},\ \bibinfo
		{pages} {072001} (\bibinfo {year} {2015})}\BibitemShut {NoStop}%
	\bibitem [{\citenamefont {Aaij}\ \emph {et~al.}(2019)\citenamefont {Aaij} \emph
		{et~al.}}]{LHCb:2019kea}%
	\BibitemOpen
	\bibfield  {author} {\bibinfo {author} {\bibfnamefont {R.}~\bibnamefont
			{Aaij}} \emph {et~al.} (\bibinfo {collaboration} {LHCb}),\ }\href {\doibase
		10.1103/PhysRevLett.122.222001} {\bibfield  {journal} {\bibinfo  {journal}
			{Phys. Rev. Lett.}\ }\textbf {\bibinfo {volume} {122}},\ \bibinfo {pages}
		{222001} (\bibinfo {year} {2019})},\ \Eprint
	{http://arxiv.org/abs/1904.03947} {arXiv:1904.03947 [hep-ex]} \BibitemShut
	{NoStop}%
	\bibitem [{\citenamefont {Aaij}\ \emph {et~al.}(2021)\citenamefont {Aaij} \emph
		{et~al.}}]{LHCb:2020jpq}%
	\BibitemOpen
	\bibfield  {author} {\bibinfo {author} {\bibfnamefont {R.}~\bibnamefont
			{Aaij}} \emph {et~al.} (\bibinfo {collaboration} {LHCb}),\ }\href {\doibase
		10.1016/j.scib.2021.02.030} {\bibfield  {journal} {\bibinfo  {journal} {Sci.
				Bull.}\ }\textbf {\bibinfo {volume} {66}},\ \bibinfo {pages} {1278} (\bibinfo
		{year} {2021})},\ \Eprint {http://arxiv.org/abs/2012.10380} {arXiv:2012.10380
		[hep-ex]} \BibitemShut {NoStop}%
	\bibitem [{\citenamefont {Aaij}\ \emph
		{et~al.}(2022{\natexlab{b}})\citenamefont {Aaij} \emph
		{et~al.}}]{LHCb:2021chn}%
	\BibitemOpen
	\bibfield  {author} {\bibinfo {author} {\bibfnamefont {R.}~\bibnamefont
			{Aaij}} \emph {et~al.} (\bibinfo {collaboration} {LHCb}),\ }\href {\doibase
		10.1103/PhysRevLett.128.062001} {\bibfield  {journal} {\bibinfo  {journal}
			{Phys. Rev. Lett.}\ }\textbf {\bibinfo {volume} {128}},\ \bibinfo {pages}
		{062001} (\bibinfo {year} {2022}{\natexlab{b}})},\ \Eprint
	{http://arxiv.org/abs/2108.04720} {arXiv:2108.04720 [hep-ex]} \BibitemShut
	{NoStop}%
	\bibitem [{\citenamefont {Aaij}\ \emph
		{et~al.}(2023{\natexlab{a}})\citenamefont {Aaij} \emph
		{et~al.}}]{LHCb:2022ogu}%
	\BibitemOpen
	\bibfield  {author} {\bibinfo {author} {\bibfnamefont {R.}~\bibnamefont
			{Aaij}} \emph {et~al.} (\bibinfo {collaboration} {LHCb}),\ }\href {\doibase
		10.1103/PhysRevLett.131.031901} {\bibfield  {journal} {\bibinfo  {journal}
			{Phys. Rev. Lett.}\ }\textbf {\bibinfo {volume} {131}},\ \bibinfo {pages}
		{031901} (\bibinfo {year} {2023}{\natexlab{a}})},\ \Eprint
	{http://arxiv.org/abs/2210.10346} {arXiv:2210.10346 [hep-ex]} \BibitemShut
	{NoStop}%
	\bibitem [{\citenamefont {Cheng}\ \emph {et~al.}(2026)\citenamefont {Cheng},
		\citenamefont {Lin}, \citenamefont {Wang},\ and\ \citenamefont
		{Zhu}}]{Cheng:2026cgo}%
	\BibitemOpen
	\bibfield  {author} {\bibinfo {author} {\bibfnamefont {J.-B.}\ \bibnamefont
			{Cheng}}, \bibinfo {author} {\bibfnamefont {Z.-Y.}\ \bibnamefont {Lin}},
		\bibinfo {author} {\bibfnamefont {J.-Z.}\ \bibnamefont {Wang}}, \ and\
		\bibinfo {author} {\bibfnamefont {S.-L.}\ \bibnamefont {Zhu}},\ }\href
	{\doibase 10.1103/v5jm-59g9} {\bibfield  {journal} {\bibinfo  {journal}
			{Phys. Rev. D}\ }\textbf {\bibinfo {volume} {113}},\ \bibinfo {pages}
		{096001} (\bibinfo {year} {2026})},\ \Eprint
	{http://arxiv.org/abs/2601.20740} {arXiv:2601.20740 [hep-ph]} \BibitemShut
	{NoStop}%
	\bibitem [{\citenamefont {Shen}\ \emph {et~al.}(2026)\citenamefont {Shen},
		\citenamefont {Liu}, \citenamefont {Liu}, \citenamefont {Shi}, \citenamefont
		{Xiao}, \citenamefont {Liang},\ and\ \citenamefont {Geng}}]{Shen:2025jmy}%
	\BibitemOpen
	\bibfield  {author} {\bibinfo {author} {\bibfnamefont {Y.-B.}\ \bibnamefont
			{Shen}}, \bibinfo {author} {\bibfnamefont {Z.-W.}\ \bibnamefont {Liu}},
		\bibinfo {author} {\bibfnamefont {M.-Z.}\ \bibnamefont {Liu}}, \bibinfo
		{author} {\bibfnamefont {R.-X.}\ \bibnamefont {Shi}}, \bibinfo {author}
		{\bibfnamefont {C.-W.}\ \bibnamefont {Xiao}}, \bibinfo {author}
		{\bibfnamefont {W.-H.}\ \bibnamefont {Liang}}, \ and\ \bibinfo {author}
		{\bibfnamefont {L.-S.}\ \bibnamefont {Geng}},\ }\href {\doibase
		10.1103/rnsq-c7j6} {\bibfield  {journal} {\bibinfo  {journal} {Phys. Rev. D}\
		}\textbf {\bibinfo {volume} {113}},\ \bibinfo {pages} {074034} (\bibinfo
		{year} {2026})},\ \Eprint {http://arxiv.org/abs/2512.24247} {arXiv:2512.24247
		[hep-ph]} \BibitemShut {NoStop}%
	\bibitem [{\citenamefont {Liu}\ \emph {et~al.}(2025{\natexlab{a}})\citenamefont
		{Liu}, \citenamefont {Pan}, \citenamefont {Liu}, \citenamefont {Wu},
		\citenamefont {Lu},\ and\ \citenamefont {Geng}}]{LIU20251}%
	\BibitemOpen
	\bibfield  {author} {\bibinfo {author} {\bibfnamefont {M.-Z.}\ \bibnamefont
			{Liu}}, \bibinfo {author} {\bibfnamefont {Y.-W.}\ \bibnamefont {Pan}},
		\bibinfo {author} {\bibfnamefont {Z.-W.}\ \bibnamefont {Liu}}, \bibinfo
		{author} {\bibfnamefont {T.-W.}\ \bibnamefont {Wu}}, \bibinfo {author}
		{\bibfnamefont {J.-X.}\ \bibnamefont {Lu}}, \ and\ \bibinfo {author}
		{\bibfnamefont {L.-S.}\ \bibnamefont {Geng}},\ }\href {\doibase
		https://doi.org/10.1016/j.physrep.2024.12.001} {\bibfield  {journal}
		{\bibinfo  {journal} {Physics Reports}\ }\textbf {\bibinfo {volume} {1108}},\
		\bibinfo {pages} {1} (\bibinfo {year} {2025}{\natexlab{a}})},\ \bibinfo
	{note} {three ways to decipher the nature of exotic hadrons: Multiplets,
		three-body hadronic molecules, and correlation functions}\BibitemShut
	{NoStop}%
	\bibitem [{\citenamefont {Chen}\ \emph
		{et~al.}(2016{\natexlab{a}})\citenamefont {Chen}, \citenamefont {Chen},
		\citenamefont {Liu},\ and\ \citenamefont {Zhu}}]{CHEN20161}%
	\BibitemOpen
	\bibfield  {author} {\bibinfo {author} {\bibfnamefont {H.-X.}\ \bibnamefont
			{Chen}}, \bibinfo {author} {\bibfnamefont {W.}~\bibnamefont {Chen}}, \bibinfo
		{author} {\bibfnamefont {X.}~\bibnamefont {Liu}}, \ and\ \bibinfo {author}
		{\bibfnamefont {S.-L.}\ \bibnamefont {Zhu}},\ }\href {\doibase
		https://doi.org/10.1016/j.physrep.2016.05.004} {\bibfield  {journal}
		{\bibinfo  {journal} {Physics Reports}\ }\textbf {\bibinfo {volume} {639}},\
		\bibinfo {pages} {1} (\bibinfo {year} {2016}{\natexlab{a}})},\ \bibinfo
	{note} {the hidden-charm pentaquark and tetraquark states}\BibitemShut
	{NoStop}%
	\bibitem [{\citenamefont {Brambilla}\ \emph {et~al.}(2020)\citenamefont
		{Brambilla}, \citenamefont {Eidelman}, \citenamefont {Hanhart}, \citenamefont
		{Nefediev}, \citenamefont {Shen}, \citenamefont {Thomas}, \citenamefont
		{Vairo},\ and\ \citenamefont {Yuan}}]{BRAMBILLA20201}%
	\BibitemOpen
	\bibfield  {author} {\bibinfo {author} {\bibfnamefont {N.}~\bibnamefont
			{Brambilla}}, \bibinfo {author} {\bibfnamefont {S.}~\bibnamefont {Eidelman}},
		\bibinfo {author} {\bibfnamefont {C.}~\bibnamefont {Hanhart}}, \bibinfo
		{author} {\bibfnamefont {A.}~\bibnamefont {Nefediev}}, \bibinfo {author}
		{\bibfnamefont {C.-P.}\ \bibnamefont {Shen}}, \bibinfo {author}
		{\bibfnamefont {C.~E.}\ \bibnamefont {Thomas}}, \bibinfo {author}
		{\bibfnamefont {A.}~\bibnamefont {Vairo}}, \ and\ \bibinfo {author}
		{\bibfnamefont {C.-Z.}\ \bibnamefont {Yuan}},\ }\href {\doibase
		https://doi.org/10.1016/j.physrep.2020.05.001} {\bibfield  {journal}
		{\bibinfo  {journal} {Physics Reports}\ }\textbf {\bibinfo {volume} {873}},\
		\bibinfo {pages} {1} (\bibinfo {year} {2020})},\ \bibinfo {note} {the XYZ
		states: experimental and theoretical status and perspectives}\BibitemShut
	{NoStop}%
	\bibitem [{\citenamefont {Guo}\ \emph {et~al.}(2018)\citenamefont {Guo},
		\citenamefont {Hanhart}, \citenamefont {Mei\ss{}ner}, \citenamefont {Wang},
		\citenamefont {Zhao},\ and\ \citenamefont {Zou}}]{RevModPhys.90.015004}%
	\BibitemOpen
	\bibfield  {author} {\bibinfo {author} {\bibfnamefont {F.-K.}\ \bibnamefont
			{Guo}}, \bibinfo {author} {\bibfnamefont {C.}~\bibnamefont {Hanhart}},
		\bibinfo {author} {\bibfnamefont {U.-G.}\ \bibnamefont {Mei\ss{}ner}},
		\bibinfo {author} {\bibfnamefont {Q.}~\bibnamefont {Wang}}, \bibinfo {author}
		{\bibfnamefont {Q.}~\bibnamefont {Zhao}}, \ and\ \bibinfo {author}
		{\bibfnamefont {B.-S.}\ \bibnamefont {Zou}},\ }\href {\doibase
		10.1103/RevModPhys.90.015004} {\bibfield  {journal} {\bibinfo  {journal}
			{Rev. Mod. Phys.}\ }\textbf {\bibinfo {volume} {90}},\ \bibinfo {pages}
		{015004} (\bibinfo {year} {2018})}\BibitemShut {NoStop}%
	\bibitem [{\citenamefont {L\"u}\ \emph {et~al.}(2020)\citenamefont {L\"u},
		\citenamefont {Chen},\ and\ \citenamefont {Dong}}]{Lu:2020cns}%
	\BibitemOpen
	\bibfield  {author} {\bibinfo {author} {\bibfnamefont {Q.-F.}\ \bibnamefont
			{L\"u}}, \bibinfo {author} {\bibfnamefont {D.-Y.}\ \bibnamefont {Chen}}, \
		and\ \bibinfo {author} {\bibfnamefont {Y.-B.}\ \bibnamefont {Dong}},\ }\href
	{\doibase 10.1140/epjc/s10052-020-08454-1} {\bibfield  {journal} {\bibinfo
			{journal} {Eur. Phys. J. C}\ }\textbf {\bibinfo {volume} {80}},\ \bibinfo
		{pages} {871} (\bibinfo {year} {2020})},\ \Eprint
	{http://arxiv.org/abs/2006.14445} {arXiv:2006.14445 [hep-ph]} \BibitemShut
	{NoStop}%
	\bibitem [{\citenamefont {Tiwari}\ \emph {et~al.}(2023)\citenamefont {Tiwari},
		\citenamefont {Rathaud},\ and\ \citenamefont {Rai}}]{Tiwari:2021tmz}%
	\BibitemOpen
	\bibfield  {author} {\bibinfo {author} {\bibfnamefont {R.}~\bibnamefont
			{Tiwari}}, \bibinfo {author} {\bibfnamefont {D.~P.}\ \bibnamefont {Rathaud}},
		\ and\ \bibinfo {author} {\bibfnamefont {A.~K.}\ \bibnamefont {Rai}},\ }\href
	{\doibase 10.1007/s12648-022-02427-8} {\bibfield  {journal} {\bibinfo
			{journal} {Indian J. Phys.}\ }\textbf {\bibinfo {volume} {97}},\ \bibinfo
		{pages} {943} (\bibinfo {year} {2023})},\ \Eprint
	{http://arxiv.org/abs/2108.04017} {arXiv:2108.04017 [hep-ph]} \BibitemShut
	{NoStop}%
	\bibitem [{\citenamefont {Liu}\ \emph {et~al.}(2021)\citenamefont {Liu},
		\citenamefont {Liu}, \citenamefont {Zhong},\ and\ \citenamefont
		{Zhao}}]{Liu:2021rtn}%
	\BibitemOpen
	\bibfield  {author} {\bibinfo {author} {\bibfnamefont {F.-X.}\ \bibnamefont
			{Liu}}, \bibinfo {author} {\bibfnamefont {M.-S.}\ \bibnamefont {Liu}},
		\bibinfo {author} {\bibfnamefont {X.-H.}\ \bibnamefont {Zhong}}, \ and\
		\bibinfo {author} {\bibfnamefont {Q.}~\bibnamefont {Zhao}},\ }\href {\doibase
		10.1103/PhysRevD.104.116029} {\bibfield  {journal} {\bibinfo  {journal}
			{Phys. Rev. D}\ }\textbf {\bibinfo {volume} {104}},\ \bibinfo {pages}
		{116029} (\bibinfo {year} {2021})},\ \Eprint
	{http://arxiv.org/abs/2110.09052} {arXiv:2110.09052 [hep-ph]} \BibitemShut
	{NoStop}%
	\bibitem [{\citenamefont {Karliner}\ and\ \citenamefont
		{Rosner}(2020{\natexlab{a}})}]{Karliner:2020dta}%
	\BibitemOpen
	\bibfield  {author} {\bibinfo {author} {\bibfnamefont {M.}~\bibnamefont
			{Karliner}}\ and\ \bibinfo {author} {\bibfnamefont {J.~L.}\ \bibnamefont
			{Rosner}},\ }\href {\doibase 10.1103/PhysRevD.102.114039} {\bibfield
		{journal} {\bibinfo  {journal} {Phys. Rev. D}\ }\textbf {\bibinfo {volume}
			{102}},\ \bibinfo {pages} {114039} (\bibinfo {year} {2020}{\natexlab{a}})},\
	\Eprint {http://arxiv.org/abs/2009.04429} {arXiv:2009.04429 [hep-ph]}
	\BibitemShut {NoStop}%
	\bibitem [{\citenamefont {Faustov}\ \emph {et~al.}(2020)\citenamefont
		{Faustov}, \citenamefont {Galkin},\ and\ \citenamefont
		{Savchenko}}]{Faustov:2020qfm}%
	\BibitemOpen
	\bibfield  {author} {\bibinfo {author} {\bibfnamefont {R.~N.}\ \bibnamefont
			{Faustov}}, \bibinfo {author} {\bibfnamefont {V.~O.}\ \bibnamefont {Galkin}},
		\ and\ \bibinfo {author} {\bibfnamefont {E.~M.}\ \bibnamefont {Savchenko}},\
	}\href {\doibase 10.1103/PhysRevD.102.114030} {\bibfield  {journal} {\bibinfo
			{journal} {Phys. Rev. D}\ }\textbf {\bibinfo {volume} {102}},\ \bibinfo
		{pages} {114030} (\bibinfo {year} {2020})},\ \Eprint
	{http://arxiv.org/abs/2009.13237} {arXiv:2009.13237 [hep-ph]} \BibitemShut
	{NoStop}%
	\bibitem [{\citenamefont {Chen}\ \emph {et~al.}(2023)\citenamefont {Chen},
		\citenamefont {Chen}, \citenamefont {Liu}, \citenamefont {Liu},\ and\
		\citenamefont {Zhu}}]{Chen:2022asf}%
	\BibitemOpen
	\bibfield  {author} {\bibinfo {author} {\bibfnamefont {H.-X.}\ \bibnamefont
			{Chen}}, \bibinfo {author} {\bibfnamefont {W.}~\bibnamefont {Chen}}, \bibinfo
		{author} {\bibfnamefont {X.}~\bibnamefont {Liu}}, \bibinfo {author}
		{\bibfnamefont {Y.-R.}\ \bibnamefont {Liu}}, \ and\ \bibinfo {author}
		{\bibfnamefont {S.-L.}\ \bibnamefont {Zhu}},\ }\href {\doibase
		10.1088/1361-6633/aca3b6} {\bibfield  {journal} {\bibinfo  {journal} {Rept.
				Prog. Phys.}\ }\textbf {\bibinfo {volume} {86}},\ \bibinfo {pages} {026201}
		(\bibinfo {year} {2023})},\ \Eprint {http://arxiv.org/abs/2204.02649}
	{arXiv:2204.02649 [hep-ph]} \BibitemShut {NoStop}%
	\bibitem [{\citenamefont {Liu}\ \emph {et~al.}(2019)\citenamefont {Liu},
		\citenamefont {Chen}, \citenamefont {Chen}, \citenamefont {Liu},\ and\
		\citenamefont {Zhu}}]{Liu:2019zoy}%
	\BibitemOpen
	\bibfield  {author} {\bibinfo {author} {\bibfnamefont {Y.-R.}\ \bibnamefont
			{Liu}}, \bibinfo {author} {\bibfnamefont {H.-X.}\ \bibnamefont {Chen}},
		\bibinfo {author} {\bibfnamefont {W.}~\bibnamefont {Chen}}, \bibinfo {author}
		{\bibfnamefont {X.}~\bibnamefont {Liu}}, \ and\ \bibinfo {author}
		{\bibfnamefont {S.-L.}\ \bibnamefont {Zhu}},\ }\href {\doibase
		10.1016/j.ppnp.2019.04.003} {\bibfield  {journal} {\bibinfo  {journal} {Prog.
				Part. Nucl. Phys.}\ }\textbf {\bibinfo {volume} {107}},\ \bibinfo {pages}
		{237} (\bibinfo {year} {2019})},\ \Eprint {http://arxiv.org/abs/1903.11976}
	{arXiv:1903.11976 [hep-ph]} \BibitemShut {NoStop}%
	\bibitem [{\citenamefont {Chen}\ \emph
		{et~al.}(2016{\natexlab{b}})\citenamefont {Chen}, \citenamefont {Chen},
		\citenamefont {Liu},\ and\ \citenamefont {Zhu}}]{Chen:2016qju}%
	\BibitemOpen
	\bibfield  {author} {\bibinfo {author} {\bibfnamefont {H.-X.}\ \bibnamefont
			{Chen}}, \bibinfo {author} {\bibfnamefont {W.}~\bibnamefont {Chen}}, \bibinfo
		{author} {\bibfnamefont {X.}~\bibnamefont {Liu}}, \ and\ \bibinfo {author}
		{\bibfnamefont {S.-L.}\ \bibnamefont {Zhu}},\ }\href {\doibase
		10.1016/j.physrep.2016.05.004} {\bibfield  {journal} {\bibinfo  {journal}
			{Phys. Rept.}\ }\textbf {\bibinfo {volume} {639}},\ \bibinfo {pages} {1}
		(\bibinfo {year} {2016}{\natexlab{b}})},\ \Eprint
	{http://arxiv.org/abs/1601.02092} {arXiv:1601.02092 [hep-ph]} \BibitemShut
	{NoStop}%
	\bibitem [{\citenamefont {Karliner}\ and\ \citenamefont
		{Rosner}(2021)}]{Karliner:2021xnq}%
	\BibitemOpen
	\bibfield  {author} {\bibinfo {author} {\bibfnamefont {M.}~\bibnamefont
			{Karliner}}\ and\ \bibinfo {author} {\bibfnamefont {J.~L.}\ \bibnamefont
			{Rosner}},\ }\href {\doibase 10.1016/j.scib.2021.04.013} {\bibfield
		{journal} {\bibinfo  {journal} {Sci. Bull.}\ }\textbf {\bibinfo {volume}
			{66}},\ \bibinfo {pages} {1257} (\bibinfo {year} {2021})},\ \Eprint
	{http://arxiv.org/abs/2104.15077} {arXiv:2104.15077 [hep-ph]} \BibitemShut
	{NoStop}%
	\bibitem [{\citenamefont {Wang}\ \emph {et~al.}(2020)\citenamefont {Wang},
		\citenamefont {Meng},\ and\ \citenamefont {Zhu}}]{Wang:2019nvm}%
	\BibitemOpen
	\bibfield  {author} {\bibinfo {author} {\bibfnamefont {B.}~\bibnamefont
			{Wang}}, \bibinfo {author} {\bibfnamefont {L.}~\bibnamefont {Meng}}, \ and\
		\bibinfo {author} {\bibfnamefont {S.-L.}\ \bibnamefont {Zhu}},\ }\href
	{\doibase 10.1103/PhysRevD.101.034018} {\bibfield  {journal} {\bibinfo
			{journal} {Phys. Rev. D}\ }\textbf {\bibinfo {volume} {101}},\ \bibinfo
		{pages} {034018} (\bibinfo {year} {2020})},\ \Eprint
	{http://arxiv.org/abs/1912.12592} {arXiv:1912.12592 [hep-ph]} \BibitemShut
	{NoStop}%
	\bibitem [{\citenamefont {Zou}(2021)}]{Zou:2021sha}%
	\BibitemOpen
	\bibfield  {author} {\bibinfo {author} {\bibfnamefont {B.-S.}\ \bibnamefont
			{Zou}},\ }\href {\doibase 10.1016/j.scib.2021.04.023} {\bibfield  {journal}
		{\bibinfo  {journal} {Sci. Bull.}\ }\textbf {\bibinfo {volume} {66}},\
		\bibinfo {pages} {1258} (\bibinfo {year} {2021})},\ \Eprint
	{http://arxiv.org/abs/2103.15273} {arXiv:2103.15273 [hep-ph]} \BibitemShut
	{NoStop}%
	\bibitem [{\citenamefont {Deng}(2022)}]{Deng:2022vkv}%
	\BibitemOpen
	\bibfield  {author} {\bibinfo {author} {\bibfnamefont {C.-R.}\ \bibnamefont
			{Deng}},\ }\href {\doibase 10.1103/PhysRevD.105.116021} {\bibfield  {journal}
		{\bibinfo  {journal} {Phys. Rev. D}\ }\textbf {\bibinfo {volume} {105}},\
		\bibinfo {pages} {116021} (\bibinfo {year} {2022})},\ \Eprint
	{http://arxiv.org/abs/2202.13570} {arXiv:2202.13570 [hep-ph]} \BibitemShut
	{NoStop}%
	\bibitem [{\citenamefont {Liu}\ \emph {et~al.}(2025{\natexlab{b}})\citenamefont
		{Liu}, \citenamefont {Pan}, \citenamefont {Liu}, \citenamefont {Wu},
		\citenamefont {Lu},\ and\ \citenamefont {Geng}}]{Liu:2024uxn}%
	\BibitemOpen
	\bibfield  {author} {\bibinfo {author} {\bibfnamefont {M.-Z.}\ \bibnamefont
			{Liu}}, \bibinfo {author} {\bibfnamefont {Y.-W.}\ \bibnamefont {Pan}},
		\bibinfo {author} {\bibfnamefont {Z.-W.}\ \bibnamefont {Liu}}, \bibinfo
		{author} {\bibfnamefont {T.-W.}\ \bibnamefont {Wu}}, \bibinfo {author}
		{\bibfnamefont {J.-X.}\ \bibnamefont {Lu}}, \ and\ \bibinfo {author}
		{\bibfnamefont {L.-S.}\ \bibnamefont {Geng}},\ }\href {\doibase
		10.1016/j.physrep.2024.12.001} {\bibfield  {journal} {\bibinfo  {journal}
			{Phys. Rept.}\ }\textbf {\bibinfo {volume} {1108}},\ \bibinfo {pages} {1}
		(\bibinfo {year} {2025}{\natexlab{b}})},\ \Eprint
	{http://arxiv.org/abs/2404.06399} {arXiv:2404.06399 [hep-ph]} \BibitemShut
	{NoStop}%
	\bibitem [{\citenamefont {Aaij}\ \emph
		{et~al.}(2023{\natexlab{b}})\citenamefont {Aaij} \emph
		{et~al.}}]{LHCb:2022sfr}%
	\BibitemOpen
	\bibfield  {author} {\bibinfo {author} {\bibfnamefont {R.}~\bibnamefont
			{Aaij}} \emph {et~al.} (\bibinfo {collaboration} {LHCb}),\ }\href {\doibase
		10.1103/PhysRevLett.131.041902} {\bibfield  {journal} {\bibinfo  {journal}
			{Phys. Rev. Lett.}\ }\textbf {\bibinfo {volume} {131}},\ \bibinfo {pages}
		{041902} (\bibinfo {year} {2023}{\natexlab{b}})},\ \Eprint
	{http://arxiv.org/abs/2212.02716} {arXiv:2212.02716 [hep-ex]} \BibitemShut
	{NoStop}%
	\bibitem [{\citenamefont {Aaij}\ \emph
		{et~al.}(2023{\natexlab{c}})\citenamefont {Aaij} \emph
		{et~al.}}]{LHCb:2022lzp}%
	\BibitemOpen
	\bibfield  {author} {\bibinfo {author} {\bibfnamefont {R.}~\bibnamefont
			{Aaij}} \emph {et~al.} (\bibinfo {collaboration} {LHCb}),\ }\href {\doibase
		10.1103/PhysRevD.108.012017} {\bibfield  {journal} {\bibinfo  {journal}
			{Phys. Rev. D}\ }\textbf {\bibinfo {volume} {108}},\ \bibinfo {pages}
		{012017} (\bibinfo {year} {2023}{\natexlab{c}})},\ \Eprint
	{http://arxiv.org/abs/2212.02717} {arXiv:2212.02717 [hep-ex]} \BibitemShut
	{NoStop}%
	\bibitem [{\citenamefont {Aaij}\ \emph {et~al.}(2024)\citenamefont {Aaij} \emph
		{et~al.}}]{LHCb:2024vfz}%
	\BibitemOpen
	\bibfield  {author} {\bibinfo {author} {\bibfnamefont {R.}~\bibnamefont
			{Aaij}} \emph {et~al.} (\bibinfo {collaboration} {LHCb}),\ }\href {\doibase
		10.1103/PhysRevLett.133.131902} {\bibfield  {journal} {\bibinfo  {journal}
			{Phys. Rev. Lett.}\ }\textbf {\bibinfo {volume} {133}},\ \bibinfo {pages}
		{131902} (\bibinfo {year} {2024})},\ \Eprint
	{http://arxiv.org/abs/2406.03156} {arXiv:2406.03156 [hep-ex]} \BibitemShut
	{NoStop}%
	\bibitem [{\citenamefont {Aaij}\ \emph {et~al.}(2025)\citenamefont {Aaij} \emph
		{et~al.}}]{LHCb:2024xyx}%
	\BibitemOpen
	\bibfield  {author} {\bibinfo {author} {\bibfnamefont {R.}~\bibnamefont
			{Aaij}} \emph {et~al.} (\bibinfo {collaboration} {LHCb}),\ }\href {\doibase
		10.1103/PhysRevLett.134.101901} {\bibfield  {journal} {\bibinfo  {journal}
			{Phys. Rev. Lett.}\ }\textbf {\bibinfo {volume} {134}},\ \bibinfo {pages}
		{101901} (\bibinfo {year} {2025})},\ \Eprint
	{http://arxiv.org/abs/2411.19781} {arXiv:2411.19781 [hep-ex]} \BibitemShut
	{NoStop}%
	\bibitem [{\citenamefont {Yang}\ \emph {et~al.}(2023)\citenamefont {Yang},
		\citenamefont {Xin},\ and\ \citenamefont
		{Wang}}]{doi:10.1142/S0217751X23500562}%
	\BibitemOpen
	\bibfield  {author} {\bibinfo {author} {\bibfnamefont {X.-S.}\ \bibnamefont
			{Yang}}, \bibinfo {author} {\bibfnamefont {Q.}~\bibnamefont {Xin}}, \ and\
		\bibinfo {author} {\bibfnamefont {Z.-G.}\ \bibnamefont {Wang}},\ }\href
	{\doibase 10.1142/S0217751X23500562} {\bibfield  {journal} {\bibinfo
			{journal} {International Journal of Modern Physics A}\ }\textbf {\bibinfo
			{volume} {38}},\ \bibinfo {pages} {2350056} (\bibinfo {year} {2023})},\
	\Eprint {http://arxiv.org/abs/https://doi.org/10.1142/S0217751X23500562}
	{https://doi.org/10.1142/S0217751X23500562} \BibitemShut {NoStop}%
	\bibitem [{\citenamefont {Lian}\ \emph {et~al.}(2024)\citenamefont {Lian},
		\citenamefont {Chen}, \citenamefont {Chen}, \citenamefont {Dai},\ and\
		\citenamefont {Steele}}]{Lian:2023cgs}%
	\BibitemOpen
	\bibfield  {author} {\bibinfo {author} {\bibfnamefont {D.-K.}\ \bibnamefont
			{Lian}}, \bibinfo {author} {\bibfnamefont {W.}~\bibnamefont {Chen}}, \bibinfo
		{author} {\bibfnamefont {H.-X.}\ \bibnamefont {Chen}}, \bibinfo {author}
		{\bibfnamefont {L.-Y.}\ \bibnamefont {Dai}}, \ and\ \bibinfo {author}
		{\bibfnamefont {T.~G.}\ \bibnamefont {Steele}},\ }\href {\doibase
		10.1140/epjc/s10052-023-12355-4} {\bibfield  {journal} {\bibinfo  {journal}
			{Eur. Phys. J. C}\ }\textbf {\bibinfo {volume} {84}},\ \bibinfo {pages} {1}
		(\bibinfo {year} {2024})},\ \Eprint {http://arxiv.org/abs/2302.01167}
	{arXiv:2302.01167 [hep-ph]} \BibitemShut {NoStop}%
	\bibitem [{\citenamefont {Chen}\ \emph {et~al.}(2017)\citenamefont {Chen},
		\citenamefont {Chen}, \citenamefont {Liu}, \citenamefont {Steele},\ and\
		\citenamefont {Zhu}}]{PhysRevD.95.114005}%
	\BibitemOpen
	\bibfield  {author} {\bibinfo {author} {\bibfnamefont {W.}~\bibnamefont
			{Chen}}, \bibinfo {author} {\bibfnamefont {H.-X.}\ \bibnamefont {Chen}},
		\bibinfo {author} {\bibfnamefont {X.}~\bibnamefont {Liu}}, \bibinfo {author}
		{\bibfnamefont {T.~G.}\ \bibnamefont {Steele}}, \ and\ \bibinfo {author}
		{\bibfnamefont {S.-L.}\ \bibnamefont {Zhu}},\ }\href {\doibase
		10.1103/PhysRevD.95.114005} {\bibfield  {journal} {\bibinfo  {journal} {Phys.
				Rev. D}\ }\textbf {\bibinfo {volume} {95}},\ \bibinfo {pages} {114005}
		(\bibinfo {year} {2017})}\BibitemShut {NoStop}%
	\bibitem [{\citenamefont {Ge}\ \emph {et~al.}(2022)\citenamefont {Ge},
		\citenamefont {Liu},\ and\ \citenamefont {Ke}}]{Ge:2022dsp}%
	\BibitemOpen
	\bibfield  {author} {\bibinfo {author} {\bibfnamefont {Y.-H.}\ \bibnamefont
			{Ge}}, \bibinfo {author} {\bibfnamefont {X.-H.}\ \bibnamefont {Liu}}, \ and\
		\bibinfo {author} {\bibfnamefont {H.-W.}\ \bibnamefont {Ke}},\ }\href
	{\doibase 10.1140/epjc/s10052-022-10923-8} {\bibfield  {journal} {\bibinfo
			{journal} {Eur. Phys. J. C}\ }\textbf {\bibinfo {volume} {82}},\ \bibinfo
		{pages} {955} (\bibinfo {year} {2022})},\ \Eprint
	{http://arxiv.org/abs/2207.09900} {arXiv:2207.09900 [hep-ph]} \BibitemShut
	{NoStop}%
	\bibitem [{\citenamefont {Liu}\ \emph {et~al.}(2020{\natexlab{a}})\citenamefont
		{Liu}, \citenamefont {Yan}, \citenamefont {Ke}, \citenamefont {Li},\ and\
		\citenamefont {Xie}}]{Liu:2020orv}%
	\BibitemOpen
	\bibfield  {author} {\bibinfo {author} {\bibfnamefont {X.-H.}\ \bibnamefont
			{Liu}}, \bibinfo {author} {\bibfnamefont {M.-J.}\ \bibnamefont {Yan}},
		\bibinfo {author} {\bibfnamefont {H.-W.}\ \bibnamefont {Ke}}, \bibinfo
		{author} {\bibfnamefont {G.}~\bibnamefont {Li}}, \ and\ \bibinfo {author}
		{\bibfnamefont {J.-J.}\ \bibnamefont {Xie}},\ }\href {\doibase
		10.1140/epjc/s10052-020-08762-6} {\bibfield  {journal} {\bibinfo  {journal}
			{Eur. Phys. J. C}\ }\textbf {\bibinfo {volume} {80}},\ \bibinfo {pages}
		{1178} (\bibinfo {year} {2020}{\natexlab{a}})},\ \Eprint
	{http://arxiv.org/abs/2008.07190} {arXiv:2008.07190 [hep-ph]} \BibitemShut
	{NoStop}%
	\bibitem [{\citenamefont {Wang}\ \emph {et~al.}(2024)\citenamefont {Wang},
		\citenamefont {Chen}, \citenamefont {Meng},\ and\ \citenamefont
		{Zhu}}]{PhysRevD.109.034027}%
	\BibitemOpen
	\bibfield  {author} {\bibinfo {author} {\bibfnamefont {B.}~\bibnamefont
			{Wang}}, \bibinfo {author} {\bibfnamefont {K.}~\bibnamefont {Chen}}, \bibinfo
		{author} {\bibfnamefont {L.}~\bibnamefont {Meng}}, \ and\ \bibinfo {author}
		{\bibfnamefont {S.-L.}\ \bibnamefont {Zhu}},\ }\href {\doibase
		10.1103/PhysRevD.109.034027} {\bibfield  {journal} {\bibinfo  {journal}
			{Phys. Rev. D}\ }\textbf {\bibinfo {volume} {109}},\ \bibinfo {pages}
		{034027} (\bibinfo {year} {2024})}\BibitemShut {NoStop}%
	\bibitem [{\citenamefont {He}\ \emph {et~al.}(2020)\citenamefont {He},
		\citenamefont {Wang},\ and\ \citenamefont {Zhu}}]{He:2020jna}%
	\BibitemOpen
	\bibfield  {author} {\bibinfo {author} {\bibfnamefont {X.-G.}\ \bibnamefont
			{He}}, \bibinfo {author} {\bibfnamefont {W.}~\bibnamefont {Wang}}, \ and\
		\bibinfo {author} {\bibfnamefont {R.}~\bibnamefont {Zhu}},\ }\href {\doibase
		10.1140/epjc/s10052-020-08597-1} {\bibfield  {journal} {\bibinfo  {journal}
			{Eur. Phys. J. C}\ }\textbf {\bibinfo {volume} {80}},\ \bibinfo {pages}
		{1026} (\bibinfo {year} {2020})},\ \Eprint {http://arxiv.org/abs/2008.07145}
	{arXiv:2008.07145 [hep-ph]} \BibitemShut {NoStop}%
	\bibitem [{\citenamefont {Karliner}\ and\ \citenamefont
		{Rosner}(2020{\natexlab{b}})}]{PhysRevD.102.094016}%
	\BibitemOpen
	\bibfield  {author} {\bibinfo {author} {\bibfnamefont {M.}~\bibnamefont
			{Karliner}}\ and\ \bibinfo {author} {\bibfnamefont {J.~L.}\ \bibnamefont
			{Rosner}},\ }\href {\doibase 10.1103/PhysRevD.102.094016} {\bibfield
		{journal} {\bibinfo  {journal} {Phys. Rev. D}\ }\textbf {\bibinfo {volume}
			{102}},\ \bibinfo {pages} {094016} (\bibinfo {year}
		{2020}{\natexlab{b}})}\BibitemShut {NoStop}%
	\bibitem [{\citenamefont {Yang}\ \emph {et~al.}(2021)\citenamefont {Yang},
		\citenamefont {Ping},\ and\ \citenamefont {Segovia}}]{PhysRevD.103.074011}%
	\BibitemOpen
	\bibfield  {author} {\bibinfo {author} {\bibfnamefont {G.}~\bibnamefont
			{Yang}}, \bibinfo {author} {\bibfnamefont {J.}~\bibnamefont {Ping}}, \ and\
		\bibinfo {author} {\bibfnamefont {J.}~\bibnamefont {Segovia}},\ }\href
	{\doibase 10.1103/PhysRevD.103.074011} {\bibfield  {journal} {\bibinfo
			{journal} {Phys. Rev. D}\ }\textbf {\bibinfo {volume} {103}},\ \bibinfo
		{pages} {074011} (\bibinfo {year} {2021})}\BibitemShut {NoStop}%
	\bibitem [{\citenamefont {Guo}\ \emph {et~al.}(2022)\citenamefont {Guo},
		\citenamefont {Li}, \citenamefont {Zhao},\ and\ \citenamefont
		{He}}]{PhysRevD.105.054018}%
	\BibitemOpen
	\bibfield  {author} {\bibinfo {author} {\bibfnamefont {T.}~\bibnamefont
			{Guo}}, \bibinfo {author} {\bibfnamefont {J.}~\bibnamefont {Li}}, \bibinfo
		{author} {\bibfnamefont {J.}~\bibnamefont {Zhao}}, \ and\ \bibinfo {author}
		{\bibfnamefont {L.}~\bibnamefont {He}},\ }\href {\doibase
		10.1103/PhysRevD.105.054018} {\bibfield  {journal} {\bibinfo  {journal}
			{Phys. Rev. D}\ }\textbf {\bibinfo {volume} {105}},\ \bibinfo {pages}
		{054018} (\bibinfo {year} {2022})}\BibitemShut {NoStop}%
	\bibitem [{\citenamefont {Wang}\ \emph {et~al.}(2021)\citenamefont {Wang},
		\citenamefont {Meng}, \citenamefont {Xiao}, \citenamefont {Oka},\ and\
		\citenamefont {Zhu}}]{Wang:2020prk}%
	\BibitemOpen
	\bibfield  {author} {\bibinfo {author} {\bibfnamefont {G.-J.}\ \bibnamefont
			{Wang}}, \bibinfo {author} {\bibfnamefont {L.}~\bibnamefont {Meng}}, \bibinfo
		{author} {\bibfnamefont {L.-Y.}\ \bibnamefont {Xiao}}, \bibinfo {author}
		{\bibfnamefont {M.}~\bibnamefont {Oka}}, \ and\ \bibinfo {author}
		{\bibfnamefont {S.-L.}\ \bibnamefont {Zhu}},\ }\href {\doibase
		10.1140/epjc/s10052-021-08978-0} {\bibfield  {journal} {\bibinfo  {journal}
			{Eur. Phys. J. C}\ }\textbf {\bibinfo {volume} {81}},\ \bibinfo {pages} {188}
		(\bibinfo {year} {2021})},\ \Eprint {http://arxiv.org/abs/2010.09395}
	{arXiv:2010.09395 [hep-ph]} \BibitemShut {NoStop}%
	\bibitem [{\citenamefont {Liu}\ \emph {et~al.}(2020{\natexlab{b}})\citenamefont
		{Liu}, \citenamefont {Xie},\ and\ \citenamefont
		{Geng}}]{PhysRevD.102.091502}%
	\BibitemOpen
	\bibfield  {author} {\bibinfo {author} {\bibfnamefont {M.-Z.}\ \bibnamefont
			{Liu}}, \bibinfo {author} {\bibfnamefont {J.-J.}\ \bibnamefont {Xie}}, \ and\
		\bibinfo {author} {\bibfnamefont {L.-S.}\ \bibnamefont {Geng}},\ }\href
	{\doibase 10.1103/PhysRevD.102.091502} {\bibfield  {journal} {\bibinfo
			{journal} {Phys. Rev. D}\ }\textbf {\bibinfo {volume} {102}},\ \bibinfo
		{pages} {091502(R)} (\bibinfo {year} {2020}{\natexlab{b}})}\BibitemShut
	{NoStop}%
	\bibitem [{\citenamefont {Kong}\ \emph {et~al.}(2021)\citenamefont {Kong},
		\citenamefont {Zhu}, \citenamefont {Song},\ and\ \citenamefont
		{He}}]{PhysRevD.104.094012}%
	\BibitemOpen
	\bibfield  {author} {\bibinfo {author} {\bibfnamefont {S.-Y.}\ \bibnamefont
			{Kong}}, \bibinfo {author} {\bibfnamefont {J.-T.}\ \bibnamefont {Zhu}},
		\bibinfo {author} {\bibfnamefont {D.}~\bibnamefont {Song}}, \ and\ \bibinfo
		{author} {\bibfnamefont {J.}~\bibnamefont {He}},\ }\href {\doibase
		10.1103/PhysRevD.104.094012} {\bibfield  {journal} {\bibinfo  {journal}
			{Phys. Rev. D}\ }\textbf {\bibinfo {volume} {104}},\ \bibinfo {pages}
		{094012} (\bibinfo {year} {2021})}\BibitemShut {NoStop}%
	\bibitem [{\citenamefont {Wang}\ and\ \citenamefont
		{Zhu}(2022)}]{Wang:2021lwy}%
	\BibitemOpen
	\bibfield  {author} {\bibinfo {author} {\bibfnamefont {B.}~\bibnamefont
			{Wang}}\ and\ \bibinfo {author} {\bibfnamefont {S.-L.}\ \bibnamefont {Zhu}},\
	}\href {\doibase 10.1140/epjc/s10052-022-10396-9} {\bibfield  {journal}
		{\bibinfo  {journal} {Eur. Phys. J. C}\ }\textbf {\bibinfo {volume} {82}},\
		\bibinfo {pages} {419} (\bibinfo {year} {2022})},\ \Eprint
	{http://arxiv.org/abs/2107.09275} {arXiv:2107.09275 [hep-ph]} \BibitemShut
	{NoStop}%
	\bibitem [{\citenamefont {Bulava}\ \emph {et~al.}(2019)\citenamefont {Bulava},
		\citenamefont {H{\"o}rz}, \citenamefont {Knechtli}, \citenamefont {Koch},
		\citenamefont {Moir}, \citenamefont {Morningstar},\ and\ \citenamefont
		{Peardon}}]{Bulava:2019iut}%
	\BibitemOpen
	\bibfield  {author} {\bibinfo {author} {\bibfnamefont {J.}~\bibnamefont
			{Bulava}}, \bibinfo {author} {\bibfnamefont {B.}~\bibnamefont {H{\"o}rz}},
		\bibinfo {author} {\bibfnamefont {F.}~\bibnamefont {Knechtli}}, \bibinfo
		{author} {\bibfnamefont {V.}~\bibnamefont {Koch}}, \bibinfo {author}
		{\bibfnamefont {G.}~\bibnamefont {Moir}}, \bibinfo {author} {\bibfnamefont
			{C.}~\bibnamefont {Morningstar}}, \ and\ \bibinfo {author} {\bibfnamefont
			{M.}~\bibnamefont {Peardon}},\ }\href {\doibase
		10.1016/j.physletb.2019.05.018} {\bibfield  {journal} {\bibinfo  {journal}
			{Phys. Lett. B}\ }\textbf {\bibinfo {volume} {793}},\ \bibinfo {pages} {493}
		(\bibinfo {year} {2019})},\ \Eprint {http://arxiv.org/abs/1902.04006}
	{arXiv:1902.04006 [hep-lat]} \BibitemShut {NoStop}%
	\bibitem [{\citenamefont {Bali}\ \emph {et~al.}(2005)\citenamefont {Bali},
		\citenamefont {Neff}, \citenamefont {Duessel}, \citenamefont {Lippert},\ and\
		\citenamefont {Schilling}}]{Bali:2005fu}%
	\BibitemOpen
	\bibfield  {author} {\bibinfo {author} {\bibfnamefont {G.~S.}\ \bibnamefont
			{Bali}}, \bibinfo {author} {\bibfnamefont {H.}~\bibnamefont {Neff}}, \bibinfo
		{author} {\bibfnamefont {T.}~\bibnamefont {Duessel}}, \bibinfo {author}
		{\bibfnamefont {T.}~\bibnamefont {Lippert}}, \ and\ \bibinfo {author}
		{\bibfnamefont {K.}~\bibnamefont {Schilling}} (\bibinfo {collaboration}
		{SESAM}),\ }\href {\doibase 10.1103/PhysRevD.71.114513} {\bibfield  {journal}
		{\bibinfo  {journal} {Phys. Rev. D}\ }\textbf {\bibinfo {volume} {71}},\
		\bibinfo {pages} {114513} (\bibinfo {year} {2005})},\ \Eprint
	{http://arxiv.org/abs/hep-lat/0505012} {arXiv:hep-lat/0505012} \BibitemShut
	{NoStop}%
	\bibitem [{\citenamefont {Kou}\ and\ \citenamefont {Chen}(2024)}]{Kou:2024dml}%
	\BibitemOpen
	\bibfield  {author} {\bibinfo {author} {\bibfnamefont {W.}~\bibnamefont
			{Kou}}\ and\ \bibinfo {author} {\bibfnamefont {X.}~\bibnamefont {Chen}},\
	}\href {\doibase 10.1016/j.physletb.2024.138942} {\bibfield  {journal}
		{\bibinfo  {journal} {Phys. Lett. B}\ }\textbf {\bibinfo {volume} {856}},\
		\bibinfo {pages} {138942} (\bibinfo {year} {2024})},\ \Eprint
	{http://arxiv.org/abs/2405.18697} {arXiv:2405.18697 [hep-ph]} \BibitemShut
	{NoStop}%
	\bibitem [{\citenamefont {Ma}\ \emph {et~al.}(2024)\citenamefont {Ma},
		\citenamefont {Wu}, \citenamefont {Meng}, \citenamefont {Chen},\ and\
		\citenamefont {Zhu}}]{Ma:2024vsi}%
	\BibitemOpen
	\bibfield  {author} {\bibinfo {author} {\bibfnamefont {Y.}~\bibnamefont
			{Ma}}, \bibinfo {author} {\bibfnamefont {W.-L.}\ \bibnamefont {Wu}}, \bibinfo
		{author} {\bibfnamefont {L.}~\bibnamefont {Meng}}, \bibinfo {author}
		{\bibfnamefont {Y.-K.}\ \bibnamefont {Chen}}, \ and\ \bibinfo {author}
		{\bibfnamefont {S.-L.}\ \bibnamefont {Zhu}},\ }\href {\doibase
		10.1103/PhysRevD.110.074026} {\bibfield  {journal} {\bibinfo  {journal}
			{Phys. Rev. D}\ }\textbf {\bibinfo {volume} {110}},\ \bibinfo {pages}
		{074026} (\bibinfo {year} {2024})},\ \Eprint
	{http://arxiv.org/abs/2408.00503} {arXiv:2408.00503 [hep-ph]} \BibitemShut
	{NoStop}%
	\bibitem [{\citenamefont {Wu}\ \emph {et~al.}(2024)\citenamefont {Wu},
		\citenamefont {Ma}, \citenamefont {Chen}, \citenamefont {Meng},\ and\
		\citenamefont {Zhu}}]{Wu:2024zbx}%
	\BibitemOpen
	\bibfield  {author} {\bibinfo {author} {\bibfnamefont {W.-L.}\ \bibnamefont
			{Wu}}, \bibinfo {author} {\bibfnamefont {Y.}~\bibnamefont {Ma}}, \bibinfo
		{author} {\bibfnamefont {Y.-K.}\ \bibnamefont {Chen}}, \bibinfo {author}
		{\bibfnamefont {L.}~\bibnamefont {Meng}}, \ and\ \bibinfo {author}
		{\bibfnamefont {S.-L.}\ \bibnamefont {Zhu}},\ }\href {\doibase
		10.1103/PhysRevD.110.094041} {\bibfield  {journal} {\bibinfo  {journal}
			{Phys. Rev. D}\ }\textbf {\bibinfo {volume} {110}},\ \bibinfo {pages}
		{094041} (\bibinfo {year} {2024})},\ \Eprint
	{http://arxiv.org/abs/2409.03373} {arXiv:2409.03373 [hep-ph]} \BibitemShut
	{NoStop}%
	\bibitem [{\citenamefont {Yang}\ \emph {et~al.}(2025)\citenamefont {Yang},
		\citenamefont {Ma}, \citenamefont {Wu},\ and\ \citenamefont
		{Zhu}}]{Yang:2025wqo}%
	\BibitemOpen
	\bibfield  {author} {\bibinfo {author} {\bibfnamefont {H.-M.}\ \bibnamefont
			{Yang}}, \bibinfo {author} {\bibfnamefont {Y.}~\bibnamefont {Ma}}, \bibinfo
		{author} {\bibfnamefont {W.-L.}\ \bibnamefont {Wu}}, \ and\ \bibinfo {author}
		{\bibfnamefont {S.-L.}\ \bibnamefont {Zhu}},\ }\href {\doibase
		10.1103/PhysRevD.111.074040} {\bibfield  {journal} {\bibinfo  {journal}
			{Phys. Rev. D}\ }\textbf {\bibinfo {volume} {111}},\ \bibinfo {pages}
		{074040} (\bibinfo {year} {2025})},\ \Eprint
	{http://arxiv.org/abs/2502.10798} {arXiv:2502.10798 [hep-ph]} \BibitemShut
	{NoStop}%
	\bibitem [{\citenamefont {Zheng}\ \emph {et~al.}(2026)\citenamefont {Zheng},
		\citenamefont {Ma},\ and\ \citenamefont {Zhu}}]{Zheng:2025uzy}%
	\BibitemOpen
	\bibfield  {author} {\bibinfo {author} {\bibfnamefont {X.-H.}\ \bibnamefont
			{Zheng}}, \bibinfo {author} {\bibfnamefont {Y.}~\bibnamefont {Ma}}, \ and\
		\bibinfo {author} {\bibfnamefont {S.-L.}\ \bibnamefont {Zhu}},\ }\href
	{\doibase 10.1103/2k7s-hm8b} {\bibfield  {journal} {\bibinfo  {journal}
			{Phys. Rev. D}\ }\textbf {\bibinfo {volume} {113}},\ \bibinfo {pages}
		{054027} (\bibinfo {year} {2026})},\ \Eprint
	{http://arxiv.org/abs/2510.01505} {arXiv:2510.01505 [hep-ph]} \BibitemShut
	{NoStop}%
	\bibitem [{\citenamefont {Wu}\ \emph {et~al.}(2025)\citenamefont {Wu},
		\citenamefont {Chen}, \citenamefont {Ma}, \citenamefont {Meng},\ and\
		\citenamefont {Zhu}}]{Wu:2025nbu}%
	\BibitemOpen
	\bibfield  {author} {\bibinfo {author} {\bibfnamefont {W.-L.}\ \bibnamefont
			{Wu}}, \bibinfo {author} {\bibfnamefont {Y.-K.}\ \bibnamefont {Chen}},
		\bibinfo {author} {\bibfnamefont {Y.}~\bibnamefont {Ma}}, \bibinfo {author}
		{\bibfnamefont {L.}~\bibnamefont {Meng}}, \ and\ \bibinfo {author}
		{\bibfnamefont {S.-L.}\ \bibnamefont {Zhu}},\ }\href {\doibase
		10.1016/j.jspc.2025.100184} {\bibfield  {journal} {\bibinfo  {journal} {J.
				Subatomic Part. Cosmol.}\ }\textbf {\bibinfo {volume} {4}},\ \bibinfo {pages}
		{100184} (\bibinfo {year} {2025})},\ \Eprint
	{http://arxiv.org/abs/2508.11161} {arXiv:2508.11161 [hep-ph]} \BibitemShut
	{NoStop}%
	\bibitem [{\citenamefont {Zhang}\ \emph
		{et~al.}(2025{\natexlab{a}})\citenamefont {Zhang}, \citenamefont {Liu},\ and\
		\citenamefont {Jia}}]{Zhang:2025dsx}%
	\BibitemOpen
	\bibfield  {author} {\bibinfo {author} {\bibfnamefont {W.-X.}\ \bibnamefont
			{Zhang}}, \bibinfo {author} {\bibfnamefont {W.-N.}\ \bibnamefont {Liu}}, \
		and\ \bibinfo {author} {\bibfnamefont {D.}~\bibnamefont {Jia}},\ }\href
	{\doibase 10.1103/7ly3-j3fl} {\bibfield  {journal} {\bibinfo  {journal}
			{Phys. Rev. D}\ }\textbf {\bibinfo {volume} {112}},\ \bibinfo {pages}
		{114039} (\bibinfo {year} {2025}{\natexlab{a}})},\ \Eprint
	{http://arxiv.org/abs/2510.04103} {arXiv:2510.04103 [hep-ph]} \BibitemShut
	{NoStop}%
	\bibitem [{\citenamefont {Liu}\ \emph {et~al.}(2025{\natexlab{c}})\citenamefont
		{Liu}, \citenamefont {Zhang},\ and\ \citenamefont {Dou}}]{Liu:2025fbe}%
	\BibitemOpen
	\bibfield  {author} {\bibinfo {author} {\bibfnamefont {W.-N.}\ \bibnamefont
			{Liu}}, \bibinfo {author} {\bibfnamefont {W.-X.}\ \bibnamefont {Zhang}}, \
		and\ \bibinfo {author} {\bibfnamefont {F.-Q.}\ \bibnamefont {Dou}},\ }\href
	{\doibase 10.1103/mwd4-l283} {\bibfield  {journal} {\bibinfo  {journal}
			{Phys. Rev. D}\ }\textbf {\bibinfo {volume} {111}},\ \bibinfo {pages}
		{114031} (\bibinfo {year} {2025}{\natexlab{c}})},\ \Eprint
	{http://arxiv.org/abs/2501.12727} {arXiv:2501.12727 [hep-ph]} \BibitemShut
	{NoStop}%
	\bibitem [{\citenamefont {Barnes}\ \emph {et~al.}(1983)\citenamefont {Barnes},
		\citenamefont {Close},\ and\ \citenamefont {de~Viron}}]{Barnes:1982tx}%
	\BibitemOpen
	\bibfield  {author} {\bibinfo {author} {\bibfnamefont {T.}~\bibnamefont
			{Barnes}}, \bibinfo {author} {\bibfnamefont {F.~E.}\ \bibnamefont {Close}}, \
		and\ \bibinfo {author} {\bibfnamefont {F.}~\bibnamefont {de~Viron}},\ }\href
	{\doibase 10.1016/0550-3213(83)90004-4} {\bibfield  {journal} {\bibinfo
			{journal} {Nucl. Phys. B}\ }\textbf {\bibinfo {volume} {224}},\ \bibinfo
		{pages} {241} (\bibinfo {year} {1983})}\BibitemShut {NoStop}%
	\bibitem [{\citenamefont {Chanowitz}\ and\ \citenamefont
		{Sharpe}(1983)}]{Chanowitz:1982qj}%
	\BibitemOpen
	\bibfield  {author} {\bibinfo {author} {\bibfnamefont {M.~S.}\ \bibnamefont
			{Chanowitz}}\ and\ \bibinfo {author} {\bibfnamefont {S.~R.}\ \bibnamefont
			{Sharpe}},\ }\href {\doibase 10.1016/0550-3213(83)90635-1} {\bibfield
		{journal} {\bibinfo  {journal} {Nucl. Phys. B}\ }\textbf {\bibinfo {volume}
			{222}},\ \bibinfo {pages} {211} (\bibinfo {year} {1983})},\ \bibinfo {note}
	{[Erratum: Nucl.Phys.B 228, 588--588 (1983)]}\BibitemShut {NoStop}%
	\bibitem [{\citenamefont {Jaffe}(1977{\natexlab{a}})}]{Jaffe:1976ig}%
	\BibitemOpen
	\bibfield  {author} {\bibinfo {author} {\bibfnamefont {R.~L.}\ \bibnamefont
			{Jaffe}},\ }\href {\doibase 10.1103/PhysRevD.15.267} {\bibfield  {journal}
		{\bibinfo  {journal} {Phys. Rev. D}\ }\textbf {\bibinfo {volume} {15}},\
		\bibinfo {pages} {267} (\bibinfo {year} {1977}{\natexlab{a}})}\BibitemShut
	{NoStop}%
	\bibitem [{\citenamefont {Jaffe}(1977{\natexlab{b}})}]{Jaffe:1976ih}%
	\BibitemOpen
	\bibfield  {author} {\bibinfo {author} {\bibfnamefont {R.~L.}\ \bibnamefont
			{Jaffe}},\ }\href {\doibase 10.1103/PhysRevD.15.281} {\bibfield  {journal}
		{\bibinfo  {journal} {Phys. Rev. D}\ }\textbf {\bibinfo {volume} {15}},\
		\bibinfo {pages} {281} (\bibinfo {year} {1977}{\natexlab{b}})}\BibitemShut
	{NoStop}%
	\bibitem [{\citenamefont {DeGrand}\ \emph {et~al.}(1975)\citenamefont
		{DeGrand}, \citenamefont {Jaffe}, \citenamefont {Johnson},\ and\
		\citenamefont {Kiskis}}]{DeGrand:1975cf}%
	\BibitemOpen
	\bibfield  {author} {\bibinfo {author} {\bibfnamefont {T.~A.}\ \bibnamefont
			{DeGrand}}, \bibinfo {author} {\bibfnamefont {R.~L.}\ \bibnamefont {Jaffe}},
		\bibinfo {author} {\bibfnamefont {K.}~\bibnamefont {Johnson}}, \ and\
		\bibinfo {author} {\bibfnamefont {J.~E.}\ \bibnamefont {Kiskis}},\ }\href
	{\doibase 10.1103/PhysRevD.12.2060} {\bibfield  {journal} {\bibinfo
			{journal} {Phys. Rev. D}\ }\textbf {\bibinfo {volume} {12}},\ \bibinfo
		{pages} {2060} (\bibinfo {year} {1975})}\BibitemShut {NoStop}%
	\bibitem [{\citenamefont {Zhang}\ \emph {et~al.}(2021)\citenamefont {Zhang},
		\citenamefont {Xu},\ and\ \citenamefont {Jia}}]{Zhang:2021yul}%
	\BibitemOpen
	\bibfield  {author} {\bibinfo {author} {\bibfnamefont {W.-X.}\ \bibnamefont
			{Zhang}}, \bibinfo {author} {\bibfnamefont {H.}~\bibnamefont {Xu}}, \ and\
		\bibinfo {author} {\bibfnamefont {D.}~\bibnamefont {Jia}},\ }\href {\doibase
		10.1103/PhysRevD.104.114011} {\bibfield  {journal} {\bibinfo  {journal}
			{Phys. Rev. D}\ }\textbf {\bibinfo {volume} {104}},\ \bibinfo {pages}
		{114011} (\bibinfo {year} {2021})},\ \Eprint
	{http://arxiv.org/abs/2109.07040} {arXiv:2109.07040 [hep-ph]} \BibitemShut
	{NoStop}%
	\bibitem [{\citenamefont {Yan}\ \emph {et~al.}(2023)\citenamefont {Yan},
		\citenamefont {Zhang},\ and\ \citenamefont {Jia}}]{Yan:2023lvm}%
	\BibitemOpen
	\bibfield  {author} {\bibinfo {author} {\bibfnamefont {T.-Q.}\ \bibnamefont
			{Yan}}, \bibinfo {author} {\bibfnamefont {W.-X.}\ \bibnamefont {Zhang}}, \
		and\ \bibinfo {author} {\bibfnamefont {D.}~\bibnamefont {Jia}},\ }\href
	{\doibase 10.1140/epjc/s10052-023-11956-3} {\bibfield  {journal} {\bibinfo
			{journal} {Eur. Phys. J. C}\ }\textbf {\bibinfo {volume} {83}},\ \bibinfo
		{pages} {810} (\bibinfo {year} {2023})},\ \Eprint
	{http://arxiv.org/abs/2304.01684} {arXiv:2304.01684 [hep-ph]} \BibitemShut
	{NoStop}%
	\bibitem [{\citenamefont {Liu}\ \emph {et~al.}(2025{\natexlab{d}})\citenamefont
		{Liu}, \citenamefont {Zhang},\ and\ \citenamefont {Dou}}]{mwd4-l283}%
	\BibitemOpen
	\bibfield  {author} {\bibinfo {author} {\bibfnamefont {W.-N.}\ \bibnamefont
			{Liu}}, \bibinfo {author} {\bibfnamefont {W.-X.}\ \bibnamefont {Zhang}}, \
		and\ \bibinfo {author} {\bibfnamefont {F.-Q.}\ \bibnamefont {Dou}},\ }\href
	{\doibase 10.1103/mwd4-l283} {\bibfield  {journal} {\bibinfo  {journal}
			{Phys. Rev. D}\ }\textbf {\bibinfo {volume} {111}},\ \bibinfo {pages}
		{114031} (\bibinfo {year} {2025}{\natexlab{d}})}\BibitemShut {NoStop}%
	\bibitem [{\citenamefont {An}\ \emph {et~al.}(2021)\citenamefont {An},
		\citenamefont {Chen}, \citenamefont {Liu},\ and\ \citenamefont
		{Liu}}]{PhysRevD.103.074006}%
	\BibitemOpen
	\bibfield  {author} {\bibinfo {author} {\bibfnamefont {H.-T.}\ \bibnamefont
			{An}}, \bibinfo {author} {\bibfnamefont {K.}~\bibnamefont {Chen}}, \bibinfo
		{author} {\bibfnamefont {Z.-W.}\ \bibnamefont {Liu}}, \ and\ \bibinfo
		{author} {\bibfnamefont {X.}~\bibnamefont {Liu}},\ }\href {\doibase
		10.1103/PhysRevD.103.074006} {\bibfield  {journal} {\bibinfo  {journal}
			{Phys. Rev. D}\ }\textbf {\bibinfo {volume} {103}},\ \bibinfo {pages}
		{074006} (\bibinfo {year} {2021})}\BibitemShut {NoStop}%
	\bibitem [{\citenamefont {Zhu}\ \emph {et~al.}(2024)\citenamefont {Zhu},
		\citenamefont {Zhang},\ and\ \citenamefont {Jia}}]{Zhu:2023lbx}%
	\BibitemOpen
	\bibfield  {author} {\bibinfo {author} {\bibfnamefont {Z.-H.}\ \bibnamefont
			{Zhu}}, \bibinfo {author} {\bibfnamefont {W.-X.}\ \bibnamefont {Zhang}}, \
		and\ \bibinfo {author} {\bibfnamefont {D.}~\bibnamefont {Jia}},\ }\href
	{\doibase 10.1140/epjc/s10052-024-12700-1} {\bibfield  {journal} {\bibinfo
			{journal} {Eur. Phys. J. C}\ }\textbf {\bibinfo {volume} {84}},\ \bibinfo
		{pages} {344} (\bibinfo {year} {2024})},\ \Eprint
	{http://arxiv.org/abs/2312.01908} {arXiv:2312.01908 [hep-ph]} \BibitemShut
	{NoStop}%
	\bibitem [{\citenamefont {Karliner}\ and\ \citenamefont
		{Rosner}(2014)}]{Karliner:2014gca}%
	\BibitemOpen
	\bibfield  {author} {\bibinfo {author} {\bibfnamefont {M.}~\bibnamefont
			{Karliner}}\ and\ \bibinfo {author} {\bibfnamefont {J.~L.}\ \bibnamefont
			{Rosner}},\ }\href {\doibase 10.1103/PhysRevD.90.094007} {\bibfield
		{journal} {\bibinfo  {journal} {Phys. Rev. D}\ }\textbf {\bibinfo {volume}
			{90}},\ \bibinfo {pages} {094007} (\bibinfo {year} {2014})},\ \Eprint
	{http://arxiv.org/abs/1408.5877} {arXiv:1408.5877 [hep-ph]} \BibitemShut
	{NoStop}%
	\bibitem [{\citenamefont {Karliner}\ and\ \citenamefont
		{Rosner}(2017{\natexlab{a}})}]{Karliner:2017elp}%
	\BibitemOpen
	\bibfield  {author} {\bibinfo {author} {\bibfnamefont {M.}~\bibnamefont
			{Karliner}}\ and\ \bibinfo {author} {\bibfnamefont {J.~L.}\ \bibnamefont
			{Rosner}},\ }\href {\doibase 10.1038/nature24289} {\bibfield  {journal}
		{\bibinfo  {journal} {Nature}\ }\textbf {\bibinfo {volume} {551}},\ \bibinfo
		{pages} {89} (\bibinfo {year} {2017}{\natexlab{a}})},\ \Eprint
	{http://arxiv.org/abs/1708.02547} {arXiv:1708.02547 [hep-ph]} \BibitemShut
	{NoStop}%
	\bibitem [{\citenamefont {Karliner}\ and\ \citenamefont
		{Rosner}(2017{\natexlab{b}})}]{Karliner:2017qjm}%
	\BibitemOpen
	\bibfield  {author} {\bibinfo {author} {\bibfnamefont {M.}~\bibnamefont
			{Karliner}}\ and\ \bibinfo {author} {\bibfnamefont {J.~L.}\ \bibnamefont
			{Rosner}},\ }\href {\doibase 10.1103/PhysRevLett.119.202001} {\bibfield
		{journal} {\bibinfo  {journal} {Phys. Rev. Lett.}\ }\textbf {\bibinfo
			{volume} {119}},\ \bibinfo {pages} {202001} (\bibinfo {year}
		{2017}{\natexlab{b}})},\ \Eprint {http://arxiv.org/abs/1707.07666}
	{arXiv:1707.07666 [hep-ph]} \BibitemShut {NoStop}%
	\bibitem [{\citenamefont {Karliner}\ and\ \citenamefont
		{Rosner}(2020{\natexlab{c}})}]{Karliner:2020vsi}%
	\BibitemOpen
	\bibfield  {author} {\bibinfo {author} {\bibfnamefont {M.}~\bibnamefont
			{Karliner}}\ and\ \bibinfo {author} {\bibfnamefont {J.~L.}\ \bibnamefont
			{Rosner}},\ }\href {\doibase 10.1103/PhysRevD.102.094016} {\bibfield
		{journal} {\bibinfo  {journal} {Phys. Rev. D}\ }\textbf {\bibinfo {volume}
			{102}},\ \bibinfo {pages} {094016} (\bibinfo {year} {2020}{\natexlab{c}})},\
	\Eprint {http://arxiv.org/abs/2008.05993} {arXiv:2008.05993 [hep-ph]}
	\BibitemShut {NoStop}%
	\bibitem [{\citenamefont {Zhang}\ \emph
		{et~al.}(2025{\natexlab{b}})\citenamefont {Zhang}, \citenamefont {Zhang},\
		and\ \citenamefont {Jia}}]{ljt6-cv33}%
	\BibitemOpen
	\bibfield  {author} {\bibinfo {author} {\bibfnamefont {K.-K.}\ \bibnamefont
			{Zhang}}, \bibinfo {author} {\bibfnamefont {W.-X.}\ \bibnamefont {Zhang}}, \
		and\ \bibinfo {author} {\bibfnamefont {D.}~\bibnamefont {Jia}},\ }\href
	{\doibase 10.1103/ljt6-cv33} {\bibfield  {journal} {\bibinfo  {journal}
			{Phys. Rev. D}\ }\textbf {\bibinfo {volume} {112}},\ \bibinfo {pages}
		{054008} (\bibinfo {year} {2025}{\natexlab{b}})}\BibitemShut {NoStop}%
	\bibitem [{\citenamefont {Liu}\ \emph {et~al.}(2023)\citenamefont {Liu},
		\citenamefont {Zhang},\ and\ \citenamefont {Jia}}]{PhysRevD.108.054019}%
	\BibitemOpen
	\bibfield  {author} {\bibinfo {author} {\bibfnamefont {X.-Y.}\ \bibnamefont
			{Liu}}, \bibinfo {author} {\bibfnamefont {W.-X.}\ \bibnamefont {Zhang}}, \
		and\ \bibinfo {author} {\bibfnamefont {D.}~\bibnamefont {Jia}},\ }\href
	{\doibase 10.1103/PhysRevD.108.054019} {\bibfield  {journal} {\bibinfo
			{journal} {Phys. Rev. D}\ }\textbf {\bibinfo {volume} {108}},\ \bibinfo
		{pages} {054019} (\bibinfo {year} {2023})}\BibitemShut {NoStop}%
	\bibitem [{\citenamefont {Zhang}\ \emph {et~al.}(2024)\citenamefont {Zhang},
		\citenamefont {Liu},\ and\ \citenamefont {Jia}}]{Zhang:2023teh}%
	\BibitemOpen
	\bibfield  {author} {\bibinfo {author} {\bibfnamefont {W.-X.}\ \bibnamefont
			{Zhang}}, \bibinfo {author} {\bibfnamefont {C.-L.}\ \bibnamefont {Liu}}, \
		and\ \bibinfo {author} {\bibfnamefont {D.}~\bibnamefont {Jia}},\ }\href
	{\doibase 10.1103/PhysRevD.109.114037} {\bibfield  {journal} {\bibinfo
			{journal} {Phys. Rev. D}\ }\textbf {\bibinfo {volume} {109}},\ \bibinfo
		{pages} {114037} (\bibinfo {year} {2024})},\ \Eprint
	{http://arxiv.org/abs/2312.12770} {arXiv:2312.12770 [hep-ph]} \BibitemShut
	{NoStop}%
	\bibitem [{\citenamefont {Ebert}\ \emph {et~al.}(2002)\citenamefont {Ebert},
		\citenamefont {Faustov}, \citenamefont {Galkin},\ and\ \citenamefont
		{Martynenko}}]{PhysRevD.66.014008}%
	\BibitemOpen
	\bibfield  {author} {\bibinfo {author} {\bibfnamefont {D.}~\bibnamefont
			{Ebert}}, \bibinfo {author} {\bibfnamefont {R.~N.}\ \bibnamefont {Faustov}},
		\bibinfo {author} {\bibfnamefont {V.~O.}\ \bibnamefont {Galkin}}, \ and\
		\bibinfo {author} {\bibfnamefont {A.~P.}\ \bibnamefont {Martynenko}},\ }\href
	{\doibase 10.1103/PhysRevD.66.014008} {\bibfield  {journal} {\bibinfo
			{journal} {Phys. Rev. D}\ }\textbf {\bibinfo {volume} {66}},\ \bibinfo
		{pages} {014008} (\bibinfo {year} {2002})}\BibitemShut {NoStop}%
	\bibitem [{\citenamefont {Brown}\ \emph {et~al.}(2014)\citenamefont {Brown},
		\citenamefont {Detmold}, \citenamefont {Meinel},\ and\ \citenamefont
		{Orginos}}]{PhysRevD.90.094507}%
	\BibitemOpen
	\bibfield  {author} {\bibinfo {author} {\bibfnamefont {Z.~S.}\ \bibnamefont
			{Brown}}, \bibinfo {author} {\bibfnamefont {W.}~\bibnamefont {Detmold}},
		\bibinfo {author} {\bibfnamefont {S.}~\bibnamefont {Meinel}}, \ and\ \bibinfo
		{author} {\bibfnamefont {K.}~\bibnamefont {Orginos}},\ }\href {\doibase
		10.1103/PhysRevD.90.094507} {\bibfield  {journal} {\bibinfo  {journal} {Phys.
				Rev. D}\ }\textbf {\bibinfo {volume} {90}},\ \bibinfo {pages} {094507}
		(\bibinfo {year} {2014})}\BibitemShut {NoStop}%
	\bibitem [{\citenamefont {Bali}(2001)}]{Bali:2000gf}%
	\BibitemOpen
	\bibfield  {author} {\bibinfo {author} {\bibfnamefont {G.~S.}\ \bibnamefont
			{Bali}},\ }\href {\doibase 10.1016/S0370-1573(00)00079-X} {\bibfield
		{journal} {\bibinfo  {journal} {Phys. Rept.}\ }\textbf {\bibinfo {volume}
			{343}},\ \bibinfo {pages} {1} (\bibinfo {year} {2001})},\ \Eprint
	{http://arxiv.org/abs/hep-ph/0001312} {arXiv:hep-ph/0001312} \BibitemShut
	{NoStop}%
	\bibitem [{\citenamefont {Friedberg}\ and\ \citenamefont
		{Lee}(1977)}]{Friedberg:1976eg}%
	\BibitemOpen
	\bibfield  {author} {\bibinfo {author} {\bibfnamefont {R.}~\bibnamefont
			{Friedberg}}\ and\ \bibinfo {author} {\bibfnamefont {T.~D.}\ \bibnamefont
			{Lee}},\ }\href {\doibase 10.1103/PhysRevD.15.1694} {\bibfield  {journal}
		{\bibinfo  {journal} {Phys. Rev. D}\ }\textbf {\bibinfo {volume} {15}},\
		\bibinfo {pages} {1694} (\bibinfo {year} {1977})}\BibitemShut {NoStop}%
	\bibitem [{\citenamefont {Theberge}\ \emph {et~al.}(1980)\citenamefont
		{Theberge}, \citenamefont {Thomas},\ and\ \citenamefont
		{Miller}}]{Theberge:1980ye}%
	\BibitemOpen
	\bibfield  {author} {\bibinfo {author} {\bibfnamefont {S.}~\bibnamefont
			{Theberge}}, \bibinfo {author} {\bibfnamefont {A.~W.}\ \bibnamefont
			{Thomas}}, \ and\ \bibinfo {author} {\bibfnamefont {G.~A.}\ \bibnamefont
			{Miller}},\ }\href {\doibase 10.1103/PhysRevD.22.2838} {\bibfield  {journal}
		{\bibinfo  {journal} {Phys. Rev. D}\ }\textbf {\bibinfo {volume} {22}},\
		\bibinfo {pages} {2838} (\bibinfo {year} {1980})},\ \bibinfo {note}
	{[Erratum: Phys.Rev.D 23, 2106 (1981)]}\BibitemShut {NoStop}%
	\bibitem [{\citenamefont {Hatsuda}\ and\ \citenamefont
		{Kunihiro}(1994)}]{Hatsuda:1994pi}%
	\BibitemOpen
	\bibfield  {author} {\bibinfo {author} {\bibfnamefont {T.}~\bibnamefont
			{Hatsuda}}\ and\ \bibinfo {author} {\bibfnamefont {T.}~\bibnamefont
			{Kunihiro}},\ }\href {\doibase 10.1016/0370-1573(94)90022-1} {\bibfield
		{journal} {\bibinfo  {journal} {Phys. Rept.}\ }\textbf {\bibinfo {volume}
			{247}},\ \bibinfo {pages} {221} (\bibinfo {year} {1994})},\ \Eprint
	{http://arxiv.org/abs/hep-ph/9401310} {arXiv:hep-ph/9401310} \BibitemShut
	{NoStop}%
	\bibitem [{\citenamefont {Erlich}\ \emph {et~al.}(2005)\citenamefont {Erlich},
		\citenamefont {Katz}, \citenamefont {Son},\ and\ \citenamefont
		{Stephanov}}]{Erlich:2005qh}%
	\BibitemOpen
	\bibfield  {author} {\bibinfo {author} {\bibfnamefont {J.}~\bibnamefont
			{Erlich}}, \bibinfo {author} {\bibfnamefont {E.}~\bibnamefont {Katz}},
		\bibinfo {author} {\bibfnamefont {D.~T.}\ \bibnamefont {Son}}, \ and\
		\bibinfo {author} {\bibfnamefont {M.~A.}\ \bibnamefont {Stephanov}},\ }\href
	{\doibase 10.1103/PhysRevLett.95.261602} {\bibfield  {journal} {\bibinfo
			{journal} {Phys. Rev. Lett.}\ }\textbf {\bibinfo {volume} {95}},\ \bibinfo
		{pages} {261602} (\bibinfo {year} {2005})},\ \Eprint
	{http://arxiv.org/abs/hep-ph/0501128} {arXiv:hep-ph/0501128} \BibitemShut
	{NoStop}%
	\bibitem [{\citenamefont {'t~Hooft}(1981)}]{tHooft:1981bkw}%
	\BibitemOpen
	\bibfield  {author} {\bibinfo {author} {\bibfnamefont {G.}~\bibnamefont
			{'t~Hooft}},\ }\href {\doibase 10.1016/0550-3213(81)90442-9} {\bibfield
		{journal} {\bibinfo  {journal} {Nucl. Phys. B}\ }\textbf {\bibinfo {volume}
			{190}},\ \bibinfo {pages} {455} (\bibinfo {year} {1981})}\BibitemShut
	{NoStop}%
	\bibitem [{\citenamefont {Weng}\ \emph {et~al.}(2019)\citenamefont {Weng},
		\citenamefont {Chen}, \citenamefont {Deng},\ and\ \citenamefont
		{Zhu}}]{Weng:2019ynv}%
	\BibitemOpen
	\bibfield  {author} {\bibinfo {author} {\bibfnamefont {X.-Z.}\ \bibnamefont
			{Weng}}, \bibinfo {author} {\bibfnamefont {X.-L.}\ \bibnamefont {Chen}},
		\bibinfo {author} {\bibfnamefont {W.-Z.}\ \bibnamefont {Deng}}, \ and\
		\bibinfo {author} {\bibfnamefont {S.-L.}\ \bibnamefont {Zhu}},\ }\href
	{\doibase 10.1103/PhysRevD.100.016014} {\bibfield  {journal} {\bibinfo
			{journal} {Phys. Rev. D}\ }\textbf {\bibinfo {volume} {100}},\ \bibinfo
		{pages} {016014} (\bibinfo {year} {2019})},\ \Eprint
	{http://arxiv.org/abs/1904.09891} {arXiv:1904.09891 [hep-ph]} \BibitemShut
	{NoStop}%
	\bibitem [{\citenamefont {Weng}\ \emph {et~al.}(2021)\citenamefont {Weng},
		\citenamefont {Chen}, \citenamefont {Deng},\ and\ \citenamefont
		{Zhu}}]{Weng:2020jao}%
	\BibitemOpen
	\bibfield  {author} {\bibinfo {author} {\bibfnamefont {X.-Z.}\ \bibnamefont
			{Weng}}, \bibinfo {author} {\bibfnamefont {X.-L.}\ \bibnamefont {Chen}},
		\bibinfo {author} {\bibfnamefont {W.-Z.}\ \bibnamefont {Deng}}, \ and\
		\bibinfo {author} {\bibfnamefont {S.-L.}\ \bibnamefont {Zhu}},\ }\href
	{\doibase 10.1103/PhysRevD.103.034001} {\bibfield  {journal} {\bibinfo
			{journal} {Phys. Rev. D}\ }\textbf {\bibinfo {volume} {103}},\ \bibinfo
		{pages} {034001} (\bibinfo {year} {2021})},\ \Eprint
	{http://arxiv.org/abs/2010.05163} {arXiv:2010.05163 [hep-ph]} \BibitemShut
	{NoStop}%
	\bibitem [{\citenamefont {C.}(1992)}]{gaoc1992}%
	\BibitemOpen
	\bibfield  {author} {\bibinfo {author} {\bibfnamefont {G.}~\bibnamefont
			{C.}},\ }\href@noop {} {\emph {\bibinfo {title} {Group Theory and its
				Application in Particle Physics}}}\ (\bibinfo  {publisher} {Higher Education
		Press},\ \bibinfo {address} {Beijing},\ \bibinfo {year} {1992})\BibitemShut
	{NoStop}%
	\bibitem [{\citenamefont {Weng}\ \emph {et~al.}(2022)\citenamefont {Weng},
		\citenamefont {Deng},\ and\ \citenamefont {Zhu}}]{Weng:2021ngd}%
	\BibitemOpen
	\bibfield  {author} {\bibinfo {author} {\bibfnamefont {X.-Z.}\ \bibnamefont
			{Weng}}, \bibinfo {author} {\bibfnamefont {W.-Z.}\ \bibnamefont {Deng}}, \
		and\ \bibinfo {author} {\bibfnamefont {S.-L.}\ \bibnamefont {Zhu}},\ }\href
	{\doibase 10.1103/PhysRevD.105.034026} {\bibfield  {journal} {\bibinfo
			{journal} {Phys. Rev. D}\ }\textbf {\bibinfo {volume} {105}},\ \bibinfo
		{pages} {034026} (\bibinfo {year} {2022})},\ \Eprint
	{http://arxiv.org/abs/2109.05243} {arXiv:2109.05243 [hep-ph]} \BibitemShut
	{NoStop}%
	\bibitem [{\citenamefont {An}\ \emph {et~al.}(2022)\citenamefont {An},
		\citenamefont {Chen},\ and\ \citenamefont {Liu}}]{An:2020vku}%
	\BibitemOpen
	\bibfield  {author} {\bibinfo {author} {\bibfnamefont {H.-T.}\ \bibnamefont
			{An}}, \bibinfo {author} {\bibfnamefont {K.}~\bibnamefont {Chen}}, \ and\
		\bibinfo {author} {\bibfnamefont {X.}~\bibnamefont {Liu}},\ }\href {\doibase
		10.1103/PhysRevD.105.034018} {\bibfield  {journal} {\bibinfo  {journal}
			{Phys. Rev. D}\ }\textbf {\bibinfo {volume} {105}},\ \bibinfo {pages}
		{034018} (\bibinfo {year} {2022})},\ \Eprint
	{http://arxiv.org/abs/2010.05014} {arXiv:2010.05014 [hep-ph]} \BibitemShut
	{NoStop}%
	\bibitem [{\citenamefont {Zhang}\ \emph {et~al.}(2023)\citenamefont {Zhang},
		\citenamefont {An},\ and\ \citenamefont {Jia}}]{Zhang:2023hmg}%
	\BibitemOpen
	\bibfield  {author} {\bibinfo {author} {\bibfnamefont {W.-X.}\ \bibnamefont
			{Zhang}}, \bibinfo {author} {\bibfnamefont {H.-T.}\ \bibnamefont {An}}, \
		and\ \bibinfo {author} {\bibfnamefont {D.}~\bibnamefont {Jia}},\ }\href
	{\doibase 10.1140/epjc/s10052-023-11845-9} {\bibfield  {journal} {\bibinfo
			{journal} {Eur. Phys. J. C}\ }\textbf {\bibinfo {volume} {83}},\ \bibinfo
		{pages} {727} (\bibinfo {year} {2023})},\ \Eprint
	{http://arxiv.org/abs/2304.14876} {arXiv:2304.14876 [hep-ph]} \BibitemShut
	{NoStop}%
	\bibitem [{\citenamefont {Jaffe}(1977{\natexlab{c}})}]{PhysRevD.15.267}%
	\BibitemOpen
	\bibfield  {author} {\bibinfo {author} {\bibfnamefont {R.~J.}\ \bibnamefont
			{Jaffe}},\ }\href {\doibase 10.1103/PhysRevD.15.267} {\bibfield  {journal}
		{\bibinfo  {journal} {Phys. Rev. D}\ }\textbf {\bibinfo {volume} {15}},\
		\bibinfo {pages} {267} (\bibinfo {year} {1977}{\natexlab{c}})}\BibitemShut
	{NoStop}%
\end{thebibliography}
\end{document}